\documentclass[structabstract]{aa}  
\usepackage{graphicx}
\usepackage{float}
\usepackage{indentfirst}
\usepackage{lscape}
\usepackage{longtable}
\usepackage{txfonts}
\begin{document}
   \title{A line confusion-limited millimeter survey of Orion KL}
   \subtitle{III. Sulfur oxide species}

   \author{G. B. Esplugues
          \inst{1},
          B. Tercero
          \inst{1},   
          J. Cernicharo  
          \inst{1},
          J. R. Goicoechea
          \inst{1},
          Aina Palau
          \inst{2}, 
          N. Marcelino
          \inst{3},
          \and
          T. A. Bell
          \inst{1}         
         }

   \institute{Centro de Astrobiolog\'ia (CSIC-INTA), Ctra. de Torrej\'on-Ajalvir, km. 4, E-28850 Torrej\'on de Ardoz, Madrid, Spain\\
              \email{espluguesbg@cab.inta-csic.es}
        \and
Institut de Ci$\grave{e}$ncies de l'Espai (CSIC-IEEC), Campus UAB-Facultat de Ciencies, Torre C5-parell 2, E-08193 Bellaterra, Barcelona, Spain.
         \and
                      National Radio Astronomy Observatory, 520 Edgemont Road, Charlottesville, VA 22903, USA. 
             }

   \date{Received ; accepted }

  \abstract   
   {We present a study of the sulfur-bearing species detected in a line confusion-limited survey towards Orion KL performed with the IRAM 30-m telescope in the frequency range 80-281 GHz.} 
   {This study is part of an analysis of the line survey divided into families of molecules. Our aim is to derive accurate physical conditions, as well as molecular abundances, in the different components of Orion KL from observed SO and SO$_2$ lines.}
   {As a starting point, we assumed LTE conditions obtain rotational temperatures. We then used a radiative transfer model, assuming either LVG or LTE excitation to derive column densities of these molecules in the different components of Orion KL.}
   {We have detected 68 lines of SO, $^{34}$SO, $^{33}$SO, and S$^{18}$O and 653 lines of SO$_2$, $^{34}$SO$_2$, $^{33}$SO$_2$, SO$^{18}$O, and SO$_2$ $\nu$$_2$=1. 
We provide column densities for all of them and also upper limits for the column densities of S$^{17}$O, $^{36}$SO, $^{34}$S$^{18}$O, SO$^{17}$O, and $^{34}$SO$_2$ $\nu$$_2$=1 and for several undetected sulfur-bearing species. In addition, we present 2$\arcmin$$\times$2$\arcmin$ maps around Orion IRc2 of SO$_2$ transitions with energies from 19 to 131 K and also maps with four transitions of SO, $^{34}$SO, and $^{34}$SO$_2$. We observe an elongation of the gas along the NE-SW direction. An unexpected emission peak appears at 20.5 km s$^{-1}$ in most lines of SO and SO$_2$. A study of the spatial distribution of this emission feature shows that it is a new component of a few arcseconds ($\sim$5$\arcsec$´´) in diameter, which lies $\sim$4$\arcsec$ west of IRc2. 
We suggest the emission from this feature is related to shocks associated to the $BN$ object.}
   {The highest column densities for SO and SO$_2$ are found in the high-velocity plateau (a region dominated by shocks) and in the hot core. These values are up to three orders of magnitude higher than the results for the ridge components. We also find high column densities for their isotopologues in both components. Therefore, we conclude that SO and SO$_2$ are good tracers, not only of regions affected by shocks, but also of regions with warm dense gas (hot cores).}

   \titlerunning{Survey towards Orion KL. III. Sulfur oxide species}
   \authorrunning{G. B. Esplugues et al.}
   \maketitle

\section{Introduction}

  The hot core phase of massive star formation shows a particularly rich chemistry that results from gas-phase chemical reactions and dust grain mantle evaporation. During cloud collapse, depletion of molecules onto dust surfaces takes place. When a new protostar forms, the surrounding gas and dust are heated and molecules sublimate from the grain mantles, giving rise to new species in the warm gas and to enhanced abundances of pre-existing species. The existence of molecular outflows and associated shocked regions also plays an important role in the chemical evolution, because they heat up the gas significantly and modify its chemistry.

Orion KL is the closest high-mass star-forming region ($\simeq$414 pc, Menten et al. 2007). It is one of the most studied regions owing its chemical complexity and high gas temperature, which lead to a dense and bright line spectrum. In the Orion KL cloud it is useful to differentiate five distinct components, characterized by different physical and chemical conditions (Blake et al. 1987, Persson et al. 2007, Tercero et al. 2010, and references therein): i) the hot core (HC) with 10$\arcsec$ diameter, which contains a high abundance of complex species (Wilson et al. 2000). It is characterized by line widths of 7$\leq$ $\Delta$v $\leq$ 15 km s$^{-1}$ at $v$$_{\mathrm{LSR}}$ $\simeq$ 5 km s$^{-1}$. It contains dense and warm gas with $T$$_{K}$$\simeq$200 K and $n$(H$_{2}$)$\simeq$10$^{7}$ cm$^{-3}$. ii) The plateau (PL), a component with 30$\arcsec$ diameter, is affected by shocks with typical line widths of $\Delta$v $\simeq$ 20-25 km s$^{-1}$ at v$_{\mathrm{LSR}}$ $\simeq$ 6 km/s. Typical temperatures and densities are $T$$_{K}$$\simeq$150 K and $n$(H$_{2}$)$\simeq$10$^{6}$ cm$^{-3}$, respectively. iii) The high velocity plateau, HVP, (component affected by shocks, with similar temperature and densitity to the PL) with line widths of $\Delta$v $\simeq$ 30-55 km s$^{-1}$ at v$_{\mathrm{LSR}}$ $\simeq$ 11 km s$^{-1}$. iv) The compact ridge (CR), with 15$\arcsec$ diameter, centered on v$_{\mathrm{LSR}}$ $\simeq$ 7.5 km s$^{-1}$ with line widths of $\sim$ 4 km s$^{-1}$. Temperatures are about 110 K and densities $\simeq$10$^{6}$ cm$^{-3}$. And v) an extended component, the extended ridge (ER) or ambient cloud, whose emission is characterized by low temperature and density (60 K and 10$^{5}$ cm$^{-3}$, respectively), and line widths similar to the compact ridge, but centered on a velocity of v$_{\mathrm{LSR}}$ $\simeq$ 9 km s$^{-1}$. 
The luminosity of the Orion Becklin-Neugebauer$/$Kleinmann-Low complex is $\sim$10$^{5}$ $L_\odot$ (Gezari et al. 1998). From the model proposed by Wynn-Williams et al. (1984) and without observational evidence, IRc2 was thought to be the main source of luminosity, heating, and dynamics within the region. However, with the detection of two radio continuum point sources, $B$ (coincident with the BN Object) and $I$ (centroid of the Orion SiO maser), it was concluded that the intrinsic luminosity of IRc2 is only a fraction ($L$$\simeq$1000 $L_\odot$) of the total luminosity of the complex (Gezari et al. 1998), with source $I$ being the main contributor.

In this paper, we continue our analysis of the line survey towards Orion IRc2 in the frequency range 80-281 GHz, first presented by Tercero et al. (2010). Here we concentrate on SO, SO$_2$, and their isotopologues; we model the different cloud components (hot core, plateau, ridge) and derive their physical and chemical conditions, such as column densities and temperatures. Since Gottlieb \& Ball (1973) discovered SO in Orion A, there have been many studies of this molecule, as well as SO$_2$, in this region, including studies of the gas kinematics (Plambeck et al. 1982), molecular abundances (Blake et al. 1987), and spatial distribution (Sutton et al. 1995). Also we find several interferometric studies of these two molecules such as those from Wright et al. (1996) and Beuther et al. (2005). sulfur-bearing species are especially sensitive to physical and chemical variations during the lifetime of a hot core (Viti et al. 2001), and therefore are considered good probes of their time evolution (Hatchell et al. 1998). As such, they can be used as tools for investigating the chemistry and physical properties of complex star-forming regions (SFRs) located in dense molecular clouds. On the other hand, it is known that some molecules (SiO, H$_2$CS, SO, SO$_2$) show increased abudances in regions affected by shocks (Bachiller et al. 1996) as a result of the action of outflows on the surrounding gas. The study of molecular lines from shocked areas provides valuable information about chemical processes and the physical conditions of the shocked components.

The observations are described in Sect. \ref{observations}. We present more than 700 detected lines of SO, SO$_2$, their isotopologues, and their vibrationally excited states. 
In Section \ref{results} we present the data and compute rotational temperatures as a first LTE approximation. In addition, we present maps of eight emission lines of SO$_2$, SO, $^{34}$SO$_2$, and $^{34}$SO in the 1.3 mm window, over a 2$\arcmin$$\times$2$\arcmin$ region around Orion IRc2 (Sect. \ref{section:maps}). 
Unlike other studies of SO, we use a non-LTE radiative transfer code (LVG) to derive physical and chemical parameters (Sect. \ref{section:analysis}). We provide column density calculations for SO and SO$_2$, and isotopic abundance ratios, which have been improved over previous works due to the much larger number of available lines and to the up-to-date information on the physical properties of the region and molecular constants.
Discussions on our results are included in Sect. \ref{section:discussion}, while Sect. \ref{section:summary} summarizes the main conclusions.

\section{Observations}
\label{observations}

  We continue our analysis of the line survey towards Orion IRc2 covering frequency ranges 80-115.5 GHz, 130-178 GHz, and 197-281 GHz, first presented by Tercero et al. (2010). The observations were carried out using the IRAM 30-m radiotelescope during September 2004 (1.3 mm and 3 mm windows), March 2005 (full 2 mm window), and April 2005 (completion of the 1.3 mm and 3 mm windows). Four SiS receivers operating at 1.3, 2, and 3 mm were used simultaneously, with image sideband rejections within $\sim$13 dB (1.3 mm receivers), 12-16 dB (2 mm receivers), and 20-27 dB (3 mm receivers). System temperatures were in the range 200-800 K for the 1.3 mm receivers, 200-500 K for the 2 mm receivers, and 100-350 K for the 3 mm receivers, depending on the particular frequency, weather conditions, and source elevation. For the spectra between 172-178 GHz, the system temperature was significantly higher, 1000-4000 K, owing proximity of the atmospheric water line at 183.31 GHz. The intensity scale was calibrated using two absorbers at different temperatures and using the atmospheric transmission model (ATM, Cernicharo 1985; Pardo et al. 2001).

\begin{table}
\caption{IRAM 30m telescope efficiency data along the covered frequency range.}               
\centering          
\label{table:tablaeficiencias}
\begin{tabular}{c c c}     
\hline\hline                    
Frequency (GHz) &  $\eta$$_{\mathrm{MB}}$ & HVPBW \\
(GHz) & & ($\arcsec$)\\
\hline                    
   86  & 0.82 & 29.0  \\
   100 & 0.79 & 22.0  \\
   145 & 0.74 & 17.0  \\
   170 & 0.70 & 14.5  \\
   210 & 0.62 & 12.0  \\
   235 & 0.57 & 10.5  \\
   260 & 0.52 & 9.5   \\
   279 & 0.48 & 9.0   \\
\hline                  
\end{tabular}
\end{table}

  Pointing and focus were regularly checked on the nearby quasars 0420-014 and 0528+134. Observations were made in the balanced wobbler-switching mode, with a wobbling frequency of 0.5 Hz and a beam throw in azimuth of $\pm$240$\arcsec$. No contamination from the off position affected our observations, except for a marginal amount at the lowest elevations (25$\degr$) for molecules showing low-$J$ emission along the extended ridge. Two filter banks with 512$\times$1 MHz channels and a correlator providing two 512 MHz bandwidths and 1.25 MHz resolution were used as backends. We pointed the observations towards IRc2 at $\alpha$(J2000)=5$^{h}$ 35$^{m}$ 14.5$^{s}$, $\delta$(J2000)=-5$\degr$ 22$\arcmin$ 30.0$\arcsec$. 

The data were processed using the IRAM GILDAS software\footnote{http://www.iram.fr/IRAMFR/GILDAS} (developed by the Institut de Radioastronomie Millim\'etrique). In our analysis we only considered lines with intensities $\geq$0.02 K, covering three or more channels.
Spectra with Gaussian line fits are shown in units of antenna temperature $T$$^{\star}_{\mathrm{A}}$ corrected for atmospheric absorption and spillover losses. Figures with results from LVG/LTE analysis are shown in units of main beam temperature $T$$_{\mathrm{MB}}$, which is defined as

\begin{equation}
T_{\mathrm{MB}}=\left(T^{\star}_{\mathrm{A}}/\eta_{\mathrm{MB}}\right),
\end{equation}

\noindent where $\eta$$_{\mathrm{MB}}$ is the main beam efficiency. Table \ref{table:tablaeficiencias} shows the half power beam width (HVPBW) and the mean beam efficiencies over the covered frequency range.
For further information about the data reduction and line identification, see Tercero et al. (2010).

We also used the 30-m telescope to map a 2$\arcmin$$\times$2$\arcmin$ region around IRc2 at 1.3 mm. In this two-dimensional (2D) line  survey (Marcelino et al., in prep.), we covered the 1.3 mm window using the nine pixel HERA receiver array (216--250 GHz) and the EMIR single-pixel heterodyne receivers (200--216 and 250-282 GHz). We also mapped a small fraction of the 3 mm band taking advantage of simultaneous observations with the E230 and E090 receivers.  Fully sampled maps over 140$\times$140 arcsec${^2}$, centered on the position of IRc2, were performed in the {\it on-the-fly} (OTF) mapping mode, scanning both in $\alpha$ and $\delta$ with a 4$\arcsec$ spacing, and using position-switching to an emission-free reference position at an offset (-600$\arcsec$, 0$\arcsec$) with respect to IRc2. The observations presented here were obtained in February and December 2008 (HERA), February 2010, and January 
2012 (EMIR). We usedlocal oscillator settings at frequencies of 109.983, 221.600, 226.100, 235.100, 239.100, and 258.000 GHz, depending on the observed transition. 
We used short Wobbler-switching observations on the central position with a slightly  different frequency for each setting, in order to remove all features arising from the image side band.
We used the WILMA spectrometer backend, with a total bandwidth of 4 GHz (EMIR), 1 GHz (HERA), and a spectral resolution of 2 MHz, corresponding to velocity resolutions of 5.4 km s$^{-1}$ at 3 mm and 2.7-2.3 at 1.3 mm. 
Weather conditions were the typically good winter conditions (with opacities $\sim0.1-0.2$ at 1.3 mm and 1.3-2 mm of 
precipitable water vapor) resulting in system temperatures of 230-250 K (EMIR) and 300-400 K (HERA), except for observations in February 2010, when conditions were $\tau$$\sim$0.3-0.4 and 5 mm of pwv. In this case, system temperatures of 150 K and 300-400 K were obtained at 110 and 239 GHz, respectively. Pointing was checked every hour on strong and nearby quasars, and found to have errors of typically less than 3-4 arcsec.

The basic data reduction consisted of fitting and removing first-order polynomial baselines, checking for image sideband contamination and emission from the reference position. HERA data needed further reduction analysis due to the different performance of each pixel in the array. Spectra from all pixels were averaged to obtain a uniform map gridding of 4$\arcsec$, taking their different flux calibration and internal pointing errors into account (see Marcelino et al. in prep. for details).

\section{Results}
\label{results}

 \begin{figure*}
   \centering
   \includegraphics[angle=-90,width=15.5cm]{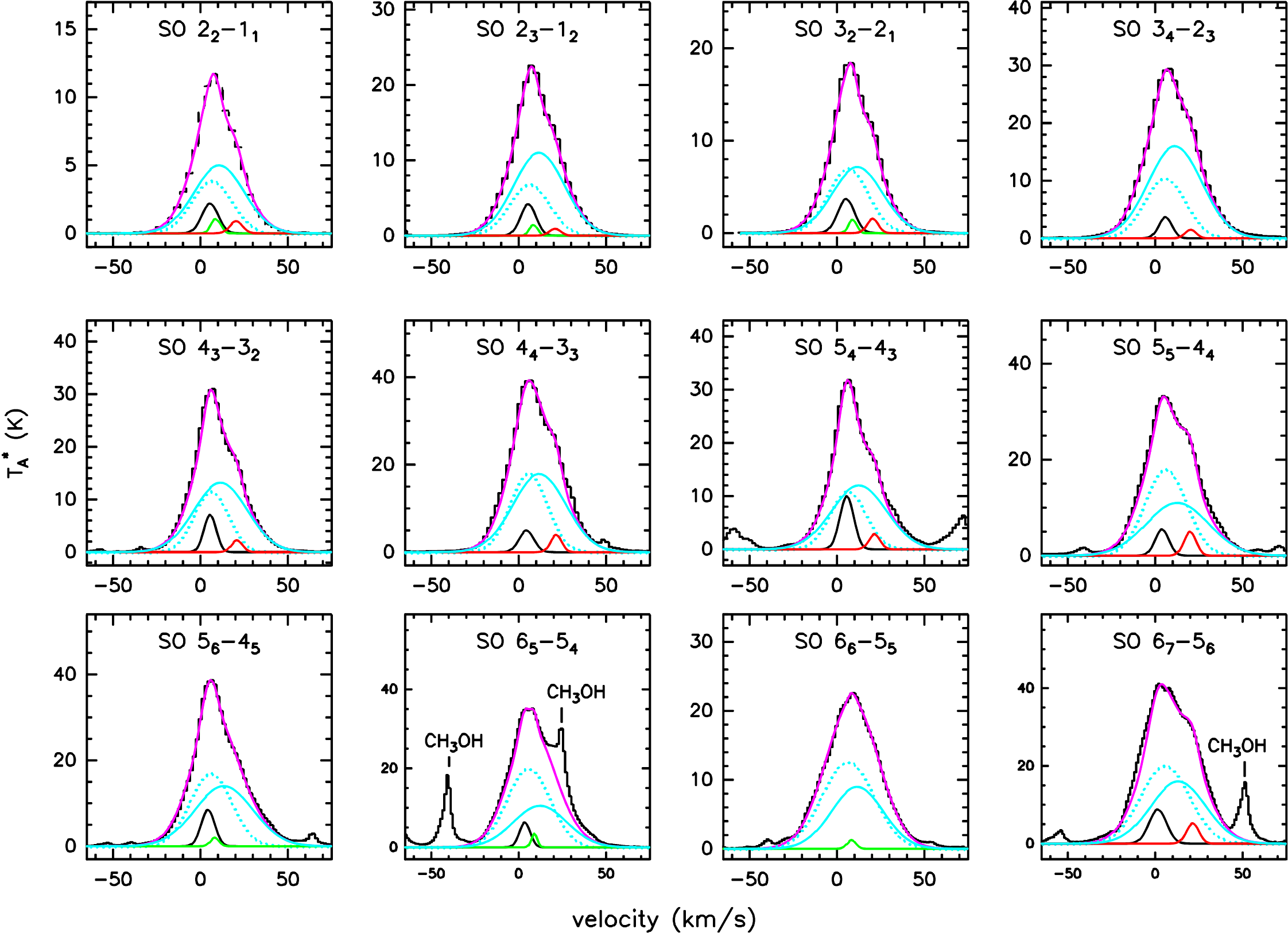}
   \caption{Gaussian fits to the observed SO lines. Dashed line for the plateau, cyan (solid line) for the high-velocity plateau, black for hot core, green for extended ridge, and red for the contribution of the component at 20.5 km s$^{-1}$. The total fit is shown in magenta. The data are the black histogram spectra.}            
   \label{figure:SO todos}
   \end{figure*}

 \begin{figure*}
   \centering
   \includegraphics[angle=0,width=16cm]{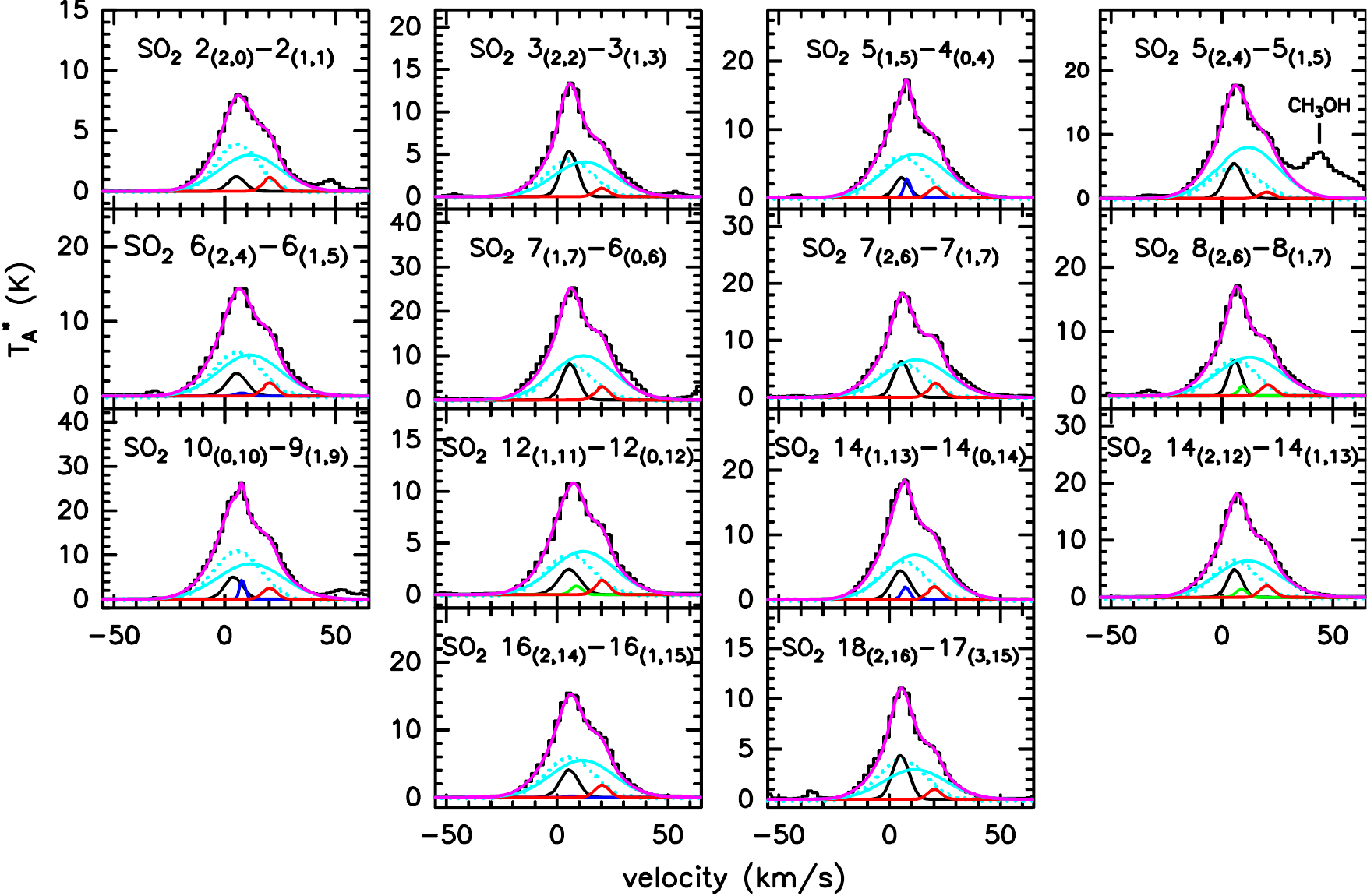}
   \caption{Gaussian fits for the SO$_{2}$ lines (2 mm data). The total fit is shown in magenta. Plateau is represented with the dashed line, high-velocity plateau in cyan (solid line), hot core in black, compact ridge in blue, extended ridge in green, and 20.5 km s$^{-1}$ component in red. The data are the black histogram spectra.}
   \label{figure:so2 2mm}
   \end{figure*}

 \begin{figure*}
   \centering
   \includegraphics[angle=0,width=16cm]{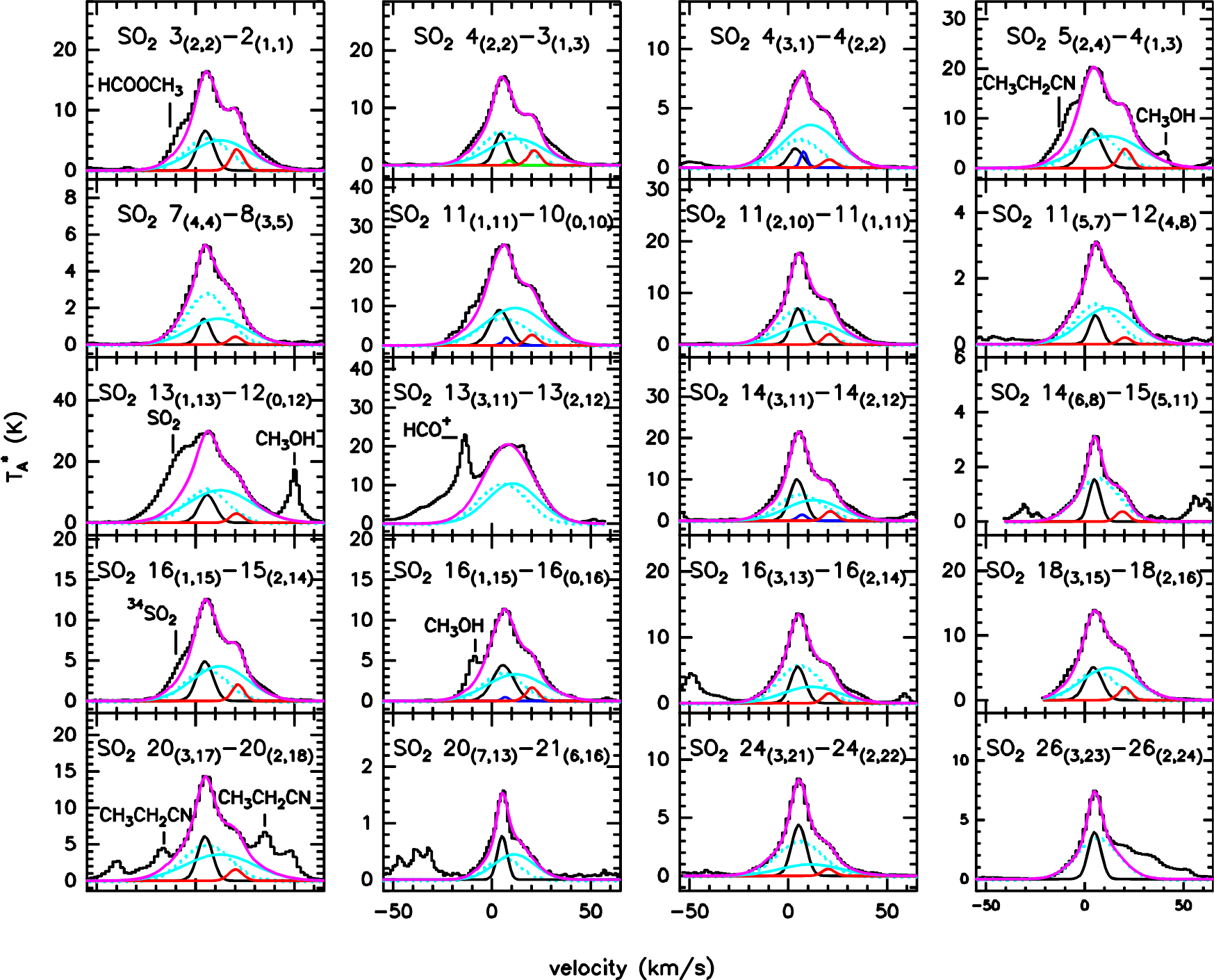}
   \caption{Gaussian fits for the SO$_{2}$ lines (1.3 mm data). The total fit is shown in magenta. Plateau is represented with the dashed line, high-velocity plateau in cyan (solid line), hot core in black, compact ridge in blue, extended ridge in green, and 20.5 km s$^{-1}$ component in red. The data are the black histogram spectra.}
   \label{figure:so2 1mm}
   \end{figure*}

 \begin{figure*}
   \centering
   \includegraphics[angle=270,width=14.0cm]{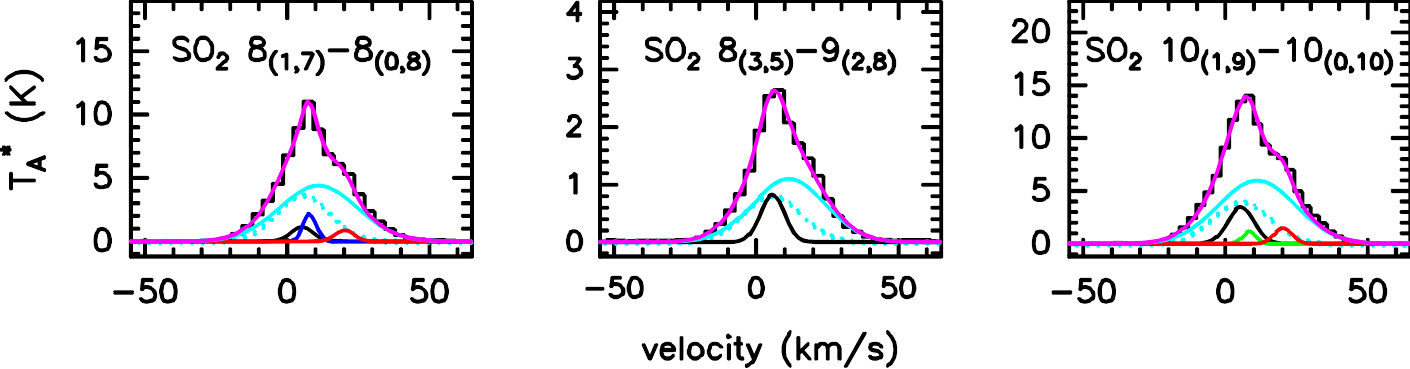}
   \caption{Gaussian fits for the SO$_{2}$ lines (3 mm data). The total fit is shown in magenta. Plateau is represented with the dashed line, high-velocity plateau in cyan (solid line), hot core in black, compact ridge in blue, extended ridge in green, and 20.5 km s$^{-1}$ component in red. The data are the black histogram spectra.}
   \label{figure:so2 3mm}
   \end{figure*}

  In total, the survey covers a bandwidth of 168 GHz, and of the 15,200 detected spectral features, about 10,700 have been identified and attributed to 45 molecules, including 191 isotopologues and vibrationally excited states (Tercero et al. 2010). 
We identify 20 lines of SO, 21 lines of $^{34}$SO, 13 lines of $^{33}$SO, and 14 lines of S$^{18}$O. We also detect 166 lines of SO$_{2}$, 129 lines of  $^{34}$SO$_{2}$, 85 lines of $^{33}$SO$_{2}$, 129 lines of SO$^{18}$O, 74 lines of SO$^{17}$O, and 78 lines of SO$_{2}$ v$_{2}$. Observed transitions of SO have a range of energy $E$$_{\mathrm{up}}$ between 16 and 100 K and a full width at half maximum (FWHM) of 40 km s$^{-1}$. In the case of SO$_{2}$ the energy range for the observed transitions is 12-1480 K and FWHM of 30-40 km s$^{-1}$ for transitions $J$$<$25 and FWHM $\sim$10-20 km s$^{-1}$ for transitions $J$$>$25. All these identifications are shown in Tables \ref{table:tab_so} and \ref{table:tab_so2} (see Appendix). Those tables list the spectroscopic parameters\footnote{Spectroscopic parameters for SO (dipole moment $\mu$=1.535 D) have been obtained from Clark \& DeLucia (1976), Tiemann (1982), Lovas et al. (1992), Cazzoli et al. (1994), Klaus et al. (1996), Bogey et al. (1997), Powell \& Lide (1964), and Martin-Drumel (2012). For $^{34}$SO and S$^{18}$O ($\mu$=1.535 D) from Tiemann (1974), Tiemann (1982), Lovas et al. (1992), Klaus et al. (1996), Bogey et al. (1982), and Powell \& Lide (1964). And for $^{33}$SO ($\mu$=1.535 D) from Klauss et al. (1996), Lovas et al. (1992), and Powell \& Lide (1964). In the case of SO$_{2}$ and $^{33}$SO$_{2}$ ($\mu$=1.633 D), the spectroscopic parameters were taken from M\"uller et al. (2000) and Patel et al. (1970). For $^{34}$SO$_{2}$ ($\mu$=1.633 D) from Belov et al. (1998) and Patel et al. (1979). For the isotopologue SO$^{18}$O ($\mu$$_{a}$=0.0328 D, $\mu$$_{b}$=1.633 D) obtained from Belov et al. (1998) and for SO$^{17}$O ($\mu$$_{a}$=0.02 D, $\mu$$_{b}$=1.633 D) from M\"uller et al. (2000). For the vibrational state SO$_{2}$ $\nu$$_{2}$=1, ($\mu$=1.626 D) from M\"uller \& Br\"unken (2005) and from Patel et al. (1979), and for $^{34}$SO$_{2}$ $\nu$$_{2}$=1, ($\mu$=1.626 D) from Maki \& Kuritsyn (1990).} of each transition, together with the observed line properties of the detected lines.

SO was the first molecule with a $^{3}$$\Sigma$ electronic ground-state detected in space by radio techniques (Gottlieb \& Ball 1973). Its rotational levels are characterized by the rotational angular momentum quantum number, $N$, and the total angular momentum quantum number, $J$, which includes the contribution of the angular momentum of two unpaired electrons.
For $^{33}$S and $^{17}$O, the nuclear quadrupolar momentum couples with the rotation to produce a hyperfine splitting of the rotational levels. Selection rules for the electric dipole transitions are: $\Delta$$N$=$\pm$1, $\Delta$$F$=0, $\pm$1, and $\Delta$$J$=0, $\pm$1, in the absence of external fields. In the case of intermediate coupling, transitions are allowed for $\Delta$$N$=$\pm$ 3. The magnetic dipole transitions occur with the selection rules: $\Delta$$N$=0, $\pm$2 and $\Delta$$J$=0, $\pm$1. However, these transitions are extremely weak compared to the electric dipole transitions. We have estimated that the magnetic dipole allowed transitions SO will have intensities $\sim$1-6 mK, i.e., lines within the confusion limit of Orion ($T$$^{\star}_{\mathrm{A}}$=20 mK).

SO$_{2}$ is an asymmetric molecule. The rotational energy levels are characterized by the three quantum numbers $J$, $K$$_{-1}$, and $K$$_{+1}$. Since triatomic molecules are planar, the dipole moment components can only occur in the a- and b-axis directions. The selection rules for a-type transitions are $\Delta$$J$=0, $\pm1$, $\Delta$$K$$_{-1}$=0, $\pm2$, and $\Delta$$K$$_{+1}$= $\pm1$, $\pm3$. For b-type transitions: $\Delta$$J$=0, $\pm1$, $\Delta$$K$$_{-1}$= $\pm1$, $\pm3$, and $\Delta$$K$$_{+1}$= $\mp1$, $\mp3$. SO$_{2}$ has its dipole moment along the b axis of the molecule. The nuclear quadrupolar momentum of $^{33}$S and $^{17}$O also couples with the rotation leading to hyperfine structure.

As a starting point, we fitted each observed line with Gaussian profiles using CLASS to derive the contribution of each cloud spectral component (see Sect. \ref{section:LTE_gauss}). We assumed that the emission is optically thin and the observed lines are thermalized at a given temperature that was derived from rotational diagrams (see Goldsmith \& Langer 1999), providing rotational temperatures for the different components of the cloud (Sect. \ref{subsection:diagrams}). In Sect. \ref{section:analysis} we use a radiative transfer code for a more advanced analysis of the LTE and non-LTE emission of SO and SO$_{2}$ species.

\subsection{Line profiles}
\label{section:LTE_gauss}

Figures \ref{figure:SO todos}, \ref{figure:so2 2mm}, \ref{figure:so2 1mm}, and \ref{figure:so2 3mm} show the line profiles of some observed transitions of SO and SO$_{2}$, together with Gaussian fit results. To avoid degeneration, we fixed radial velocities ($v$$_{\mathrm{LSR}}$) considering the characteristic values of each component of Orion KL. And we left the line width, the integrated intensity, and the antenna temperature as free parameters in the fits (in the results, we took the typical ranges of line widths into account for each component found in the bibliography, discarding those with large differences). In addition to the contribution from the usual components listed above, we also observe an unexpected emission peak centered on a velocity of $\simeq$20.5 km s$^{-1}$. We discuss its origin in Sects. \ref{section:feature} and \ref{section:nature}.

We have detected 20 rotational transitions of SO, eight of which are blended with lines of other species. Figure \ref{figure:SO todos} shows the contribution of the different cloud components to the emerging profile. 
We tried to fit the lines by considering only one plateau instead of two (at high and low velocity). First we centered this single plateau component on a velocity around 6-7 km s$^{-1}$, but with this we could not fit the part of the line profiles covering high ($>$20 km s$^{-1}$) velocities. With an increase in the line width of the fit, we reproduced this part of the profiles, but we overestimated the part of the lines that covers negative velocities. We found the opposite behavior if we fixed the single plateau component at higher velocities. Therefore we deduced that the best fits were obtained by considering two plateau components: one at low velocity, PL, ($\sim$6.5 km s$^{-1}$) and the other at high velocity, HVP, ($\sim$12 km s$^{-1}$).
We observe that the emission mainly arises from these both plateau components. For transitions with angular momentum quantum number $N$$\lesssim$5, the strongest emission is found from the HVP, while for $N$$\gtrsim$5 PL presents the highest contribution to the emission. The contribution from the ER is very small. 
In this LTE analysis of SO, we have not considered the compact ridge component since its contribution is difficult to distinguish from the contribution from the HC.
The parameters obtained from Gaussian fits for some of the SO lines are listed in Table \ref{table:so parameters gaussian} (see Appendix).

For SO$_{2}$ we have clearly identified 166 rotational transitions (see Figs. \ref{figure:so2 2mm}, \ref{figure:so2 1mm}, and \ref{figure:so2 3mm}). As indicated by their broad line profiles, we see that most of the emission comes from the plateau components, especially for transitions $J$$<$20 where the HVP plays an important role. However, for transitions $J$$>$20 the HC dominates. The parameters obtained for each component from Gaussian fits for SO$_{2}$ lines are listed in Tables \ref{table:so2 parameters gaussianI} and \ref{table:so2 parameters gaussianII} (see Appendix).
Comparing this with SO lines, we see that the line profiles of both species are very similar.

\subsection{The 20.5 km s$^{-1}$ feature and absorption at 15 km s$^{-1}$}
\label{section:feature}

\begin{figure}[h!]
   \centering 
   \includegraphics[angle=-90,width=7cm]{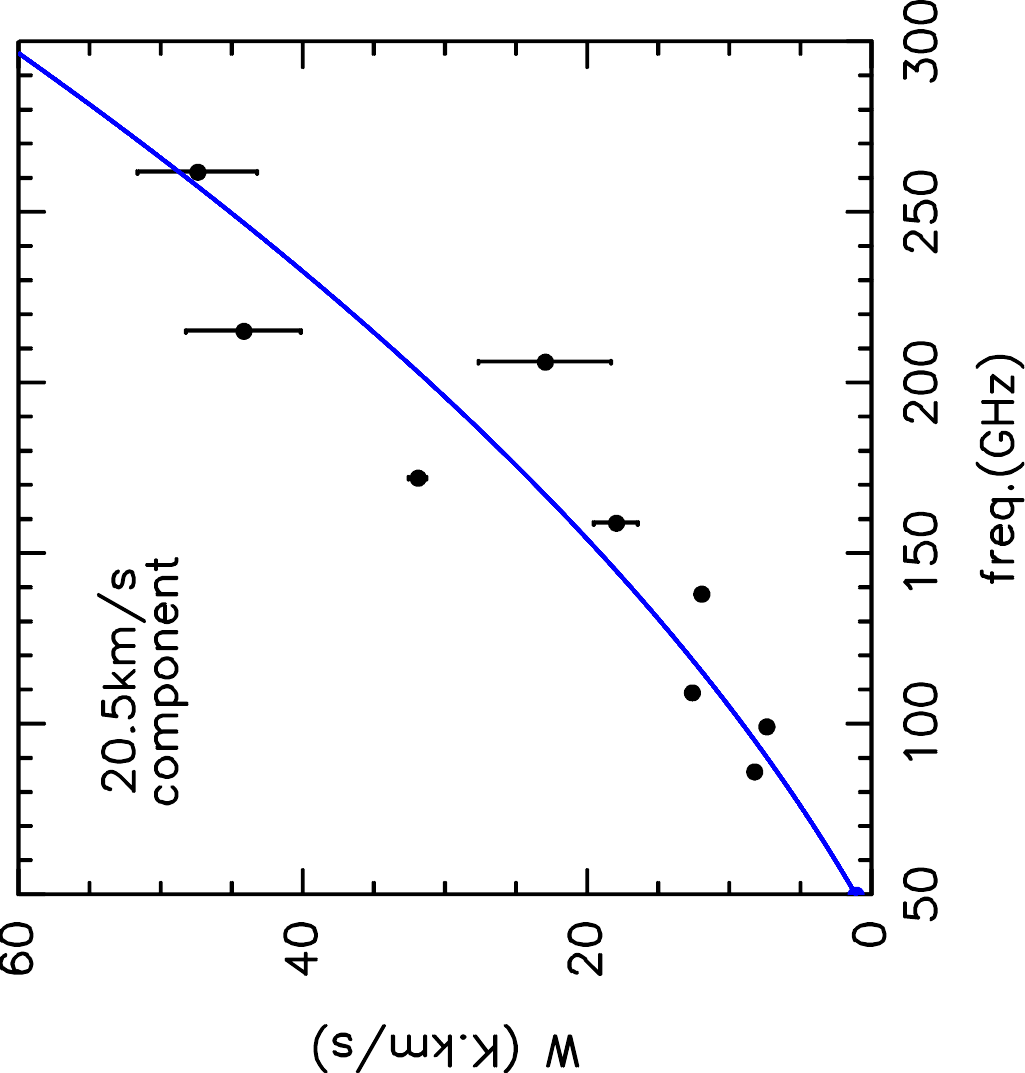}
   \caption{Integrated line intensities of the 20.5 km s$^{-1}$ component as a function of the line frequency for SO transitions.}
   \label{figure: 20.5 study}
   \end{figure}

Unlike the other species detected in the survey, we clearly observe an emission peak at $\simeq$20.5 km s$^{-1}$ in most SO and SO$_2$ lines. From Gaussian fits, we find that the line widths are $\sim$7.5 km s$^{-1}$. This feature presents an increase in the line intensity with frequency (i.e., with smaller beam). Figure \ref{figure: 20.5 study} shows the integrated intensity, $W$, as a function of the frequency for the 20.5 km s$^{-1}$ component for SO lines. We see that, for the same line width, $W$ increases with frequency with a dependence as $\nu$$^{2}$. This means that the size of the region responsible for this velocity componenent is only a few arcseconds in diameter ($<$9$\arcsec$, smaller than the IRAM 30-m beam at the highest frequencies). The emission of this component could come from the interaction of the outflow with the ambient cloud; however, it should be noted that while in Orion-KL the $v$$_{\mathrm{LSR}}$ of the different components varies approximately between 3 and 10 km s$^{-1}$, Scoville et al. (1983) inferred that the $BN$ (Becklin-Neugebauer) object in Orion presents a significantly higher LSR velocity of around 21 km s$^{-1}$, therefore another possibility is that this feature located at 20.5 km s$^{-1}$ arises from the $BN$ source.

On the other hand, between the main emission peak of the line profiles of SO and SO$_2$ and this emission peak at 20.5 km s$^{-1}$, we observe a dip at $\simeq$15 km s$^{-1}$. Tercero et al. (2011) also find a velocity component at 15.5 km s$^{-1}$ in SiS emission lines ($\nu$=0), and one component of the SiO maser emission ($\nu$=1). Since the opacity is high for some lines of SO and SO$_2$, as well as for SiO, another possibility is that this dip may be the result of self-absorption. This could suggest that the SO and SO$_2$ dips at 15.5 km s$^{-1}$ are produced by the same gas observed in SiS. In Section \ref{section:discussion} we draw firmer conclusions about the origin of the dip at 15.5 km s$^{-1}$ and of the peak at 20.5 km s$^{-1}$ from maps of SO and SO$_2$.

\subsection{Rotational diagrams}
\label{subsection:diagrams}

The results from the Gaussian fits have been used to build rotational diagrams for each species. This method involves the representation of the upper level column density normalized by the statistical weight of each rotational level versus the upper level energy, assuming optically thin emission filling the beam (see, e.g., Goldsmith \& Langer et al 1999). The expression used to obtain the rotational diagrams, taking an optical depth ($C$$_{\mathrm{\tau}}$) different to unity into account, is

\textbf{
\begin{equation}
\mathrm{ln}({\gamma_{\mathrm{u}}W/g_{\mathrm{u}}})=\mathrm{ln}(N)-\mathrm{ln}(C_{\mathrm{\tau}})-\mathrm{ln}(Z)-(E_{\mathrm{u}}/kT),
\end{equation}
}

\noindent where $W$ is the integrated line intensity, $g$$_{\mathrm{u}}$ is the statistical weight of each level, $N$ the column density, $Z$ the partition function, $E$$_{\mathrm{u}}$ the level energy, $k$ the Boltzman constant, $T$ the temperature considering local thermodynamic equilibrium, and $\gamma$$_{\mathrm{u}}$ a constant that depends on the transition frequency and the Einstein coefficient $A$$_{\mathrm{ul}}$ (see Goldsmith \& Langer 1999 for more details).
Each cloud component is considered separately in the analysis. The rotational diagrams, shown in Figs. \ref{figure:rotational diagrams I} and \ref{figure:rotational diagrams II} (see Appendix), were obtained considering only lines without contamination from other species. The rotational temperatures obtained from SO$_{2}$ lines are plateau (PL)=120$\pm$20 K, hot core (HC)=190$\pm$60 K, high-velocity plateau (HVP)=110$\pm$20 K, compact ridge (CR)=80$\pm$30 K, extended ridge (ER)=83$\pm$40 and 20.5 km s$^{-1}$ component=90$\pm$20 K. The results from SO lines are (PL)=130$\pm$20 K, (HC)=288$\pm$90 K, (HVP)=111$\pm$15 K, (ER)=107$\pm$40 K, and 20.5 km s$^{-1}$ component=51$\pm$10 K. These results are shown in Table \ref{table:rotational} (see Appendix), together with the derived column densities and the optical depths. For each component, the obtained rotational temperatures for both molecules are consistent with each other, except for the HC where we obtain a large difference between both temperatures. 
This could indicate a temperature gradient in the hot core, or simply that the obtained rotational temperature could be overestimated due to the high scatter in the SO data. It would be necessary to have values of this species at higher energies in order to obtain firmer conlcusions.  
We observe that HVP presents a similar rotational temperature to the component of the plateau with lower velocity (PL). 
For the CR, we obtained a low temperature, probably due to the beam dilution, which is not corrected for in the rotational diagrams. 
From the diagrams, we also deduce that the new component at 20.5 km s$^{-1}$ is not a very warm region.

\subsection{2\arcmin$\times$2\arcmin maps around IRc2}
\label{section:maps}

\begin{figure*}
   \centering 
   \includegraphics[angle=-90,width=18.2cm]{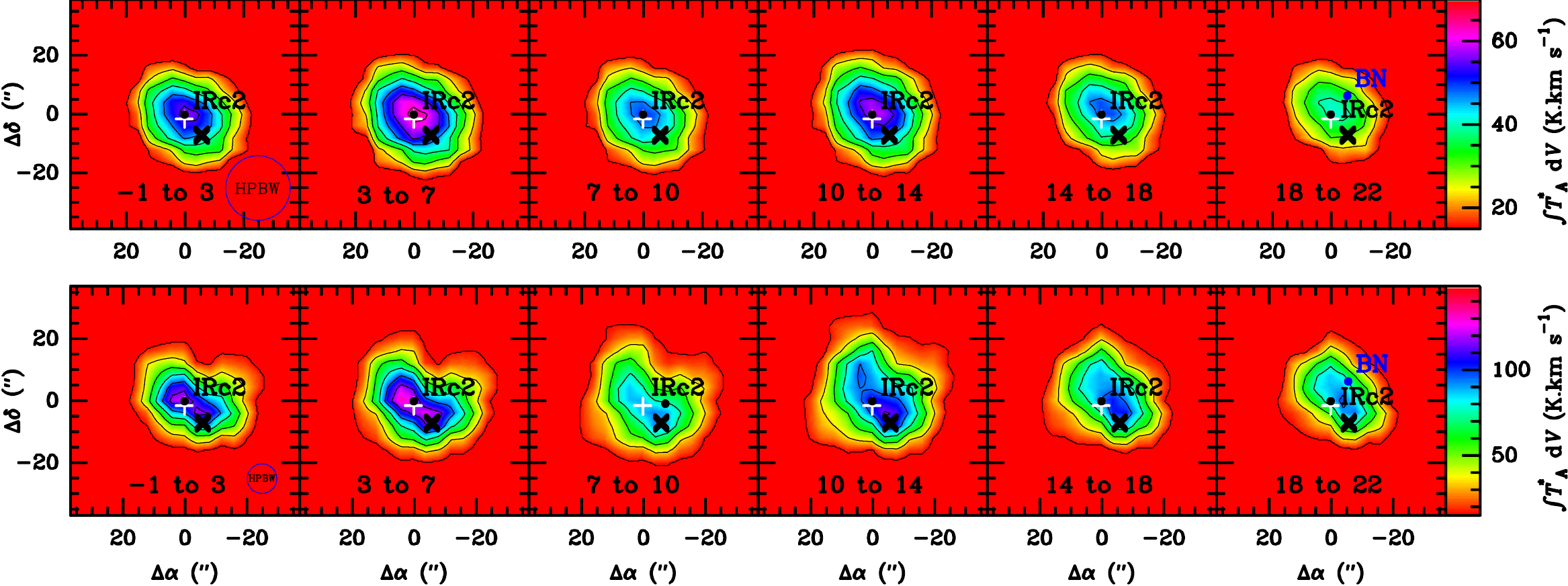}
   \caption{SO-integrated intensity maps over different velocity ranges (indicated at the bottom of each panel in km s$^{-1}$). Row 1 shows the transition 3$_{2}$-2$_{1}$ ($E$$_{up}$=21.1 K, $A$$_{\mathrm{ul}}$=1.1x10$^{-5}$ s$^{-1}$, $S$=1.5). The interval between contours is 8 K km s$^{-1}$ and the minimum contour is 15 K km s$^{-1}$. Row 2 shows the transition 6$_{6}$-5$_{5}$ ($E$$_{\mathrm{up}}$=56.5 K, $A$$_{\mathrm{ul}}$=2.2x10$^{-4}$ s$^{-1}$, $S$=5.8). The interval between contours is 20 K km s$^{-1}$ and the minimum contour is 15 K km s$^{-1}$. The white cross indicates the position of the hot core and the black cross the position of the compact ridge.}
    \label{figure:so_maps}
   \end{figure*}

\begin{figure*}
   \centering 
   \includegraphics[angle=-90,width=18.2cm]{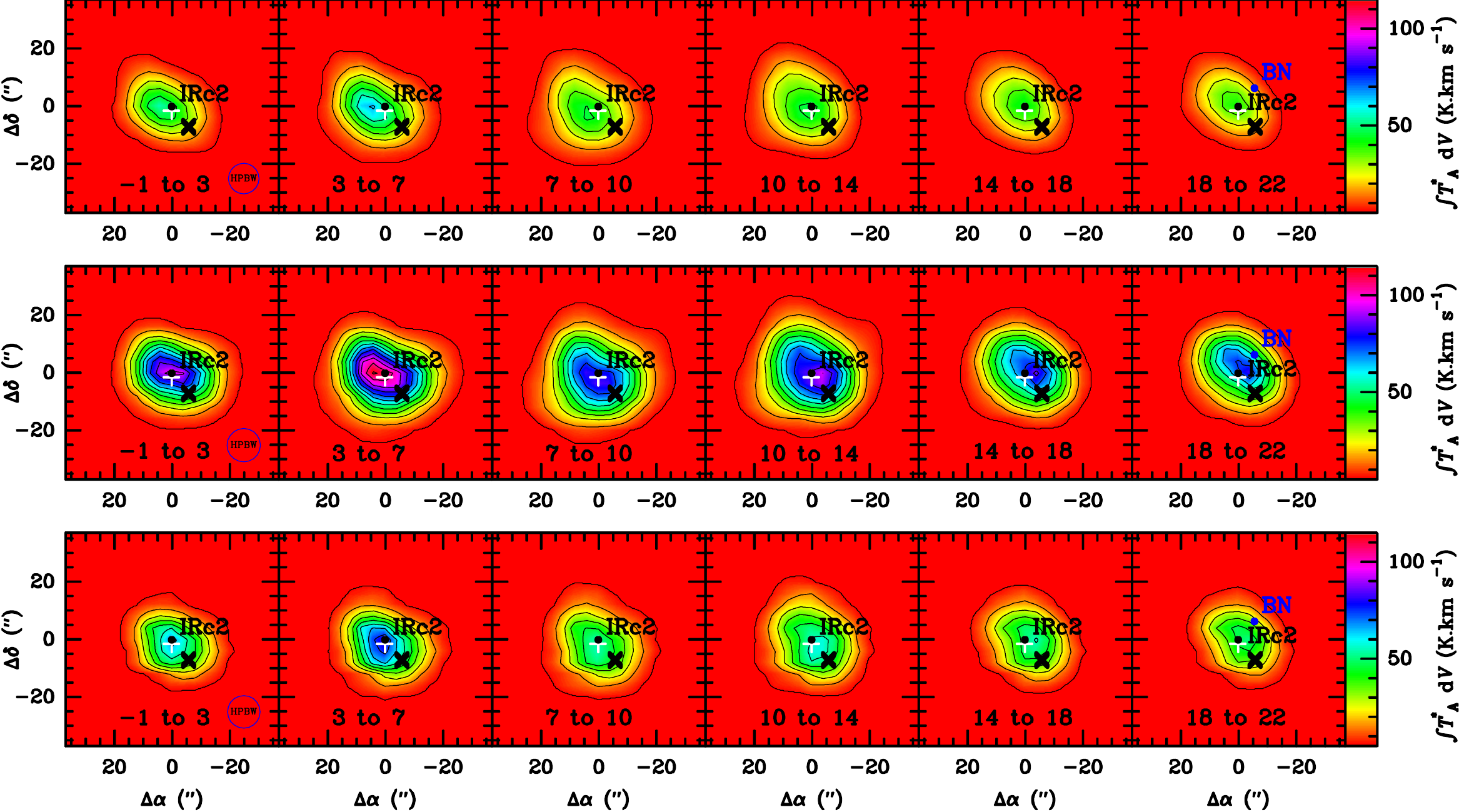}
   \caption{SO$_{2}$-integrated intensity maps over different velocity ranges (indicated at the bottom of each panel in km s$^{-1}$). Row 1 shows the transition 4$_{(2,2)}$-3$_{(1,3)}$ (energy $E$$_{\mathrm{up}}$=19 K, Einstein coefficient $A$$_{\mathrm{ul}}$=7.7x10$^{-5}$ s$^{-1}$, and line strength $S$=1.7), row 2 the transition 11$_{(1,11)}$-10$_{(0,10)}$ ($E$$_{\mathrm{up}}$=60.4 K, $A$$_{\mathrm{ul}}$=1.1x10$^{-4}$ s$^{-1}$, $S$=7.7), and the last row shows the transition 14$_{(3,11)}$-14$_{(2,12)}$ ($E$$_{\mathrm{up}}$=119 K, $A$$_{\mathrm{ul}}$=1.1x10$^{-4}$ s$^{-1}$, $S$=8.1). The interval between contours is 10 K km s$^{-1}$, the minimum contour is 5 K km s$^{-1}$ and the maximum 155 K km s$^{-1}$. The white cross indicates the position of the hot core and the black cross the position of the compact ridge.}
    \label{figure:so2_maps}
   \end{figure*}

\begin{figure*}
   \centering 
   \includegraphics[angle=-90,width=18.2cm]{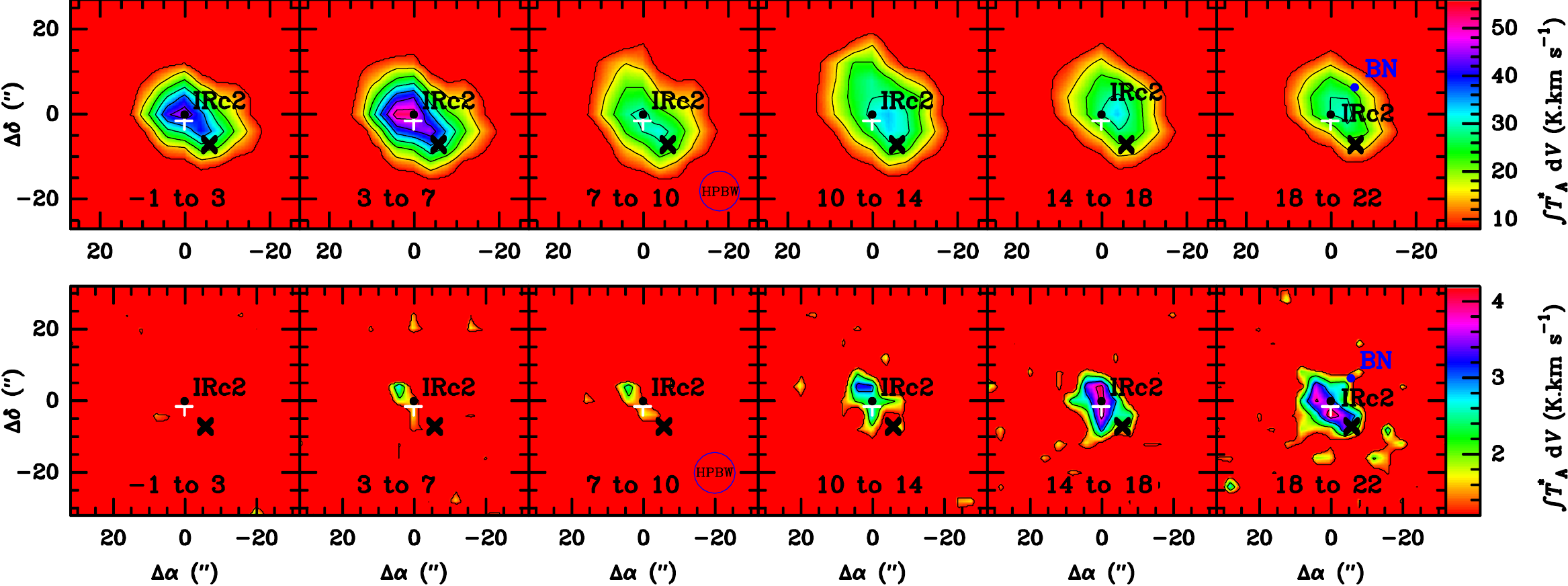}
   \caption{$^{34}$SO and $^{34}$SO$_{2}$-integrated intensity maps over different velocity ranges (indicated at the bottom of each panel in km s$^{-1}$). Row 1 shows the transition 6$_{7}$-5$_{6}$ of $^{34}$SO ($E$$_{\mathrm{up}}$=46.7 K, $A$$_{\mathrm{ul}}$=2.2x10$^{-4}$ s$^{-1}$). The contour interval is 7 K km s$^{-1}$ and the minimum contour is 8 K km s$^{-1}$. Row 2 shows the transition 28$_{(3,25)}$-28$_{(2,26)}$ of $^{34}$SO$_{2}$ ($E$$_{\mathrm{up}}$=402.1 K, $A$$_{\mathrm{ul}}$=1.5x10$^{-4}$ s$^{-1}$). The interval between contours is 0.9 K km s$^{-1}$, and the minimum contour is 1.2 K km s$^{-1}$. The white cross indicates the position of the hot core and the black cross the position of the compact ridge.}
   \label{figure:34so_34so2_maps}
   \end{figure*}

From the 2D line survey data of Orion KL, (Marcelino et al. in prep.), we produced integrated intensity maps of several SO and SO$_2$ transitions over different velocity ranges. Figure \ref{figure:so2_maps} shows the transitions 4$_{(2,2)}$-3$_{(1,3)}$, 11$_{(1,11)}$-10$_{(0,10)}$, and 14$_{(3,11)}$-14$_{(2,12)}$ of SO$_2$; Fig. \ref{figure:so_maps} shows the transitions 6$_{6}$-5$_{5}$ and 3$_{2}$-2$_{1}$ of SO; and Fig. \ref{figure:34so_34so2_maps} shows the transition 6$_{7}$-5$_{6}$ of $^{34}$SO and the transition 28$_{(3,25)}$-28$_{(2,26)}$ of $^{34}$SO$_{2}$. Velocity intervals in the figures have been chosen to represent different source components.
For all species and transitions, the strongest contribution arises from the velocity ranges 3-7 km s$^{-1}$ and 10-14 km s$^{-1}$, belonging to the HC and the HVP, respectively. The range 3-7 km s$^{-1}$ also includes PL velocities.
These maps show elongated emission along the direction NE-SW. This agrees with Plambeck et al. (2009), who find (from SiO $J$=2-1 observations with an angular resolution of 0.45$\arcsec$) that the strongest emission arises from a bipolar outflow covering velocities from -13 to 16 km s$^{-1}$ along the NE-SW direction. This distribution is clearly seen in the maps of $^{34}$SO and of SO (see lower panel in Fig. \ref{figure:so_maps}), especially in the ranges 7-10 km s$^{-1}$ (ridge) and 10-14 km s$^{-1}$ (HVP).  
Since the line widths corresponding to the HVP are the widest, with FWHM$\sim$ 30-40 km s$^{-1}$, altogether this suggests that the gas of the high velocity plateau is expanding in the direction NE-SW. 
On the other hand, the spatial distribution of $^{34}$SO$_2$ is less extended and usually shows a peak to the NE of IRc2 (also seen in SO$_2$ and SO at velocities 3-7 km s$^{-1}$). For this species, the NE-SW distribution is better traced by the 20.5 component. This different behavior should be due to the high energy level ($E$$_{\mathrm{up}}$=402.1 K) compared to the other mapped transitions, revealing the most compact and hottest regions in the KL cloud.
Also evident on the maps is the new component at 20.5 km s$^{-1}$. From Figs. \ref{figure:so2_maps} and \ref{figure:so_maps} we observe that its emission peak is located between the HC and BN positions.

\section{Analysis}
\label{section:analysis}

\subsection{LVG models of the SO lines}
\label{section:LVG}

In this section we analyze the non-LTE excitation and radiative transfer of SO lines. Following the study started by Tercero et al. (2010), we use an LVG (large velocity gradient) code, MADEX, developed by Cernicharo (2012) assuming that the width of the lines is due to the existence of large velocity gradients across the cloud, so that the radiative coupling between two relatively close points is negligible, and the excitation problem is local. The LVG models are based on the Goldreich \& Kwan (1974) formalism. The final considered fit is the one that reproduces more line profiles better from transitions covering a wide energy range, within a $\sim$30$\%$ of the uncertainty in line intensity. 
For each component of the cloud, we assume uniform physical conditions (kinetic temperature, density, line width, radial velocity, and source size) that we choose taking into account the parameters obtained from the Gaussians fits of the line profiles, the rotational diagrams (with the derived rotational temperatures), and the mapped transitions. 

Only for the HC do we consider LTE, which means that most transitions will be thermalized to the same temperature ($T$$_{\mathrm{rot}}$$\simeq$$T$$_{\mathrm{K}}$). If this condition was not satisfied, but we kept considering LTE approxmation, the temperatures would be overestimated, and this would produce a variation in the column densities. We cannot estimate whether they would be overestimated or underestimated since we do not know the population of each level.
However, the HC of Orion presents a condition of temperature ($T$$_{\mathrm{K}}$$>$200 K) and density ($n$(H$_{2}$)$\simeq$10$^{7}$ cm$^{-3}$), which make the LTE assumption in this component feasible.
Corrections for beam dilution are also applied for each line depending on the different beam sizes at different frequencies. Therefore, we fix all the above parameters (see Table \ref{table:components LVG}) leaving only as a free parameter the column density fo each component. 
For the densitiy, $n$(H$_{2}$), we have adopted fixed values taken from typical values quoted in the literature. In order to determine the uncertainty of the values of hydrogen density and of temperature ($T$$_{\mathrm{K}}$), we have run several models varying only the values for these parameters and fixing the rest. 

Comparing the intensity differences between the spectra and the obtained line profiles for each case of $T$ and $n$, we deduce an uncertainty of 20 and 15$\%$ for the temperature and the hydrogen density, respectively.
Although the parameter which could introduce higher uncertainty in the line profiles is the considered source size for each component, due to it varies depending on the molecular emission used for its determination. We fixed this parameter, as well as the hydrogen density, taking into account also the values used by Tercero et al. (2011) in her models of SiO and SiS.
Other sources of uncertainty in the model predictions arise from the spatial overlap of the different cloud components. However, it has been possible to model their contribution thanks to the wide range of frequency and to the large number of lines from different isotopologues. We also find as a source of uncertainty the modest angular resolution of any single-dish line survey, pointing errors (errors as small as 2$\arcsec$ could introduce important changes in the contribution from each cloud component to the observed line profiles, especially at 1.3 mm), and line opacity effects. This last source of uncertainty becomes important when lines arising from the plateau are optically thick, causing an underestimation of the column densities of the components that are surrounded by the plateau (compact ridge, hot core, and the 20.5 km s$^{-1}$ component) along the line of sight (Schultz et al. 1999).
In Tercero et al. (2010), these sources of uncertainty are explained in more detail. We estimate an uncertainty in our model intensity predictions of 25$\%$ for SO and $^{34}$SO, and 35$\%$ for SO$_{2}$, $^{34}$SO$_{2}$, $^{33}$SO, S$^{18}$O, $^{33}$SO$_{2}$, and SO$^{18}$O lines (higher uncertainty for SO$_{2}$ with respect to SO because of considering LTE instead of LVG approximation).

\subsubsection{SO}

\begin{table*}
\caption{Physical parameters adopted for the Orion KL cloud components.}             
\centering   
\begin{tabular}{c c c c c c c}     
\hline\hline       
Component & Source diameter & Offset (IRc2) & $n$(H$_{2}$) & $T$$_{\mathrm{K}}$ & $\triangle$$v$$_{\mathrm{FWHM}}$ & $v$$_{\mathrm{LSR}}$ \\
  & ($\arcsec$) & ($\arcsec$) & cm$^{-3}$ & (K) & (km s$^{-1}$) & (km s$^{-1}$) \\  
\hline 
                   
   Extended ridge (ER) & 120 & 0 & 10$^{5}$ & 60 & 4 & 8.5 \\  
   Compact ridge (CR) & 15 & 7 & 10$^{6}$  & 110 & 3 & 8 \\
   High-velocity plateau (HVP) & 30 & 4 & 10$^{6}$ & 100 & 30 & 11\\
   Plateau (PL) & 20 & 0 & 5$\times$10$^{6}$ & 150 & 25 & 6 \\
   Hot core (HC) & 10 & 2 & 1.5$\times$10$^{7}$ & 220 & 10 & 5.5 \\
   20.5 km s$^{-1}$ component & 5 & 2 & 5$\times$10$^{6}$ & 90 & 7.5 & 20.5 \\
\hline
\label{table:components LVG}                  
\end{tabular}
\end{table*}

\begin{figure*}
   \centering
   \includegraphics[angle=0,width=15cm]{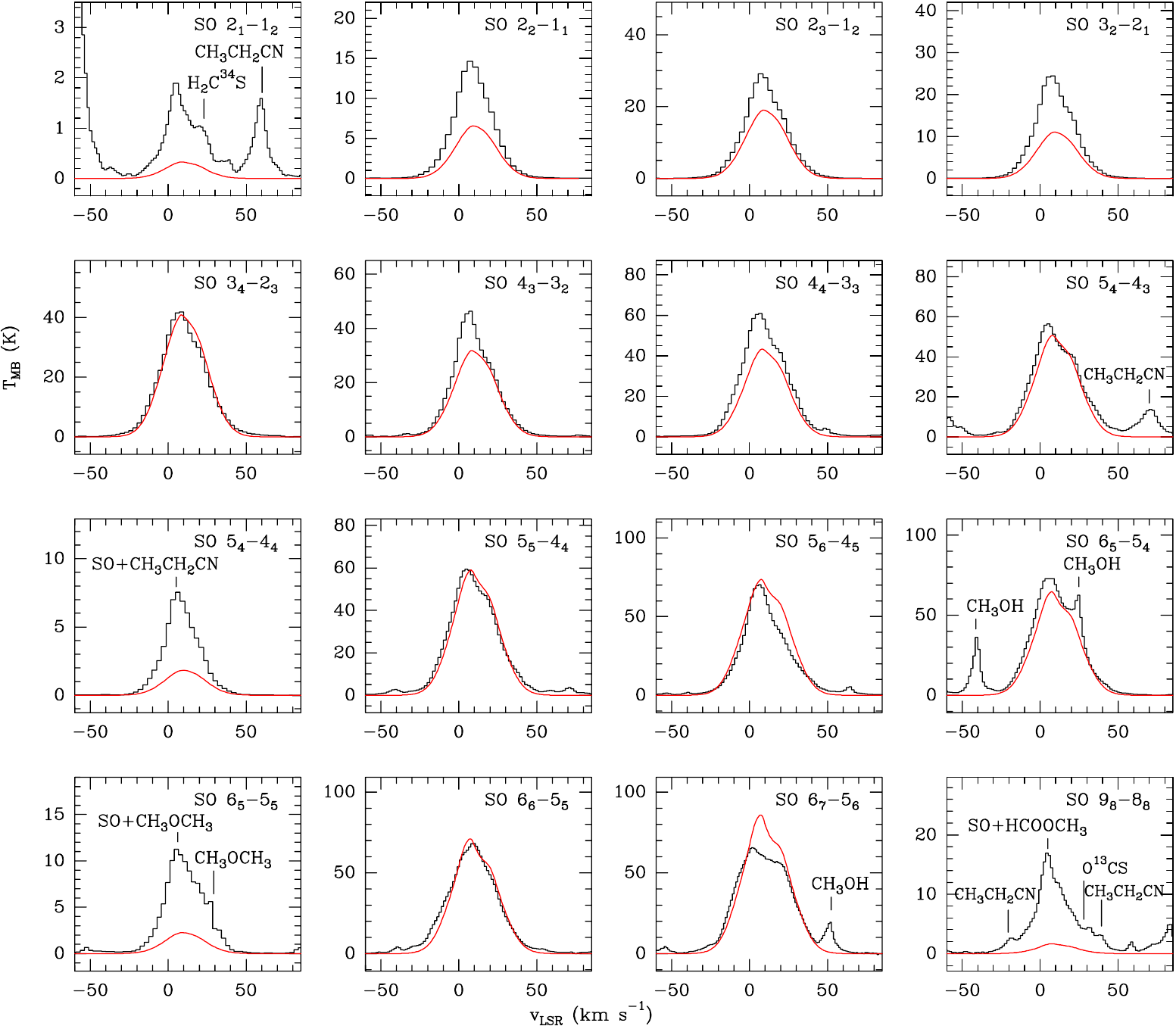}
   \caption{Observed lines of SO (black histogram) and best fit LVG model results (red).}
   \label{figure:so LVG}
   \end{figure*}

To model the rotational lines of SO (listed in Table \ref{table:tab_so}, see Appendix), we used the collisional rates from Lique et al. (2006) for collisions with H$_{2}$. Figure \ref{figure:so LVG} shows our best fit model.
The component with the highest SO column density is the HVP, with $N$(SO)=(5$\pm$1)$\times$10$^{16}$ cm$^{-2}$ (see Table \ref{table:so column densities}), although in the HC we also find a high column density with $N$(SO)=(9$\pm$3)$\times$10$^{15}$ cm$^{-2}$. We find that the PL and the 20.5 km s$^{-1}$ component also contribute to the emission, but with a column density that is one order of magnitude lower than the HVP.
We did not need to consider any contribution from the CR for SO. 
We could fit the narrow component by considering only the contribution from a single ridge (the extended ridge) at a temperature $T$$_{\mathrm{K}}$=60 K. This agrees with the previous SO analysis in Sect. \ref{section:LTE_gauss}.

The model indicates that some SO lines with emission coming mainly from the high-velocity plateau are optically thick. The optical depths are $\tau$=1.1-1.4 for the transitions 6$_{5}$-5$_{4}$ and 6$_{6}$-5$_{5}$, and $\tau$=1.4-1.8 for the transitions 5$_{6}$-4$_{5}$ and 6$_{7}$-5$_{6}$.
This means that the column densities obtained for the HC and 20.5 km s$^{-1}$ components, which are surrounded by the plateau, must be considered as lower limits, because the gas in the plateau components can absorb the emission from them. The HVP column density also has to be considered as a lower limit due to this opacity effect. 
We have also calculated the SO column density from the column density of $^{34}$SO, whose lines are optically thin, and from the the solar abundance ratio $^{32}$S/$^{34}$S=23 (Anders \& Grevesse 1989). As we expected, Table \ref{table:so column densities} shows that the column densities of SO obtained from $^{34}$SO are higher than those obtained from the fits. This confirms the presence of opacity effects on SO lines. We notice that the SO lines 2$_{1}$-1$_{2}$, 5$_{4}$-4$_{4}$, 6$_{5}$-$_{}$, and 9$_{8}$-8$_{8}$ present a poor fit. This is because of the low Einstein coefficients $A$$_{\mathrm{ul}}$ and, mainly, to the low line strengths S ($<$4$\times$10$^{-6}$ s$^{-1}$ and $<$0.35, respectively) of these lines, which provides small fits. The profile lines that we observe for these transitions are probably caused by stronger emission from other species.

Compared with values obtained in previous studies, such as those of Turner et al. (1991) or Blake et al. (1987), who derived (from source-averaged) $N$(SO)$\sim$3$\times$10$^{16}$cm$^{-2}$ in the plateau and N$\sim$3$\times$10$^{15}$cm$^{-2}$ in the HC, we see our results agree. 
If we compare these results with our obtained column densities from $^{34}$SO, we see that the plateau remains in agreement with these previous studies, but our column density in the HC is one order of magitude higher.

From the spatial distribution of the SO emission (Fig. \ref{figure:so_maps}), we find the integrated intensity peak to be in the velocity ranges 3-7 km s$^{-1}$ and 10-14 km s$^{1}$ for both transitions. These ranges correspond to emission arising mainly from the HC and the HVP, respectively, which agrees with the results obtained for the column densities in these regions. The transition with lowest energy (upper panel) shows a concentric emission distribution around IRc2, while for the transition with higher energy, the emission distribution is elongated toward NE-SW direction. We also observe that, for high energies, the emission peak in the velocity range 3-7 km s$^{1}$ is shifted toward the NE of the HC.

\subsubsection{$^{34}$SO, $^{33}$SO, and S$^{18}$O}

Figure \ref{figure:$^{34}$SO LVG} shows our best fit model for several rotational lines of $^{34}$SO, which are listed in Table \ref{table:tab_so} (see Appendix). 
We find that the HVP is also responsible for most of the emission, together with the PL and the 20.5 km s$^{-1}$ component. According to our models, all transitions are optically thin with $\tau$$<$0.4, therefore this result is not considered to be a lower limit. The 20.5 km s$^{-1}$ component presents a similar column density to the HVP, and as for SO, its contributon is greater for higher $J$. 
In the HC we also find a strong contribution to the emission, however in the ER we find the lowest column density with $N$($^{34}$SO)=(7$\pm$2)$\times$10$^{12}$ cm$^{-2}$.

For the case of $^{33}$SO we observe in Fig. \ref{figure:$^{33}$SO LVG} (see Appendix) that lines are partially blended with other species, which produces a large uncertainty in the derived fit. In addition, the hyperfine structure is noticeable in these transitions, adding more complexity to the line profiles.
The fits for this isotopologue were done by adopting the calculated frequencies, intensities, and energies for hyperfine levels up to $N$=30 provided by the CDMS catalogs.
We obtain similar column densities for all components, with $N$($^{33}$SO)=(3-6)$\times$10$^{14}$ cm$^{-2}$ (see Table \ref{table:so column densities}).

As was the case for $^{33}$SO, some lines of S$^{18}$O are blended with other species (Fig. \ref{figure:S$^{18}$O LVG} and Table \ref{table:tab_so}, see Appendix). In this case we also find similar column densities for all the components, $N$(S$^{18}$O)=(1-5)$\times$10$^{14}$ cm$^{-2}$, except for the CR, the ER, and the 20.5 km s$^{-1}$ component, where we do not find contribution to the emission.

 \begin{figure*}
   \centering
   \includegraphics[angle=0,width=11.7cm]{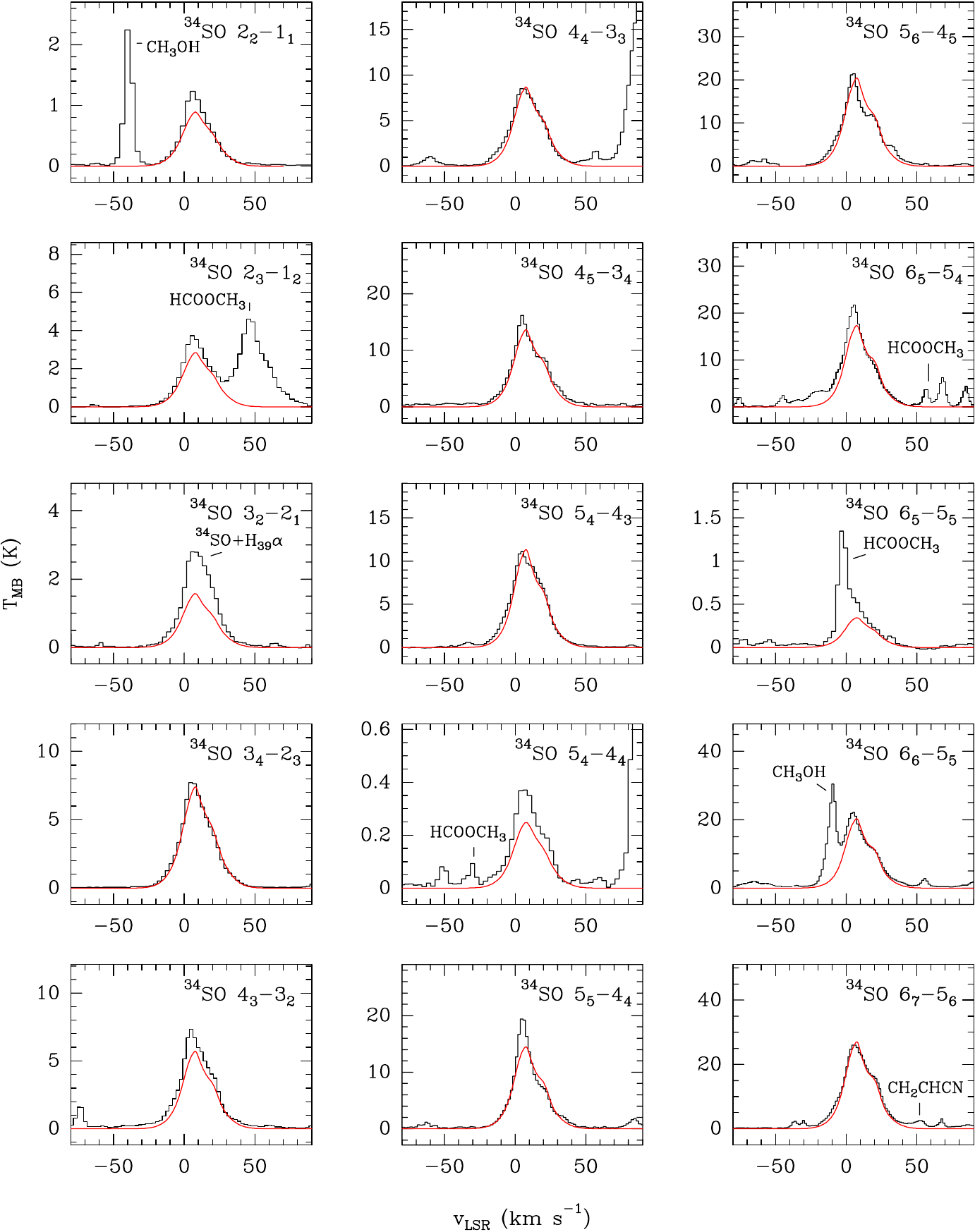}
   \caption{Observed lines of $^{34}$SO (black histogram) and best fit  LVG model results (red).}
   \label{figure:$^{34}$SO LVG}
   \end{figure*}

\begin{table*}
\caption{Column densities, $N$, for SO and its isotopologues obtained from LVG model fits.}              
\begin{center}
\begin{tabular}{c c c c c c}     
\hline\hline       
Component & SO  & $^{34}$SO & SO$^{(a)}$ & $^{33}$SO & S$^{18}$O\\
          & $N$$\times$10$^{15}$(cm$^{-2}$) & $N$$\times$10$^{15}$(cm$^{-2}$) & $N$$\times$10$^{15}$(cm$^{-2}$) & $N$$\times$10$^{15}$(cm$^{-2}$) & $N$$\times$10$^{15}$(cm$^{-2}$)\\ 
\hline 
                   
   Extended ridge (ER)           & 0.018$\pm$0.005   & 0.007$\pm$0.002  & 0.16$\pm$0.05   & ...           & ...\\  
   High-velocity plateau (HVP)    & 45$\pm$10         & 4$\pm$1          & 92$\pm$23       & 0.25$\pm$0.06 & 0.24$\pm$0.08\\
   Plateau (PL)                  & 5$\pm$1           & 3.0$\pm$0.8      & 69$\pm$19       & 0.5$\pm$0.2   & 0.10$\pm$0.04\\
   Hot core (HC)                  & 9$\pm$3          & 2.2$\pm$0.5      & 20$\pm$11       & 0.4$\pm$0.1   &  0.5$\pm$0.2 \\
   20.5 km s$^{-1}$ component    & 5$\pm$1           & 3.5$\pm$0.8      & 81$\pm$19       & 0.6$\pm$0.2   & ... \\
\hline 
\label{table:so column densities}                 
\end{tabular}
\end{center}
\tablefoot{
\tablefoottext{a}{Values calculated from the solar abundance ratio $^{32}$S / $^{34}$S=23 and from the column densities obtained for $^{34}$SO (optically thin lines).}
}
\end{table*}

\subsubsection{S$^{17}$O, $^{36}$SO, and $^{34}$S$^{18}$O}

Owing to the presence of other more intense species, we have detected neither S$^{17}$O nor $^{36}$SO in this survey, but from our data we derive upper limits for their column densities of $N$(S$^{17}$O)$<$1.3$\times$10$^{14}$cm$^{-2}$ and $N$($^{36}$SO)$<$1.6$\times$10$^{14}$cm$^{-2}$, respectively. We note that C$^{36}$S was detected by  Mauersberger et al. (1996) in Orion. They found $^{32}$S/$^{36}$S$\simeq$3500 which is consistent with our upper limit.
We have not detected $^{34}$S$^{18}$O either; however, assuming the same physical conditions as those for the main isotopologue, we obtain an upper limit for its column density of $N$($^{34}$S$^{18}$O)$<$1.4$\times$10$^{14}$cm$^{-2}$.

\subsection{LTE models of the SO$_{2}$ lines}

\begin{figure*}
   \centering
   \includegraphics[angle=0,width=15cm]{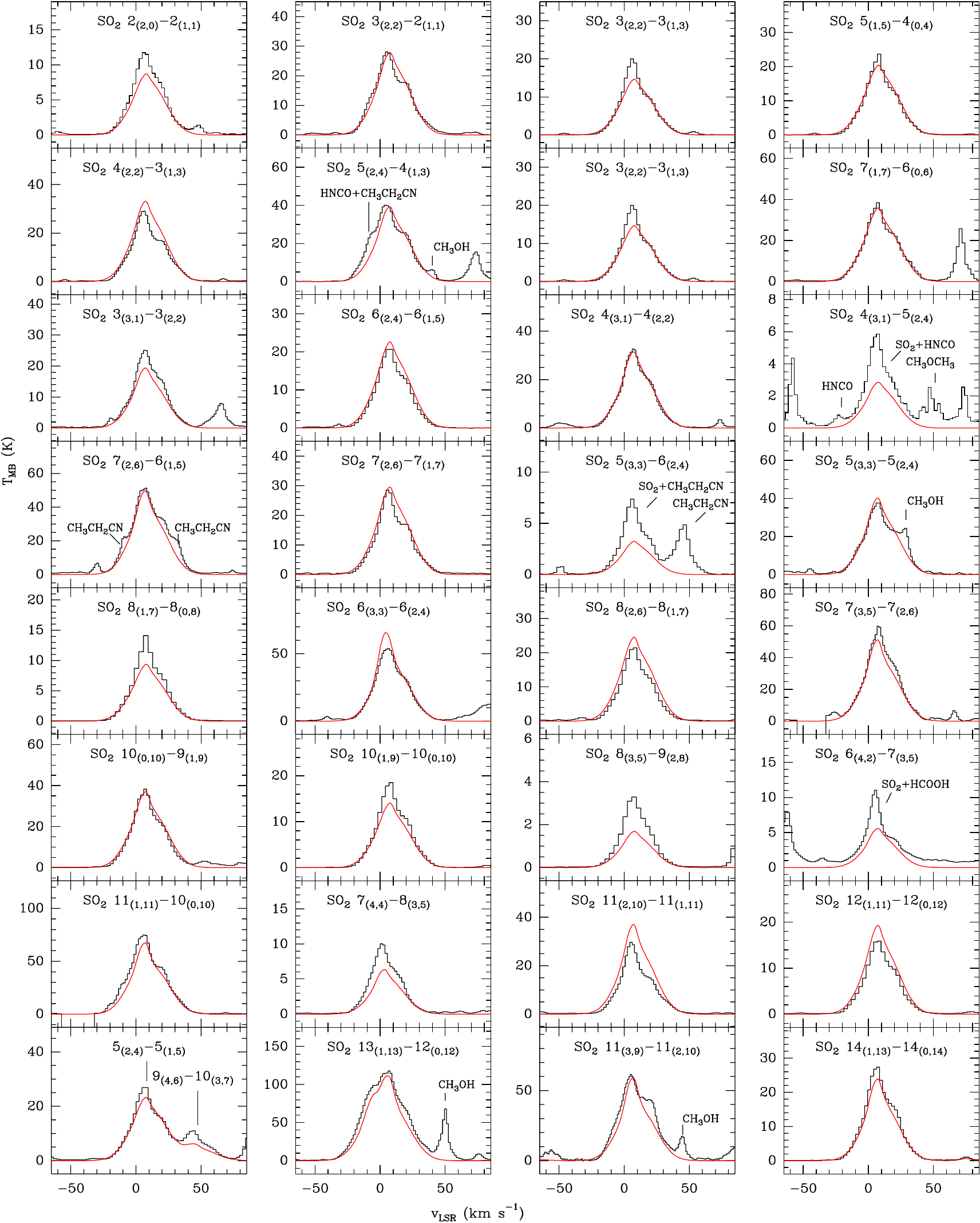}
   \caption{Observed lines (black histogram) of SO$_{2}$ with upper state energies lower than 400 K, ordered by increasing energy from top left to bottom right. Best fit LTE model results are overlaid in red.}
   \label{figure:so2 1mmLVG}
   \end{figure*}

\begin{figure*}
   \centering
   \includegraphics[angle=0,width=14cm]{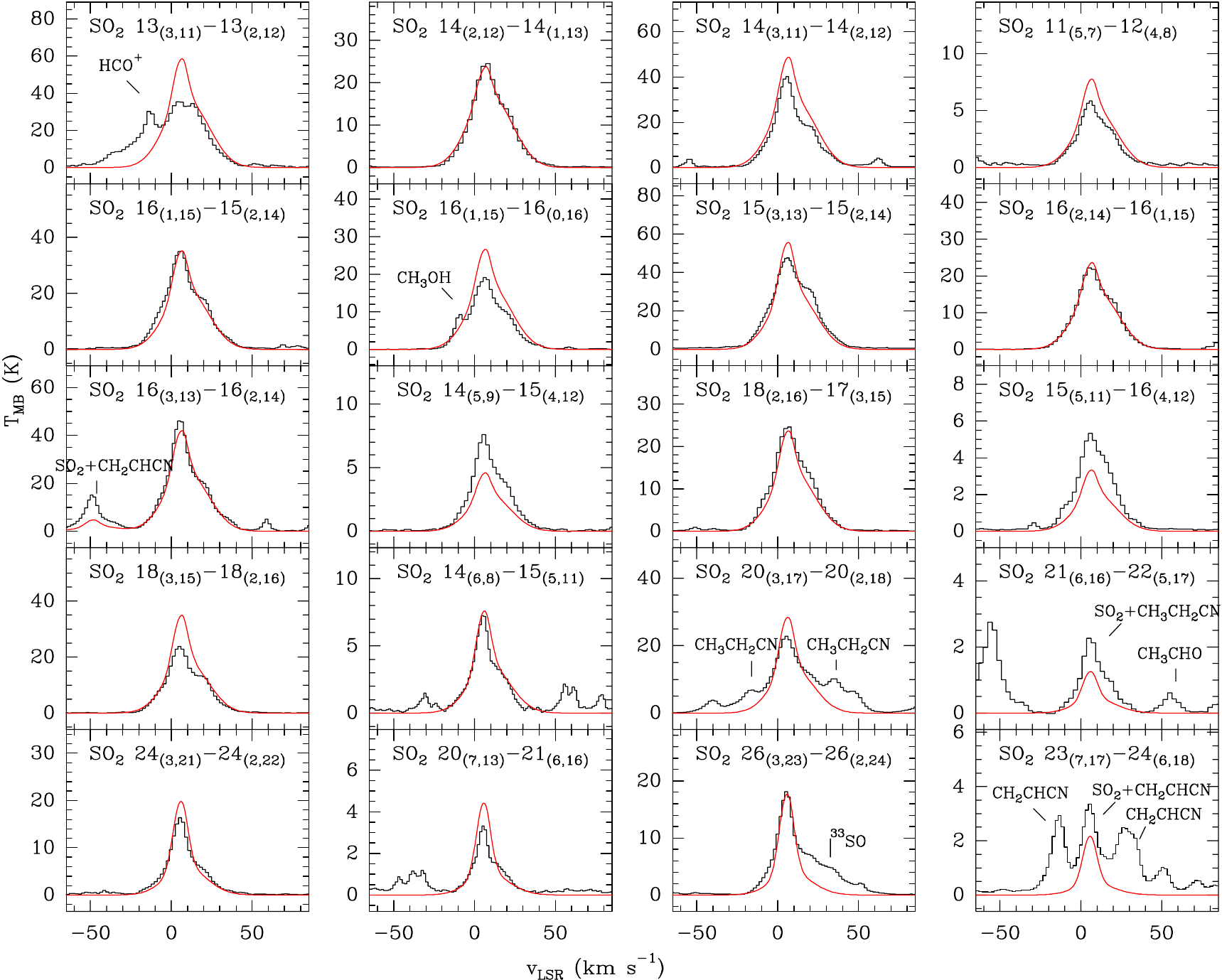}  
   \caption{Observed lines of SO$_{2}$ (black histogram) with upper state energies lower than 400 K (continued), ordered by increasing energy from top left to bottom right. Best fit LTE model results are overlaid in red.}
   \label{figure:so2 2mmLVG}
   \end{figure*}

\begin{figure*}
   \centering
   \includegraphics[angle=0,width=14cm]{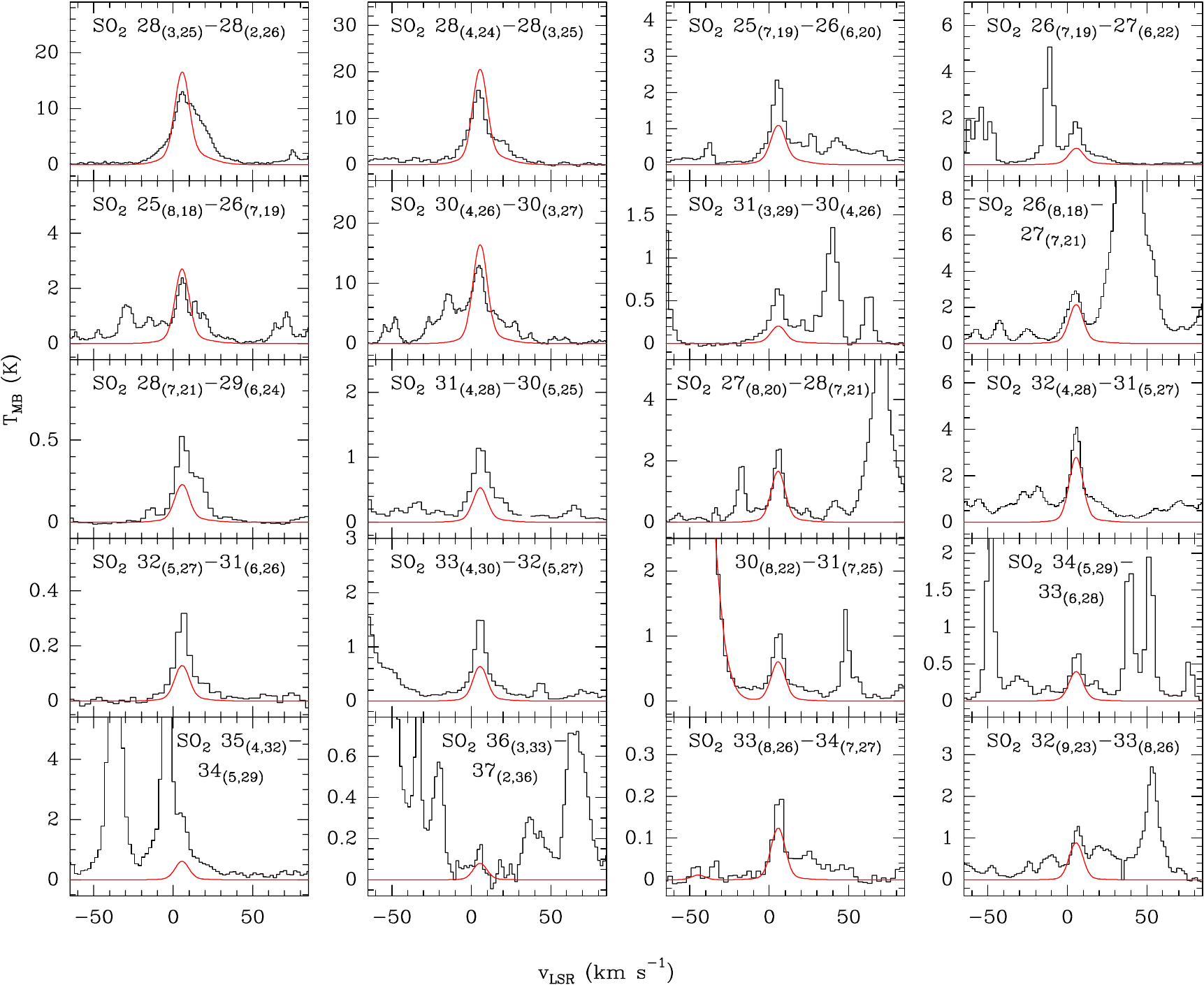}  
   \caption{Observed lines of SO$_{2}$ (black histogram) with upper state energies between 400 K and 700 K, ordered by increasing energy from top left to bottom right. Best fit LTE model results are overlaid in red.}
   \label{figure:so2 3mmLVG}
   \end{figure*}

\begin{figure*}
   \centering
   \includegraphics[angle=0,width=16.5cm]{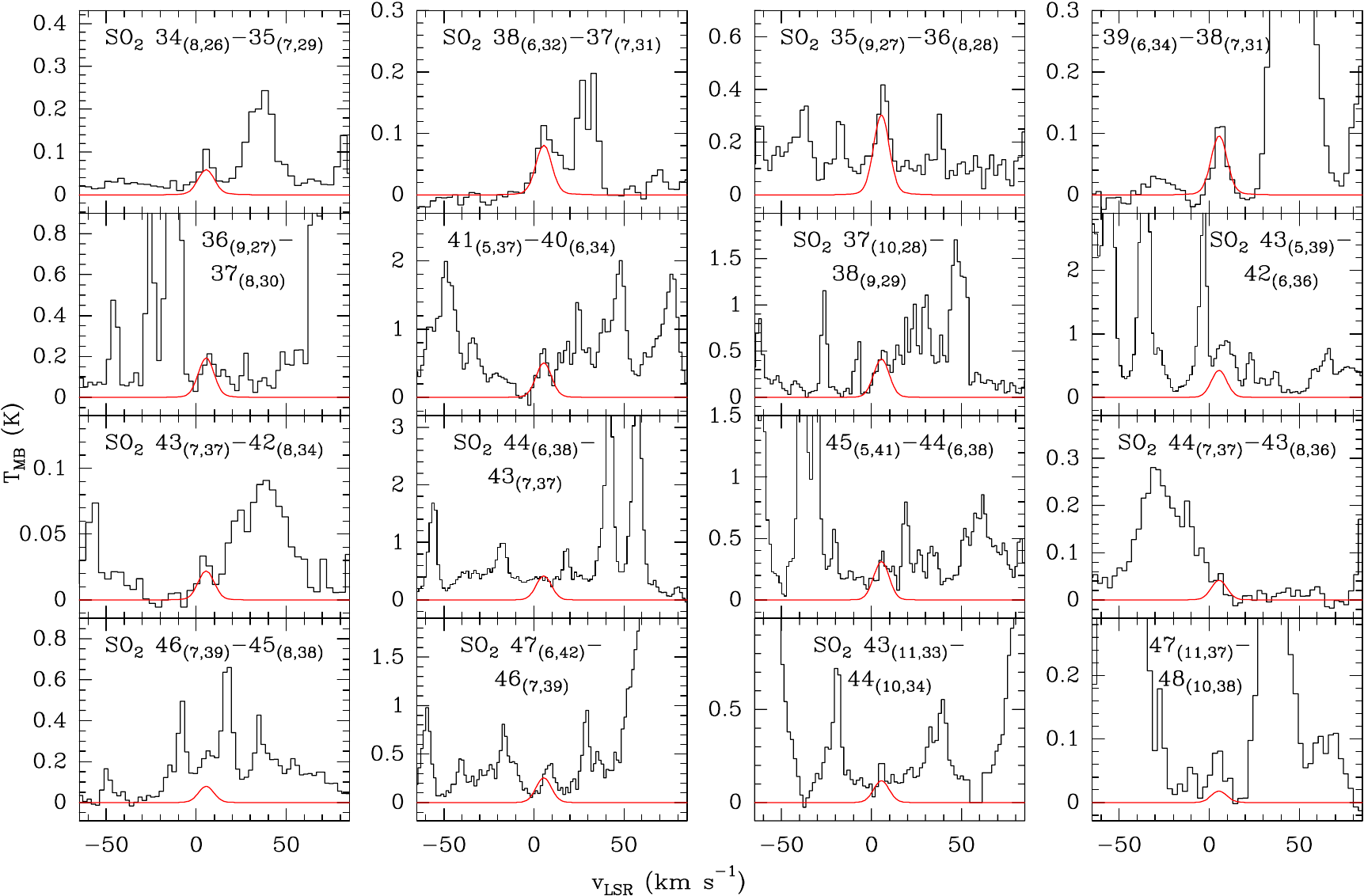}  
   \caption{Observed lines of SO$_{2}$ (black histogram) with energies higher than 700 K, ordered by increasing energy from top left to bottom right. Best fit LTE model results are overlaid in red.}
   \label{figure:so2 4mmLVG}
   \end{figure*}

Due to the lack of collisional rates for levels with energies higher than 90 K, we have assumed LTE excitation to derive the SO$_{2}$ column densities. As stated before, this can overestimate or underestimate the calculated column densities. Given that we have collisional rates for SO, we ran our SO models again but considering LTE approximation, in order to compare the results. We observed that the new fits underestimated the line profiles, especially for low transitions. Probably this also happens with the SO$_{2}$ case, so we should consider our SO$_{2}$ column densities as lower limits. 
Table \ref{table:tab_so2} (see Appendix) lists the 166 observed rotational lines. Figures \ref{figure:so2 1mmLVG}, \ref{figure:so2 2mmLVG}, \ref{figure:so2 3mmLVG}, and \ref{figure:so2 4mmLVG} show a sample of 90 observed lines (ordered by increasing energy) with our best fits overlaid. 

The main contribution to the emission of SO$_{2}$ comes from the HVP (affecting mainly the lines with energies $E$$<$400 K), with a column density of $N$(SO$_{2}$)=(1.3$\pm$0.5)$\times$10$^{17}$cm$^{-2}$. The hottest region (the HC) also presents a similar high value, $N$(SO$_{2}$)=(1.0$\pm$0.4)$\times$10$^{17}$cm$^{-2}$. 
We find that the CR presents a column density of $N$(SO$_{2}$)$\sim$10$^{15}$cm$^{-2}$, whereas the 20.5 km s$^{-1}$ component presents the lowest contribution to the emission (Table \ref{table:so2 column densities}).

If we compare our results for SO$_{2}$ with those of SO, we find that SO$_{2}$ column densities are about one order of magintude larger in all components, except in the 20.5 km s$^{-1}$ component, where SO presents a higher contribution to the emission.
As for SO, we also calculated the SO$_{2}$ column densities from $^{34}$SO$_{2}$ (optically thin lines) and the $^{32}$S/$^{34}$S solar abundance ratio (Table \ref{table:so2 column densities}). Except for the plateau components, we obtain that the column densities of SO$_{2}$ obtained from $^{34}$SO$_{2}$ are larger than those obtained from fits, suggesting they are opacity effects on the SO$_{2}$  lines.
Comparing with previous results, Blake et al. (1987) obtained (source-averaged) that SO$_{2}$ presents a column density in the plateau, $N$$\sim$10$^{16}$cm$^{-2}$. Schilke et al. (2001) derived (beam-averaged) for the same region $N$(SO$_{2}$)=9.7$\times$10$^{16}$cm$^{-2}$, which is very similar to our result in the high velocity plateau. For the HC, Sutton et al. (1995) found (also source-averaged) a column density, $N$(SO$_{2}$)$\sim$9$\times$10$^{16}$cm$^{-2}$. Our results agree with these values; however, the large number of transitions that we observed let us determine more accurately that it is the lower temperature plateau component, i.e. the HVP, which contributes more to the emission of SO$_{2}$.

From the spatial distribution of the SO$_{2}$ emission (Fig. \ref{figure:so2_maps}), we find the maximum integrated line intensity between the velocity ranges 3-7 km s$^{-1}$ and 10-14 km s$^{-1}$, as well as for SO. These velocity ranges correspond to emission of the HC and the HVP, respectively, which agrees with the column density results obtained above. The emission peak is located towards the NE of IRc2 for the range 3-7 km s$^{-1}$, while the emission peak for the HVP range is located $\sim$4$\arcsec$ to the SW of IRc2.
Emission from the 20.5 km s$^{-1}$ component presents similar integrated intensity to the ridge (range of 7-10 km s$^{-1}$).
On the other hand, the map of $^{34}$SO$_{2}$ (Fig. \ref{figure:34so_34so2_maps}) shows the strongest peak emission located northeast of IRc2 for all velocity ranges, although extended emission is seen to the southwest of IRc2, between 10 km s$^{-1}$ and 22 km s$^{-1}$. The $^{34}$SO$_{2}$ transition shown in Fig. \ref{figure:34so_34so2_maps} has an upper energy level of $\sim$400 K and therefore may trace only the hottest component of the gas.

 \begin{figure*}
   \centering
   \includegraphics[angle=0,width=15.5cm]{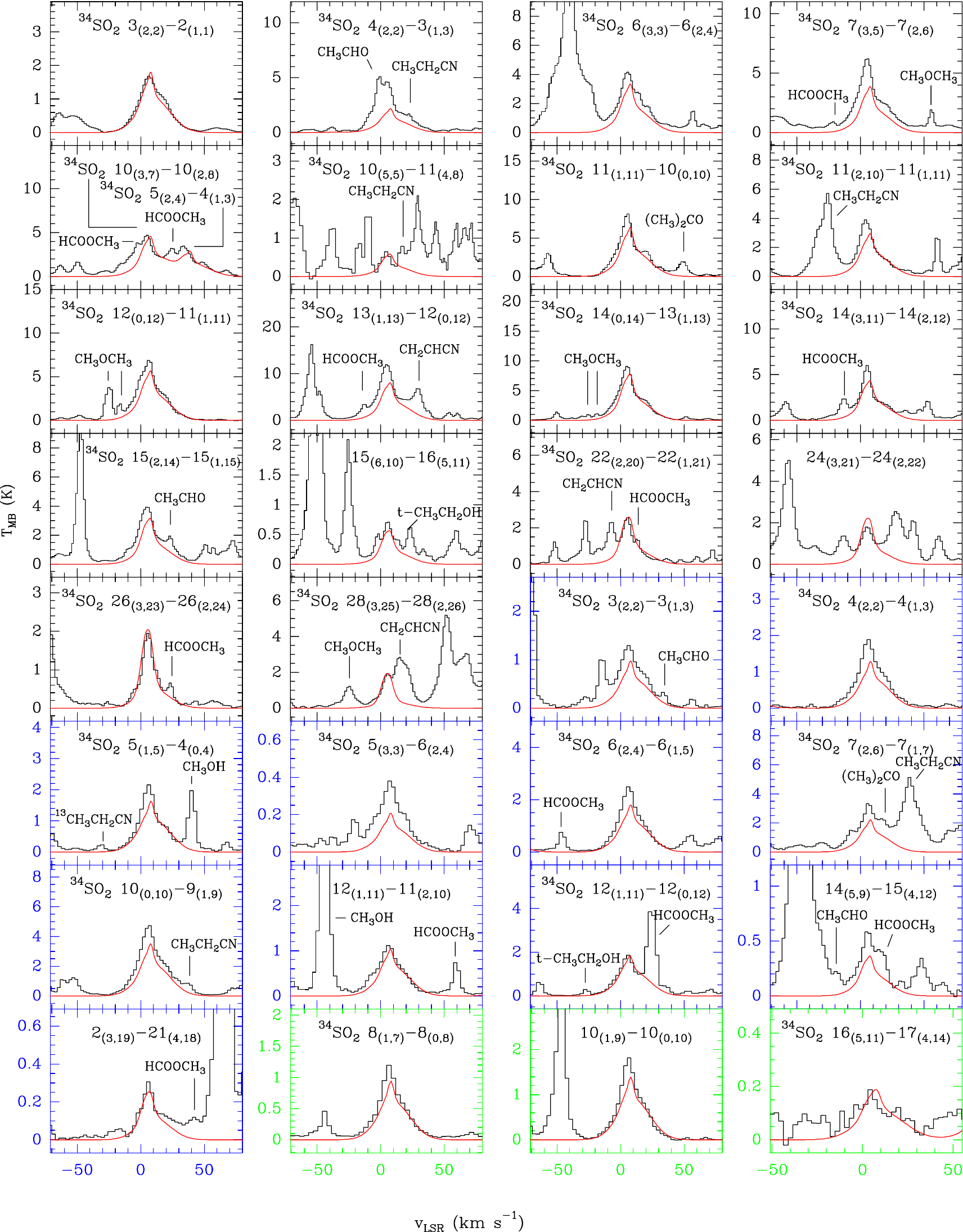}
   \caption{Observed lines of $^{34}$SO$_{2}$ (black histogram). Best fit LTE model results are shown in red. Boxes with black, blue and green borders correspond to lines observed at 1.3, 2, and 3 mm, respectively.}
   \label{figure:$^{34}$so2 LVG}
   \end{figure*}

 \begin{figure*}
   \centering
   \includegraphics[angle=0,width=15cm]{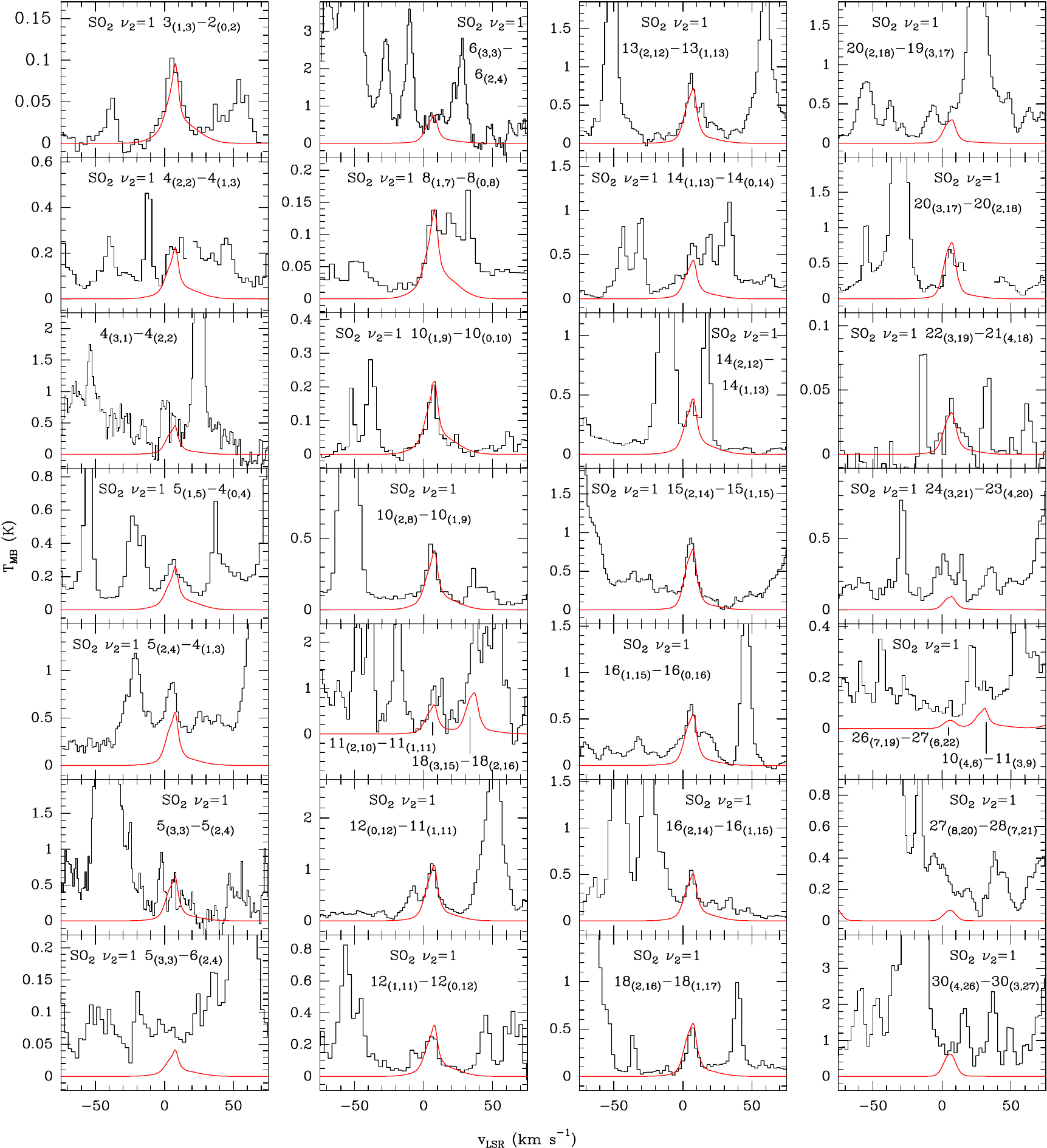}
   \caption{Observed lines of SO$_{2}$ $\nu$$_{2}$=1 (black histogram). Best fit LTE model results are shown in red.}
   \label{figure:so2v2 LVG}
   \end{figure*}

\subsubsection{$^{34}$SO$_{2}$}

We have detected 130 rotational lines of $^{34}$SO$_{2}$ (Table \ref{table:tab_so2} in Appendix). A sample of more than 30 lines is shown in Fig. \ref{figure:$^{34}$so2 LVG}.

This isotopologue, whose lines are optically thin, has the highest contribution to its emission from the HC, with $N$($^{34}$SO$_{2}$)=(1.0$\pm$0.4)$\times$10$^{16}$cm$^{-2}$. The other components with high column densities are the HVP and the PL with $N$($^{34}$SO$_{2}$)=(7$\pm$2)$\times$10$^{15}$cm$^{-2}$ and $N$($^{34}$SO$_{2}$)=(6$\pm$2)$\times$10$^{14}$cm$^{-2}$, respectively. 
Particulary interesting is the column density found in the ER, which presents the same order of magnitude as the PL. The same behavior is also observed in the isotopologues $^{33}$SO$_{2}$ and SO$^{18}$O, where the column densities in the ER are only 3 to 12 times lower than in PL. This could be due to the strong emission emerging from the HVP and HC affecting the excitation of the energy levels of this isotopologue in the ER, which means that line photons emitted from the inner components will be scattered by the lower density gas in the ER component (radiative scattering). 
For the compact ridge, we find the emission contributes mainly to the lines at low frequencies. The column density for $^{34}$SO$_{2}$ obtained in the 20.5 km s$^{-1}$ component is about 100 times lower than for $^{34}$SO (see Table \ref{table:so2 column densities}).

\subsubsection{$^{33}$SO$_{2}$, SO$^{18}$O, and SO$^{17}$O}

Figure \ref{figure:$^{33}$so2 LVG} (see Appendix) shows transitions of $^{33}$SO$_{2}$ in the frequency range covered at 1.3, 2, and 3 mm. Several of these are blended with other species, which makes the derived fits a bit biased. As for $^{33}$SO, hyperfine structure affects the line profiles; however, we did not consider it in the $^{33}$SO$_{2}$ model, so the derived column densities could be underestimated.

The most important contribution to the emission of $^{33}$SO$_{2}$ comes from the hot core, with $N$($^{33}$SO$_{2}$)=(4$\pm$1)$\times$10$^{15}$ cm$^{-2}$, and from the HVP.
The lowest contribution to the emission arises from the 20.5 km s$^{-1}$ component, with $N$($^{33}$SO$_{2}$)=(9$\pm$3)$\times$10$^{12}$ cm$^{-2}$.

For SO$^{18}$O (Fig. \ref{figure:SO$^{18}$O LVG}, see Appendix), the highest column density is also found in the HC, with $N$(SO$^{18}$O)=(1.5$\pm$0.5)$\times$10$^{15}$ cm$^{-2}$. The high-velocity plateau also presents an important contribution to the emission across the frequency range, whereas the PL mainly affects lines at 2 mm. The weakest contribution is from the ER, with $N$(SO$^{18}$O)$\sim$10$^{13}$ cm$^{-2}$.

Figure \ref{figure:SO$^{17}$O_LVG} shows the spectra that contain some frequencies of SO$^{17}$O, together with our best model. The maximum contribution to the emission of this isotopologue arises from the hot core with $N$(SO$^{17}$O)=(8$\pm$3)$\times$10$^{14}$ cm$^{-2}$. We find that the PL and HVP also contribute to the emission, but about four to eight times less than the HC. However, because all lines are weak ($T$$_{\mathrm{MB}}$$<$0.3 K) and blended with other species, we should consider these results as upper limits.

\subsubsection{SO$_{2}$ $\nu$$_{2}$=1}

Figure \ref{figure:so2v2 LVG} shows some of the detected lines of vibrationally excited SO$_{2}$ $\nu$$_{2}$, which are also listed in Table \ref{table:tab_so2} (see Appendix). 
The hot core is responsible for most of the emission of this vibrational mode, with a column densitiy one order of magnitude higher than in the other components.
From the column densities in the HC for SO$_{2}$ in its ground and vibrationally excited states, we can estimate a vibrational temperature, considering that

\begin{equation}
\frac{\exp(-E_{\mathrm{\nu_{x}}}/T_{\mathrm{vib}})}{f_{\mathrm{\nu}}}=\frac{N(SO_{2} \nu_{x})}{N(SO_{2})},
\end{equation}

\noindent where $E$$_{\mathrm{\nu_{x}}}$ is the energy of the vibrational state ($E$$_{\nu_{2}}$=745.1 K), $T$$_{\mathrm{vib}}$ is the vibrational temperature, $N$(SO$_{2}$ $\nu$$_{\mathrm{x}}$) is the column density of SO$_{2}$ in the excited vibrational state, $N$(SO$_{2}$) the total column density, and $f$$_{\mathrm{\nu}}$ is the vibrational partition function, given by

\begin{equation}
f_{\mathrm{\nu}}=1+\exp(-E_{\mathrm{\nu_{3}}}/T_{\mathrm{vib}})+2\exp(-E_{\mathrm{\nu_{2}}}/T_{\mathrm{vib}})+\exp(-E_{\mathrm{\nu_{1}}}/T_{{\mathrm{vib}}}).
\end{equation}

\noindent Taking into account that $N$(SO$_{2}$)=$N$$_{\mathrm{(ground)}}$$\times$$f$$_{\mathrm{\nu}}$, we only need the energy of the vibrational state and the calculated column densities to derive the vibrational temperature.

We obtain $T$$_{\mathrm{vib}}$=(230$\pm$40) K for SO$_{2}$ $\nu$$_{2}$=1. This value is similar to the kinetic temperature we assumed for the HC component (220 K). It is unlikely that the $\nu$$_{2}$=1 level at 745 K above the ground is excited by collisions. The main pumping mechanism could be IR radiation from the HC. A similar situation was found by Tercero et al. (2010) for OCS and other species.

\subsubsection{$^{34}$SO$_{2}$ $\nu$$_{2}$=1}

Figure \ref{figure:34so2v2 LVG} (see Appendix) shows some observed lines of $^{34}$SO$_{2}$ $\nu$$_{2}$=1. The strongest emission comes from the hot core with $N$($^{34}$SO$_{2}$ $\nu$$_{2}$=1)=(7$\pm$2)$\times$10$^{14}$ cm$^{-2}$, although we also find a small contribution from the HVP and the CR, with $N$($^{34}$SO$_{2}$ $\nu$$_{2}$=1)$\sim$5$\times$10$^{13}$ cm$^{-2}$ for both. These contributions mainly affect the lines at 2 mm. Since all the lines we detect are very weak and mixed with other species, we should consider these results as upper limits.

\begin{table*}
\caption{Column densities, $N$, for SO$_{2}$ and its isotopologues, obtained from LTE model analysis.}             
\begin{center}  
\begin{tabular}{c c c c c c c c c}     
\hline\hline       
Component & SO$_{2}$  & $^{34}$SO$_{2}$ & $^{(a)}$SO$_{2}$ & $^{33}$SO$_{2}$ & SO$^{18}$O & SO$^{17}$O & SO$_{2}$ $\nu$$_{2}$=1 & $^{34}$SO$_{2}$ $\nu$$_{2}$=1\\
 & $N$$\times$10$^{15}$ & $N$$\times$10$^{15}$ & $N$$\times$10$^{15}$ & $N$$\times$10$^{15}$ & $N$$\times$10$^{15}$ & $N$$\times$10$^{15}$ & $N$$\times$10$^{15}$ & $N$$\times$10$^{15}$ \\
  & (cm$^{-2}$) & (cm$^{-2}$) & (cm$^{-2}$) & (cm$^{-2}$) & (cm$^{-2}$) & (cm$^{-2}$) & (cm$^{-2}$) & (cm$^{-2}$)  \\
\hline 
                   
Extended ridge            & 0.23$\pm$0.06 & 0.10$\pm$0.04 & 2.3$\pm$0.7  & 0.04$\pm$0.001  & 0.020$\pm$0.007 & 0.007$\pm$0.003 & 0.013$\pm$0.003       & ... \\  
Compact ridge             & 1.2$\pm$0.4   & 0.5$\pm$0.2   & 12$\pm$2     & 0.07$\pm$0.02   & 0.03$\pm$0.01 & 0.007$\pm$0.003 & 0.20$\pm$0.05 & 0.05$\pm$0.02\\
High-velocity plateau     & 130$\pm$50    & 7$\pm$2       & 161$\pm$46   & 1.0$\pm$0.3     & 0.9$\pm$0.3   & 0.10$\pm$0.04   & 0.4$\pm$0.1 & 0.06$\pm$0.02 \\
Plateau                   & 10$\pm$3     & 0.6$\pm$0.2   & 14$\pm$3     & 0.5$\pm$0.2     & 0.06$\pm$0.02 & 0.03$\pm$0.01   &  ...          & ...\\
Hot core                  & 100$\pm$40   & 10$\pm$4     & 230$\pm$60   & 4$\pm$1           & 1.5$\pm$0.5   & 0.9$\pm$0.3     & 4$\pm$1   & 0.7$\pm$0.2 \\
20.5 km s$^{-1}$ comp.       & 0.17$\pm$0.06  & 0.04$\pm$0.01   & 0.8$\pm$0.2 & 0.009$\pm$0.003 & ...           & ...             & ...       & ...\\
\hline
\label{table:so2 column densities}                  
\end{tabular}
\end{center}
\tablefoot{
\tablefoottext{a}{Values calculated from the solar abundance ratio $^{32}$S / $^{34}$S=23 and from the column densities obtained for $^{34}$SO$_{2}$ (optically thin lines).}
}
\end{table*}

\subsection{Isotopic and molecular abundances}

From the derived column densities of SO, SO$_{2}$, and their isotopologues, we can estimate abundance ratios. These are shown in Table \ref{table:abundancesI} (see Appendix). We compare these ratios with solar isotopic abundance values from Anders \& Grevesse (1989). 

$^{32}$S/$^{34}$S: from the SO$_{2}$ lines we obtain a column density ratio for the PL of $^{32}$S/$^{34}$S=16$\pm$10, in agreement with previous studies ($^{32}$S/$^{34}$S$\simeq$16 by Johansson et al. 1984, $^{32}$S/$^{34}$S$\simeq$13-16 by Blake et al. 1987, and $^{32}$S/$^{34}$S=15$\pm$5 by Tercero et al. 2010). For the HVP, $^{32}$S/$^{34}$S=20$\pm$13, which is similar to the value for the solar isotopic abundance ratio and to the result deduced by Persson et al. (2007) where $^{32}$S/$^{34}$S$\simeq$23$\pm$7. 
From the SO lines, with the exception of the HVP, we obtain low ratios for $^{32}$S/$^{34}$S in comparison with the solar abundance, which could be due to opacity effects on the SO lines.

$^{32}$S/$^{33}$S: from $N$(SO)/$N$($^{33}$SO) we obtain $^{32}$S/$^{33}$S=180$\pm$80 for the HVP. This value agrees with the solar isotopic abundance ratio, 127, from Anders \& Grevesse (1989). 
For the other three components, the obtained ratios are very low compared to the solar abundance, probably also due to the opacity effects for SO lines. These values should be considered as lower limits. We find similar behavior for the ratio $N$(SO$_{2}$)/$N$($^{33}$SO$_{2}$).

$^{34}$S/$^{33}$S: from $N$($^{34}$SO$_{2}$)/$N$($^{33}$SO$_{2}$) we obtain $^{34}$S/$^{33}$S=4$\pm$3 for the 20.5 km s$^{-1}$ component and $^{34}$S/$^{33}$S=3-7 for the ridge (ER and CR), the HVP, and the HC. These values are similar to the solar abundance ratio (5.5). From $N$($^{34}$SO)/$N$($^{33}$SO) the obtained ratio is $^{34}$S/$^{33}$S=6$\pm$3 for the HC and both plateau components.

$^{16}$O/$^{18}$O: our results for this ratio in the plateau agree with those obtained by Tercero et al. (2010), who derived $^{16}$O/$^{18}$O=250$\pm$135 in the plateau from a study of OCS in this region. The compact ridge also presents similar ratio to that obtained by them. However, all these values are lower than the solar isotopic abundance ($\simeq$500).

SO/SO$_{2}$: in Fig. \ref{figure:ratios_abundances} we present the ratio  $N$(SO)/$N$(SO$_2$) for the different components, as well as for the different isotopologues of SO and SO$_2$. We find that SO$_{2}$ is more abundant than SO in all components, except in the 20.5 km s$^{-1}$ component. 
In the HVP, SO$_{2}$ is three times more abundant than SO, while in the HC is up to 11 times more. However, in the 20.5 km s$^{-1}$ component, SO is $\sim$30 times more abundant than SO$_{2}$.

$^{34}$SO/$^{34}$SO$_{2}$: in the region affected by shocks, this ratio implies that $^{34}$SO$_{2}$ is more abundant ($\sim$1.7 times) than $^{34}$SO. In the hot core, we also find that $^{34}$SO$_{2}$ is more abundant (5 times) than $^{34}$SO, whereas in the ER the ratio is much larger ($^{34}$SO$_{2}$ is 14 times more abundant). As was found for SO/SO$_{2}$, the main difference is in the 20.5 km s$^{-1}$ component, where $^{34}$SO is $\sim$100 times more abundant than $^{34}$SO$_{2}$.

Table \ref{table:molecular_abundances} shows the molecular abundances, $X$, of SO and SO$_2$ with respect to hydrogen in each component. They were derived using H$_2$ column density by means of the C$^{18}$O column density, from the isotopic abundance $^{16}$O/$^{18}$O, and assuming that CO is a good tracer of H$_2$ and therefore their abundance ratio is roughly constant. 
The column densities for H$_2$ are 7.5$\times$10$^{22}$, 7.5$\times$10$^{22}$, 2.1$\times$10$^{23}$, 6.2$\times$10$^{22}$, 4.2$\times$10$^{23}$ cm$^{-2}$, and 1.0$\times$10$^{23}$ for the ER, CR, PL, HVP, HC, and the 20.5 km s$_{-1}$ component, respectively (see Tercero et al. 2011).
We observe that the highest abundance of SO is obtained in the HVP, whereas in the HC and in the PL this abundance is $\sim$30 times lower. The extended ridge presents the lowest abundance of sulfur monoxide. The abundance of this molecule in the 20.5 km s$_{-1}$ component is about twice larger than in the HC.
SO$_2$ is also more abundant in the HVP (between 8 and 600 times more abundant than in the rest of components). With respect to hydrogen, SO$_2$ is about one order of magnitude more abundant than SO in the HC and in the ER, while in both plateaus sulfur dioxide is only two-three times more abundant than SO.

\subsection{Other sulfur-bearing molecules}

We provide here upper limits for the column densities of several sulfur-bearing molecules not detected in our survey. We have assumed the same spectral components as for SO and SO$_{2}$ (HC, PL, HVP, CR, and ER) and an LTE approximation, due to the lack of available collision rates. Table \ref{table:components LVG} shows the adopted temperature values, among other parameters, for each component, and Table \ref{table:upper_limits} (see Appendix) shows the results obtained, the dipole moment of each species, and references for the spectroscopic constants.
The upper limit of column density for each species were obtained summing the contribution of all the components.

SO$^{+}$: first detected in the interstellar medium towards the supernova remnant IC443 (Turner et al. 1992), it was proposed as a tracer of dissociative shocks, although later surveys carried out in dark clouds, star-forming regions, and high velocity molecular outflows suggest that this reactive ion is not associated with shock chemistry (Turner et al. 1994). SO$^{+}$ presents a high abundance in PDRs like NGC 7023 and in the Orion Bar (Fuente et al. 2003). In Orion KL, we obtain an upper limit to its column density of $N$(SO$^{+}$)$\le$2.5$\times$10$^{14}$ cm$^{-2}$, providing an abundance ratio of $N$(SO)/$N$(SO$^{+}$)$\ge$2080. This result implies that UV radiation does not play an important role in this region.

($cis$)-HOSO$^{+}$: it is the most stable isomeric form of this ion. This species has not yet been detected in the interstellar medium, but its large dipole moment (1.74 D), its easy formation through H$^{+}_{3}$ reacting with SO$_{2}$ and the fact that it does not react with H$_{2}$ make this ion an excellent candidate for being detected, mainly in hot regions where the parent SO$_{2}$ is very abundant. The upper limit calculated for this ion is $N$($cis$-HOSO$^{+}$)$\le$3.6$\times$10$^{13}$ cm$^{-2}$.

SSO: this molecule has not been detected yet in the interstellar medium, but it is a plausible candidate, since the oxides SO and SO$_{2}$ are particulary abundant, especially in star-forming regions. Disulfur monoxide (SSO) was spectroscopically studied first by Meschi \& Myers (1959) who detected rotational transitions in the ground vibrational state and in the $\nu$$_{2}$=1 state. Later, Thorwirth et al. (2006) carried out a millimeter and submillimeter wave investigation of SSO in the ground vibrational state to frequencies as high as 470 GHz. We have not detected SSO in our line survey, but we obtain an upper limit for its column density of $N$(SSO)$\le$7.6$\times$10$^{14}$ cm$^{-2}$, providing an abundance ratio of $N$(SO)/$N$(SSO)$\ge$1155.

OSiS: silicon oxysulfide was first characterized in the gas phase at high spectral resolution by Thorwirth et al. (2011). It prosseses a large dipole moment ($\mu$$_{a}$=1.47 D) and its bond distances are very short in comparison with those of SiO and SiS. It has not been detected yet in the interstellar medium, and we obtained an upper limit for the column density of this molecule in Orion KL of $N$(OSiS)$\le$6.3$\times$10$^{13}$ cm$^{-2}$. Tercero et al. (2010) found that in Orion KL the total column density for SiS in the ground state is $N$(SiS)=(1.4$\pm$0.4)$\times$10$^{15}$ cm$^{-2}$. This result provides an abundance ratio of $N$(SiS)/$N$(OSiS)$\ge$22.

S$_{3}$: thiozone is a bent chain with a bond to the apex S whose rotational spectrum was first measured by McCarthy et al. (2004). S$_{3}$ has not yet been observed in the interstellar medium; however, it is an excellent candidate for astronomical detection in rich interstellar sources. In addition, S$_{3}$ may also exist in the atmosphere of Io, where S$_{2}$ has already been detected in the ultraviolet. Owing to the presence of more intense lines from other species we have not detected S$_{3}$ in our line survey. We provide an upper limit for its column density of $N$(S$_{3}$)$\le$1$\times$10$^{15}$ cm$^{-2}$.

S$_{4}$: tetrasulfur is a singlet planar trapezoid whose rotational spectrum was observed for the first time by McCarthy et al. (2004). S$_{4}$ has a substantial dipole moment, 0.87 D, hence an intense rotational spectrum across the entire radio band. The upper limit column density we calculated for this molecule is $N$(S$_{4}$)$\le$7$\times$10$^{14}$ cm$^{-2}$.

CH$_{3}$SOCH$_{3}$: Barnes et al. (1994) obtained this molecule in the laboratory while investigating the gas-phase reaction of OH with the oxidation of dimethyl sulfide at room temperature. Dimethyl sulfoxide has not been observed yet in the interstellar medium, but we provide an upper limit for its column density of $N$(CH$_{3}$SOCH$_{3}$)$\le$1$\times$10$^{14}$ cm$^{-2}$.

H$_{2}$CSO: sulfine was first identified in 1976 as a product of the pyrolysis of a variety of sulfur-bearing precursors. H$_{2}$CSO is a planar molecule of C$_{s}$ symmetry. Joo et al. (1995) analyzed its infrared spectrum at high resolution. We provide an upper limit for the column density for this undetected molecule in the interstellar medium of $N$(H$_{2}$CSO)$\le$3$\times$10$^{13}$ cm$^{-2}$.

HNSO: thionylimide is a semi-stable molecule that adopts a cis-planar structure of C$_{s}$ symmetry in the ground state. HNSO is the simplest molecule in the group of organic nitrogen-sulfur compounds. We calculated an upper limit for its column density of $N$(HNSO)$\le$4.1$\times$10$^{14}$ cm$^{-2}$, which provides an abundance ratio $N$(SO)/$N$(HNSO)$\ge$2600.

o-H$_{2}$S$_{2}$: the rotational spectrum of disulfane (H$_{2}$S$_{2}$) has been measured in the far-infrared, millimeter, and submillimeter (Winnewisser et al. 1966). The density of its spectrum is enhanced by the presence of low-lying torsional and S-S stretching modes. We did not observe this molecule in our line survey but we provide an upper limit for its column density of $N$(o-H$_{2}$S$_{2}$)$\le$1.6$\times$10$^{14}$ cm$^{-2}$.

SSH and H$_{2}$SO$_{4}$: these molecules have not yet been detected in the interstellar medium. The calculated upper limits for their column densities are $N$(SSH)$\le$7.1$\times$10$^{13}$ cm$^{-2}$ and $N$(H$_{2}$SO$_{4}$)$\le$1.3$\times$10$^{14}$ cm$^{-2}$, respectively.

CH$_{3}$SSH: methyl hydrodisulfide has not been observed in the interstellar medium. Tyblewski et al. (1986) studied its rotational spectrum, together with that of CH$_{3}$SSD, between 18 and 40 GHz providing rotational constants. We derive an upper limit for its column density of $N$(CH$_{3}$SSH)$\le$2.6$\times$10$^{14}$ cm$^{-2}$.

($tr$)-HCSSH: the spectrum of dithioformic acid has been studied by Bak et al. (1978), who assigned the rotational transitions of this species in its ground state to a $trans$ and $cis$ rotamer. We provide an upper limit for the column densitiy of ($tr$)-HCSSH (more stable) $N$$\le$3.6$\times$10$^{13}$ cm$^{-2}$.

\section{Discussion}
\label{section:discussion}

\begin{figure*}
   \centering
   \includegraphics[angle=0,width=16.5cm]{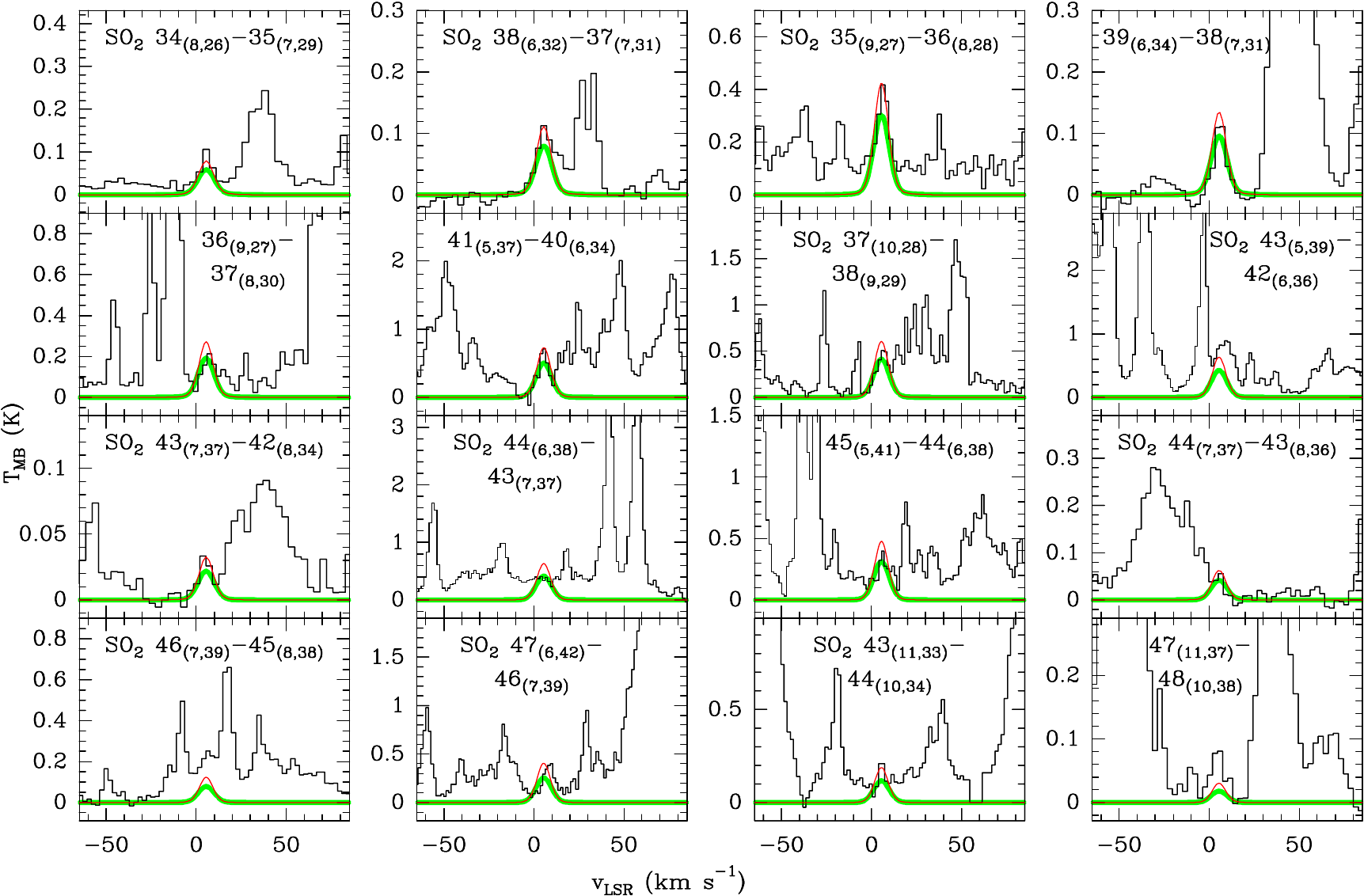}  
   \caption{Observed lines of SO$_{2}$ (black histogram) with energies higher than 700 K, ordered by increasing energy from top left to bottom right. Best fit LTE model results considering a hot core at $T$$_{\mathrm{K}}$=220 K (green curve) and with a hot core at $T$$_{\mathrm{K}}$=280 K (red curve).}
   \label{figure:so2_test_referee}
   \end{figure*}

There have been many spectral line surveys of Orion KL aimed at determining the physical and chemical structure of this region (e.g., Blake et al. 1987, Sutton et al. 1995, Schilke et al. 2001). The survey analyzed here was first presented by Tercero et al. (2010), covers the widest frequency range of all of them (80-281 GHz). Due to this wide range and to the large number of observed transitions of SO, and particularly of SO$_{2}$, it has been possible not only to determine the structure of the cloud (gas temperature, gas density, size of components, etc.) with better accuracy (the 3 mm window shows best the coldest regions, such as the ER, whereas the 1.3 mm window probes the warmest regions), but also to demonstrate the need for considering a density and temperature gradient in the HC of Orion KL. 
From rotational diagrams, we found a large difference ($\sim$100 K) between the rotational temperatures of SO and SO$_{2}$, indicating the possibility of a hotter inner region to the HC. 
We draw the same conclusion from the fits for SO$_{2}$ lines with energies $E$$>$400 K. To obtain better fits for lines with energies $E$$>$700 K, we considered a high column density for the HC, which overestimates in the fit for the lines with energy around 400 K (see Figs. \ref{figure:so2 2mmLVG} and \ref{figure:so2 3mmLVG}). But by considering an additional inner component to the HC with higher temperature, it would probably be possible to obtain a better fit to the lines with high energies, while avoiding overestimation of lines with intermediate energies.
To test this possibility, we fit the SO$_{2}$ line profiles with energies $E$$>$700 K considering a hotter ($T$$_{\mathrm{K}}$=280 K) HC (affecting only the highest energy transitions), with $n$(H$_{2}$)=5$\times$10$^{6}$ cm$^{-3}$, located at 5.5 km s$^{-1}$, and with $\triangle$$v$$_{\mathrm{FWHM}}$=7 km s$^{-1}$. For a SO$_{2}$ column density of (4$\pm$1)$\times$10m$^{16}$ cm$^{-2}$, we improve the line profile fits of these high energies transitions (see Fig. \ref{figure:so2_test_referee}). This shows the existence of a temperature and density gradient in the HC of Orion KL. However, its structure should be determined accurately with observations of SO and SO$_{2}$ (and mainly of its isotopologues\footnote{The energies of the observed isotoplogues in this survey are $<$680 K.}, which are optically thin) at higher frequencies with telescopes such as APEX.

On the other hand, the large number of observed lines of the $^{34}$S and $^{33}$S isotopologues has allowed us to calculate column densities and isotopic and molecular abundances that are key to understanding the chemical evolution of this region.

\subsection{SO and SO$_2$ as tracers of shocks and hot gas}
\label{tracers}

\begin{figure*}
   \centering
   \includegraphics[angle=270,width=10cm]{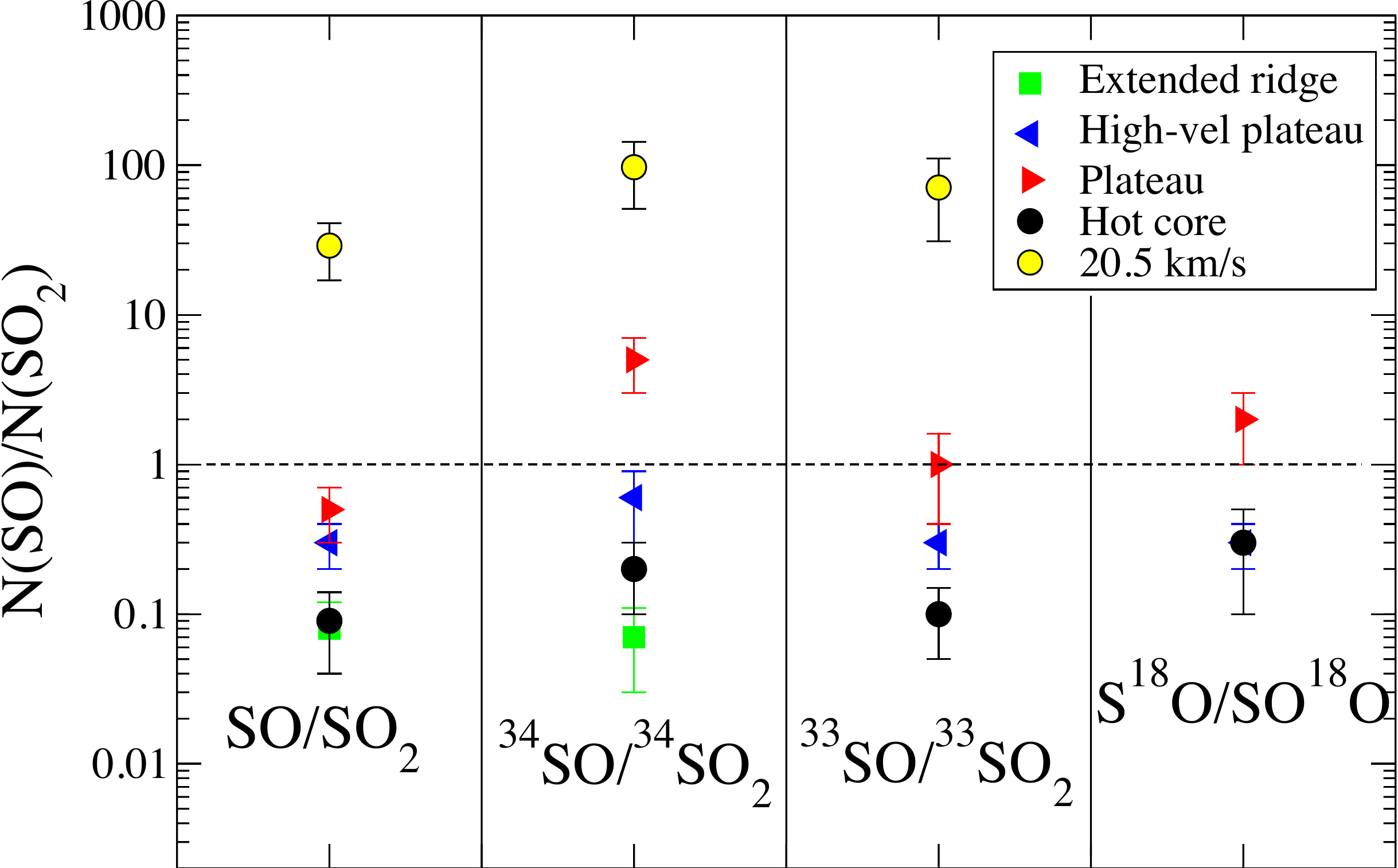}
   \caption{Ratio $N$(SO)/$N$(SO$_2$) for each component and for the different isotopologues of SO and SO$_2$.}
   \label{figure:ratios_abundances}
   \end{figure*}

In Section \ref{section:analysis} we showed that an important contribution to the emission of SO and SO$_{2}$ comes from the HC component of Orion KL. 
In Fig. \ref{figure:ratios_abundances} we plotted the ratio of $N$(SO)/$N$(SO$_2$) for the different components, as well as for the different isotopologues of SO and SO$_2$. The figure shows that the column density of SO$_{2}$ in the HC is higher than that for SO. 
We should take into account that our column densities for SO$_2$ may have been slightly underestimated because of using an LTE model to infer the column density (instead of LVG, as was used for SO). For that reason, we considered a higher uncertainty (35$\%$) in the model intensity predictions for SO$_2$, than for SO (20$\%$), as said previously. Moreover, the opacity may affect SO and SO$_2$ differently, which would in turn affect their column density ratio. But if we consider the result of the ratio $^{34}$SO/$^{34}$SO$_2$ in the hot core, we observe that SO$_2$ continues to be more abundant than SO. This could indicate that SO$_{2}$ is a better tracer of warm gas than SO.

Our results are consistent with predictions from chemical models of hot cores (e.g., Hatchell et al. 1998). 
Viti et al. (2004a) modeled the evaporation of ices near massive stars and found that SO$_{2}$ becomes more abundant than SO in the hot core from 31.500 years after the formation of a high-mass star.
For shorter timescales, SO is much more abundant than SO$_{2}$. Thus, the SO/SO$_2$ ratio could be regarded as a chemical clock (which should decrease with time), and our results showing a lower SO/SO$_2$ ratio for the HC component seem to suggest a late stage for the hot core evolution in Orion KL.

On the other hand, the ratio of SO/SO2 is higher in the PL and the HVP than in the HC and the ER (see Fig. \ref{figure:ratios_abundances}), and in particular for the PL, this ratio reaches values close to or even higher than 1.
Since SO is a well-known outflow tracer (e.g., Chernin et al. 1994; Codella \& Scappini 2003; Lee et al. 2010; Tafalla et al. 2010), and SO seems to be more enhanced than SO2 in shocks (from an observational point of view, e.g., Codella \& Bachiller (1999), Jim\'enez-Serra et al. (2005), and from a theoretical point of view, e.g., Viti et al. (2004b); Benedettini et al. (2006), for timescales $\sim$10,000 yr), it seems very feasible that the high SO/SO$_{2}$ ratios measured in the (high velocity) plateau are the consequence of a definite enhancement of the SO abundance with respect to SO$_{2}$, due to shocks propagating into the surrounding medium of Orion KL.

\subsection{Nature of the 15 km s$^{-1}$ dip and 20.5 km s$^{-1}$ velocity component}
\label{section:nature}

To properly fit the SO and SO$_2$ spectra in Orion KL, a new velocity component at 20.5 km s$^{-1}$ had to be included in the model (see Section \ref{section:feature}). In addition, a possible dip at 15.5 km s$^{-1}$ in the SO and SO$_2$ spectra has been identified. The dip at 15.5 km s$^{-1}$ could be self-absorption due to the high opacity of the observed transitions. However, in all cases, the line strength $S$ is uncorrelated with the amount of absorption. For example, the transition 11$_{(1,11)}$-10$_{(0,10)}$ of SO$_2$, with a line strength of $S$=7.7 and Einstein coefficient $A$$_{\mathrm{ul}}$=1.1x10$^{-4}$ s$^{-1}$, would be expected to display more self-absorption and therefore a lower integrated intensity than the transition 4$_{(2,2)}$-3$_{(1,3)}$ with $S$=1.7 and $A$$_{\mathrm{ul}}$=7.7x10$^{-5}$ s$^{-1}$, but we see the opposite (especially in the range at 14.5-16.5 km s$^{-1}$). In addition, we obtain in Fig. \ref{figure:so2_maps} a high integrated intensity in the velocity range 14-18 km s$^{-1}$, when the integrated intensity in the ranges 10-14 km s$^{-1}$ and 18-22 km s$^{-1}$ is also large. 
Altogether, this suggests that the emission at 15 km s$^{-1}$ is only the sum of contributions of emission coming from the HVP and the 20.5 km s$^{-1}$ component.

With respect to the nature of the emission at 20.5 km s$^{-1}$,  Fig. \ref{figure:ratios_abundances} shows that the column density ratio of SO/SO$_2$ of this component is about two to three orders of magnitude higher than the other velocity components in Orion KL. This is also true for the $^{34}$SO/$^{34}$SO$_2$ and $^{33}$SO/$^{33}$SO$_2$ ratios, suggesting that it is not an opacity effect. Such a high ratio could be due in part to filling factor problems, if the 20.5 km s$^{-1}$ component is much more compact in SO$_2$ than in SO\footnote{In fact, interferometric maps of $^{34}$SO in Orion KL reveal emission only from 1 to 15 km s$^{-1}$ (Beuther et al. 2005), indicating that the emission from the 20 km s$^{-1}$ component has probably been filtered out by the interferometer. Given the minimum baseline of the interferometric observations (Beuther et al. 2005), the largest angular scale detectable is $\sim$6$\arcsec$ (following the Appendix in Palau et al. 2010), similar to the size adopted in this work for the 20.5 km s$^{-1}$ component.}, and/or could be the result of applying a different method (LVG for SO vs LTE for SO$_2$) to infer the column densities. However, the large difference compared to the other components suggests that it is a definite chemical effect, and since the SO/SO$_2$ ratio is higher in regions associated with shocks such as the HVP and the PL (Sect. \ref{tracers}), the 20.5 km s$^{-1}$ component could be related to shocks as well, maybe associated with the explosive dynamical interaction that took place in Orion KL (G\'omez et al. 2005, Zapata et al. 2011a) and more especifically to shocks associated with the BN object. This is consistent with the fact that the BN object in Orion presents significantly high CO and $^{13}$CO emission at $\sim$20 km s$^{-1}$ (Scoville et al. 1983). 
Observations combining both single-dish and interferometric data are required to definitely identify the spatial region in Orion KL emitting the bulk of emission at 20 km s$^{-1}$ .

\section{Summary and Conclusions}
\label{section:summary}

This study is part of a series of papers with the goal of analyzing the physical and chemical conditions of Orion KL. The study is divided into different molecular families, and here we have focused on the emission lines of SO and SO$_{2}$ and their isotopologues. We have analyzed the IRAM 30-m line survey of Orion KL observed by Tercero et al. (2010), which covers the frequency range 80-281 GHz. 
We identified more than 700 rotational transitions of these molecules, including lines from the vibrational state $\nu$$_{2}$=1 of SO$_{2}$ and the isotopologue SO$^{17}$O, detected for the first time in the interstellar medium. This large sample has let us improve our knowledge about the physical and chemical conditions in Orion KL, especially due to the observation of a large number of SO$_{2}$ transitions at high energies. 
The analysis of SO and SO$_{2}$ was carried out using an LTE and LVG radiative transfer model, taking the physical structure of the source  into account (hot core, compact ridge, extended ridge, and plateau components).

First, we fit SO and SO$_{2}$ lines with Gaussian profiles to obtain an approximate $T$$_{\mathrm{rot}}$ value in each component. We detected a dip at $\sim$15 km s$^{-1}$ in most of the lines and an emission peak centered on 20.5 km s$^{-1}$. For the dip at 15 km s$^{-1}$, we discarded self-absorption as a possible cause, concluding instead that the weak emission is due to the sum of small contributions coming from the high velocity plateau and from the 20.5 km s$^{-1}$ feature, which corresponds to an unresolved component ($\sim$5$\arcsec$´´ diameter), with line width of $\sim$7.5 km s$^{-1}$ and with an especially high column density of SO in comparison to SO$_{2}$. Its rotational temperature is 50$\pm$10 K from SO lines and 90$\pm$20 K from SO$_{2}$ lines.
For the rest of the components, the rotational temperatures obtained from SO$_{2}$ lines are: plateau (PL)=120$\pm$20 K, hot core (HC)=190$\pm$60 K, high velocity plateau (HVP)=110$\pm$20 K, compact ridge (CR)=80$\pm$30 K, and extended ridge (ER)=83$\pm$40. The results from SO lines are: (PL)=130$\pm$20 K, (HC)=288$\pm$90 K, (HVP)=111$\pm$15 K, and (ER)=107$\pm$40 K.

The second part of the analysis was carried out using a radiative transfer code.  
For the case of SO, we analyzed its non-LTE excitation, however, for SO$_{2}$ we assumed LTE excitation due to the lack of collisional rates for energies higher than 90 K (we observe SO$_{2}$ lines with energies up to 1500 K). We found that most of the emission of SO$_{2}$ and SO arises from the high-velocity plateau, with column densities of $N$(SO$_{2}$)=(1.3$\pm$0.3)$\times$10$^{17}$ cm$^{-2}$ and $N$(SO)=(5$\pm$1)$\times$10$^{16}$ cm$^{-2}$, respectively, and from the hot core, in particular in the case of SO$_{2}$, whose column density is similar to that obtained in the high-velocity plateau. These values are up to three orders of magnitude higher than the column densities obtained for the ridge components. 
These results let us conclude that SO and SO$_{2}$ are not good tracers only of shock-affected areas, but also of hot dense gas.
In addition, from the ratios $^{34}$SO/$^{34}$SO$_{2}$, $^{33}$SO/$^{33}$SO$_{2}$, and S$^{18}$O/SO$^{18}$O$_{2}$ in the different components of the cloud, we observe that in the high-velocity plateau (region affected by shocks) sulfur dioxide is up to five times more abundant than SO. The same trend is found in the hot core.

We have also carried out 2$\arcmin$$\times$2$\arcmin$ mapping around Orion IRc2 in a number of lines of SO, SO$_{2}$, and their $^{34}$S isotopologues. In Sect. \ref{section:maps} we presented maps of three transitions of SO$_{2}$ (Fig. \ref{figure:so2_maps}), two transitions of SO (Fig. \ref{figure:so_maps}), one transition of $^{34}$SO and one of $^{34}$SO$_{2}$ (Fig. \ref{figure:34so_34so2_maps}). We plotted different velocity ranges for each transition to explore the spatial distribution of the emission. In agreement with our column density results, we found the maximum integrated intensities in the range containing the hot core (3-7 km s$^{-1}$) and in the range 10-14 km s$^{-1}$ (corresponding to the high velocity plateau), whose emission peak is centered approximately 4$\arcsec$ to the southwest of IRc2.
In all mapped transitions, but especially in those of SO and $^{34}$SO, we observe an elongation of the gas along the NE-SW direction. 
In these maps, we also detected a strong emission in the velocity range located at 20.5 km s$^{-1}$. From the spatial distribution of this feature and from the analysis of the line profiles, we suggest that this emission is probably related to shocks associated to the $BN$ source or to a gas cloudlet ejected in the explosive event that could have taken place in Orion KL.

In this paper, we have also demonstrated the need to consider a temperature and density gradient in the hot core of Orion KL, with a comparison between fits of SO$_{2}$ line profiles at high energies, assuming two different temperatures ($T$$_{\mathrm{K}}$=220 K and $T$$_{\mathrm{K}}$=280 K) in the hot core. Only with the low temperature it was not possible to obtain good line fits for $E$$>$700 K, without avoiding overestimation for lines with intermediate energies. In addition, the large difference between the rotational temperatures in the hot core and the need to consider a large contribution to the SO$_{2}$ isotopologue emission in the extended ridge support the conclusion of the presence of temperature and density gradients in Orion KL. However, it would be necessary to also consider emission lines (mainly from isotopologues) spanning a wider frequency range with observations from other telescopes, such as APEX, in order to determine these gradients accurately.
Moreover, to describe this molecular cloud in greater detail while avoiding spectral confusion would require interferometric observations with higher spectral resolution and higher sensitivity (such as those provided by ALMA).

\begin{acknowledgements}

We thank the Spanish MICINN for funding support through grants AYA2006-14876, AYA2009-07304, and CSD2009-00038. J.R.G. is supported by a Ram\'on y Cajal research contract. A.P. is supported by a JAE-Doc CSIC fellowship co-funded with the European Social Fund under the program ''Junta para la Ampliaci\'on de Estudios'', by the Spanish MICINN grant AYA2011-30228-C03-02 (co-funded with FEDER funds), and by the AGAUR grant 2009SGR1172 (Catalonia). T. A. B. is supported by a JAE-Doc research contract.
  
\end{acknowledgements}

\begin{appendix}
\section{Figures and tables}

\pagebreak

\pagebreak

\begin{landscape}
\begin{table}
\caption{SO emission line parameters obtained from Gaussian fits.}
\begin{center}
\begin{tabular}{llllllllllllllll}
\hline 
\hline
Species/ & \multicolumn{3}{c}{Plateau} & \multicolumn{3}{c}{High-velocity plateau} & \multicolumn{3}{c}{Hot core} & \multicolumn{3}{c}{Extended ridge} & \multicolumn{3}{c}{20.5 km s$^{-1}$ component}\\ 
Transition & $v$$_{\mathrm{LSR}}$ & $\Delta$$v$ & $T$$^{\star}_{\mathrm{A}}$  &
$v$$_{\mathrm{LSR}}$ & $\Delta$$v$ & $T$$^{\star}_{\mathrm{A}}$ &
$v$$_{\mathrm{LSR}}$ & $\Delta$$v$ & $T$$^{\star}_{\mathrm{A}}$ & $v$$_{\mathrm{LSR}}$ & $\Delta$$v$ & $T$$^{\star}_{\mathrm{A}}$ & $v$$_{\mathrm{LSR}}$ & $\Delta$$v$ & $T$$^{\star}_{\mathrm{A}}$\\
 & (km s$^{-1}$) & (km s$^{-1}$) & (K) & (km s$^{-1}$) & (km s$^{-1}$) & (K) & (km s$^{-1}$) & (km s$^{-1}$) & (K) & (km s$^{-1}$) & (km s$^{-1}$) & (K) & (km s$^{-1}$) & (km s$^{-1}$) & (K) \\
\hline
\\
   SO 2$_{2}$-1$_{1}$ &  6.8 & 27$\pm$3 & 3.86 & 10.8 & 34$\pm$1 & 5.00 & 5.5 & 11.2$\pm$0.3 & 2.20 & 8.8 & 6.5$\pm$0.4 & 1.10 & 20.5 & 8.5$\pm$0.5 & 0.91 \\
   SO 2$_{3}$-1$_{2}$ &  6.6 & 27$\pm$1 & 6.70 & 11.7 & 35$\pm$1 & 11.0 & 5.5 & 10.5$\pm$0.5 & 4.20 & 8.5 & 5.6$\pm$0.2 & 1.51 & 21.0 & 7.5$\pm$0.4 & 0.93\\
   SO 3$_{2}$-2$_{1}$ & 6.8 & 27$\pm$5 & 7.00 & 11.8 & 34$\pm$2 & 7.15 & 5.3 & 11.5$\pm$0.8 & 3.70 & 9.0 & 5.4$\pm$0.2 & 1.50 & 20.5 & 7.5$\pm$0.4 & 1.58 \\
   SO 3$_{4}$-2$_{3}$ & 6.0 & 27$\pm$3 & 10.45 & 11.4 & 37$\pm$4 & 16.0 & 5.9 & 9.0$\pm$0.3 & 3.70 & ... & ... & ... & 20.5 & 7.5$\pm$0.3 & 1.50 \\
   SO 4$_{3}$-3$_{2}$ & 6.6 & 23$\pm$1 & 11.50 & 11.0 & 36$\pm$4 & 13.2 & 5.5 & 8.4$\pm$0.5 & 7.10 & ... & ... & ... & 21.6 & 7.5$\pm$0.4 & 2.25\\
   SO 4$_{4}$-3$_{3}$ & 6.0 & 25$\pm$2 & 18.00 & 11.9 & 36$\pm$3 & 17.9 & 4.5 & 10.0$\pm$0.2 & 5.00 & ... & ... & ... & 21.5 & 7.5$\pm$0.3 & 4.00\\
   SO 5$_{4}$-4$_{3}$ & 6.7 & 24$\pm$5 & 10.93 & 12.7 & 37$\pm$1 & 12.0 & 5.7 & 9.5$\pm$0.4 & 10.0 & ... & ... & ... & 21.5 & 7.5$\pm$0.5 & 2.88\\
   SO 5$_{5}$-4$_{4}$ & 6.4 & 28$\pm$4 & 18.00 & 12.9 & 42$\pm$4 & 11.0 & 3.9 & 9.4$\pm$0.5 & 5.50 & ... & ... & ... & 20.0 & 8.3$\pm$0.6 & 5.00\\ 
   SO 5$_{6}$-4$_{5}$ & 6.3 & 28$\pm$4 & 16.95 & 13.5 & 39$\pm$2 & 14.0 & 4.3 & 9.3$\pm$0.4 & 8.50 & 8.3 & 5.9$\pm$0.3 & 2.00 & ... & ... & ...\\
   SO 6$_{5}$-5$_{4}$ & 6.0 & 29$\pm$3 & 20.00 & 12.5 & 36$\pm$1 & 10.5 & 3.5 & 7.5$\pm$0.6 & 6.30 & 9.0 & 4.0$\pm$0.2 & 3.49 & ... & ... & ...\\
   SO 6$_{6}$-5$_{5}$ & 6.7 & 35$\pm$3 & 12.62 & 11.7 & 35$\pm$2 & 8.99 & ... & ... & ... & 8.5 & 6.1$\pm$0.5 & 1.26 & ... & ... & ... \\
   SO 6$_{7}$-5$_{6}$ & 6.0 & 33$\pm$4 & 20.00 & 13.3 & 37$\pm$4 & 16.0 & 1.5 & 12.0$\pm$0.5 & 8.80 & ... & ... & ... & 21.5 & 8.5$\pm$0.5 & 5.24\\
 
\hline                  
\end{tabular}
\end{center}
\onecolumn
\tablefoot{The fit errors are provided by CLASS. $v$$_{\mathrm{LSR}}$ is the LSR central velocity, $\Delta$$v$ is the line width, and $T$$^{\star}_{\mathrm{A}}$ is the antenna temperature.}\\
\label{table:so parameters gaussian}
\end{table}
\end{landscape}

\begin{landscape}
\begin{table}
\caption{SO$_{2}$ parameters from Gaussian fits (lines at 1.3 mm).}
\begin{center}
\begin{tabular}{lllllllllllllllllll}
\hline 
\hline
 & \multicolumn{3}{c}{Plateau} & \multicolumn{3}{c}{High-velocity} & \multicolumn{3}{c}{Hot} & \multicolumn{3}{c}{Compact} & \multicolumn{3}{c}{Extended} & \multicolumn{3}{c}{20.5 km s$^{-1}$}\\ 
Transition &  & & & \multicolumn{3}{c}{plateau} & \multicolumn{3}{c}{core} & \multicolumn{3}{c}{ridge} & \multicolumn{3}{c}{ridge} & \multicolumn{3}{c}{component}\\
 & $v$$_{\mathrm{LSR}}$ & $\Delta$$v$ & $T$$^{\star}_{\mathrm{A}}$ &
$v$$_{\mathrm{LSR}}$ & $\Delta$$v$ & $T$$^{\star}_{\mathrm{A}}$ &
$v$$_{\mathrm{LSR}}$ & $\Delta$$v$ & $T$$^{\star}_{\mathrm{A}}$ & 
$v$$_{\mathrm{LSR}}$ & $\Delta$$v$ & $T$$^{\star}_{\mathrm{A}}$ & 
$v$$_{\mathrm{LSR}}$ & $\Delta$$v$ & $T$$^{\star}_{\mathrm{A}}$ & 
$v$$_{\mathrm{LSR}}$ & $\Delta$$v$ & $T$$^{\star}_{\mathrm{A}}$\\
 & (km s$^{-1}$) & (km s$^{-1}$) & (K) &  &  & &  & & &  &  & &  & &  &  &  & \\
\hline
1.3 mm\\
\hline
\\
   3$_{(2,2)}$-2$_{(1,1)}$ & 6.4 & 26$\pm$1 & 5.50 & 12.9 & 35$\pm$3 & 5.00 & 5.0 & 10.0$\pm$0.2 & 6.60 & ... & ... & ... & ... & ... & ... & 21.0 & 8.0$\pm$0.4 & 3.50 \\  
   4$_{(2,2)}$-3$_{(1,3)}$ & 6.0 & 25$\pm$1 & 6.00 & 12.6 & 33$\pm$5 & 4.60 & 4.5 & 8.0$\pm$0.4 & 5.47 & ... & ... & ... & 9.0 & 4.1$\pm$0.4 & 0.83 & 22 & 7.5$\pm$0.5 & 2.60\\
   4$_{(3,1)}$-4$_{(2,2)}$ & 6.0 & 25$\pm$1 & 2.40 & 11.5 & 32$\pm$3 & 3.60 & 3.7 & 8.9$\pm$0.4 & 1.61 & 7.8 & 4.1$\pm$0.5 & 1.37 & ... & ... & ... & 21.0 & 8.0$\pm$0.3 & 0.69 \\
   5$_{(2,4)}$-4$_{(1,3)}$ & 5.9 & 26$\pm$1 & 7.00 & 12.8 & 38$\pm$5 & 6.40 & 3.8 & 12.2$\pm$0.3  & 7.90 & ... & ... & ... & ... & ... & ... & 20.5 & 8.0$\pm$0.3 & 3.86  \\
   6$_{(4,2)}$-7$_{(3,5)}$ & 6.0 & 24$\pm$2 & 1.00 & 11.5 & 29$\pm$1 & 1.40 & 5.0 & 7.0$\pm$0.5 & 3.30 & ... & ... & ... & ... & ... & ... & 20.5 & 8.0$\pm$0.3 & 0.14  \\
   7$_{(2,6)}$-6$_{(1,5)}$ & 6.0 & 25$\pm$1 & 5.00 & 12.8 & 35$\pm$2 & 3.50 & 5.0 & 11.0$\pm$0.2 & 2.50 & ... & ... & ... & 8.9 & 5.0$\pm$0.3 & 0.87 & 20.5 & 7.5$\pm$0.4 & 2.00  \\   
   7$_{(3,5)}$-7$_{(2,6)}$ & 6.0 & 23$\pm$2 & 3.00 & 12.9 & 33$\pm$2 & 3.30 & 4.5 & 9.6$\pm$0.3 & 0.55 & 7.5 & 5.0$\pm$0.3 & 1.31 & ... & ... & ... & 20.5 & 7.5$\pm$0.4 & 0.60  \\
   7$_{(4,4)}$-8$_{(3,5)}$ & 6.3 & 24$\pm$1 & 2.80 & 11.6 & 33$\pm$1 & 1.42 & 4.4 & 7.7$\pm$0.5 & 1.40 & ... & ... & ... & ... & ... & ... & 20.5 & 7.5$\pm$0.4 & 0.43 \\
   11$_{(1,11)}$-10$_{(0,10)}$ & 6.0 & 30$\pm$4 & 7.00 & 12.0 & 35$\pm$1 & 9.50 & 4.0 & 11.9$\pm$0.2 & 9.00 & 7.7 & 4.9$\pm$0.3 & 2.00 & ... & ... & ... & 20.5 & 7.5$\pm$0.5 & 2.70  \\
   11$_{(2,10)}$-11$_{(1,11)}$ & 6.0 & 25$\pm$3 & 7.00 & 12.9 & 33$\pm$4 & 4.40 & 5.0 & 8.4$\pm$0.3 & 7.00 & ... & ... & ... & ... & ... & ... & 21.0 & 7.5$\pm$0.4 & 2.00\\
   11$_{(3,9)}$-11$_{(2,10)}$ & 6.0 & 25$\pm$2 & 3.56 & 12.7 & 35$\pm$6 & 4.30 & 5.7 & 9.0$\pm$0.3 & 3.02 & ... & ... & ... & ... & ... & ... & 21.0 & 8.0$\pm$0.3 & 1.31  \\
   11$_{(5,7)}$-12$_{(4,8)}$ & 6.2 & 25$\pm$2 & 1.23 & 11.7 & 31$\pm$1 & 1.10 & 5.5 & 7.5$\pm$0.5 & 0.88 & ... & ... & ... & ... & ... & ... & 20.5 & 7.0$\pm$0.5 & 0.21  \\
   13$_{(1,13)}$-12$_{(0,12)}$ & 6.0 & 25$\pm$3 & 11.00 & 12.7 & 38$\pm$3 & 10.6 & 5.9 & 10.0$\pm$0.2 & 9.15 & ... & ... & ... & ... & ... & ... & 21.0 & 8.0$\pm$0.2 & 2.99\\
   13$_{(3,11)}$-13$_{(2,12)}$ & 6.0 & 28$\pm$1 & 10.50 & 10.8 & 30$\pm$1 & 10.4 & ... & ... & ... & ... & ... & ... & ... & ... & ... & ... & ... & ... \\
   14$_{(3,11)}$-14$_{(2,12)}$ & 6.6 & 27$\pm$4 & 6.50 & 12.9 & 32$\pm$1 & 5.0 & 4.5 & 9.3$\pm$0.2 & 10.0 & 7.2 & 5.5$\pm$0.2 & 1.46 & ... & ... & ... & 21.5 & 7.5$\pm$0.3 & 2.30 \\  
   14$_{(6,8)}$-15$_{(5,11)}$ & 6.9 & 25$\pm$2 & 1.60 & ... & ... & ... & 5.0 & 7.2$\pm$0.4 & 1.54 & ... & ... & ... & ... & ... & ... & 19.3 & 7.5$\pm$0.3 & 0.37 \\
   15$_{(3,13)}$-15$_{(2,14)}$ & 6.0 & 24$\pm$2 & 4.20 & 11.6 & 35$\pm$1 & 3.90 & 5.5 & 8.6$\pm$0.2 & 1.92 & ... & ... & ... & ... & ... & ... & 21.0 & 7.5$\pm$0.4 & 1.30\\
   16$_{(1,15)}$-15$_{(2,14)}$ & 5.9 & 25$\pm$3 & 4.00 & 12.6 & 33$\pm$2 & 4.30 & 4.8 & 10.0$\pm$0.1 & 4.89 & ... & ... & ... & ... & ... & ... & 21.5 & 7.5$\pm$0.3 & 2.05\\
   16$_{(1,15)}$-16$_{(0,16)}$ & 6.0 & 24$\pm$2 & 3.50 & 12.3 & 34$\pm$3 & 3.30 & 5.7 & 13.0$\pm$0.5 & 4.42 & 7.0 & 4.0$\pm$0.5 & 0.49 & ... & ... & ... & 20.5 & 7.5$\pm$0.4 & 1.67 \\
   16$_{(3,13)}$-16$_{(2,14)}$ & 6.0 & 25$\pm$3 & 5.79 & 12.0 & 34$\pm$3 & 2.50 & 4.9 & 9.0$\pm$0.3 & 5.57 & ... & ... & ... & ... & ... & ... & 21.0 & 7.5$\pm$0.4 & 1.50\\
   18$_{(3,15)}$-18$_{(2,16)}$ & 6.0 & 24$\pm$1 & 4.50 & 12.3 & 32$\pm$1 & 5.00 & 4.5 & 10.3$\pm$0.2 & 5.10 & ... & ... & ... & ... & ... & ... & 20.5 & 7.5$\pm$0.3 & 1.98\\
   20$_{(3,17)}$-20$_{(2,18)}$ & 6.0 & 28$\pm$3 & 5.00 & 12.5 & 41$\pm$3 & 3.60 & 4.8 & 9.1$\pm$0.3 & 6.09 & ... & ... & ... & ... & ... & ... & 20.5 & 8.5$\pm$0.2 & 1.54\\
   20$_{(7,13)}$-21$_{(6,16)}$ & 6.0 & 20$\pm$1 & 0.40 & 11.5 & 24$\pm$2 & 0.45 & 5.2 & 6.0$\pm$0.6 & 0.77 & ... & ... & ... & ... & ... & ... & ... & ... & ... \\
   24$_{(3,21)}$-24$_{(2,22)}$ & 6.0 & 27$\pm$3 & 3.00 & 11.5 & 36$\pm$1 & 0.98 & 5.3 & 8.8$\pm$0.3 & 4.38 & ... & ... & ... & ... & ... & ... & 20.5 & 7.5$\pm$0.3 & 0.60\\
   26$_{(3,23)}$-26$_{(2,24)}$ & 6.9 & 25$\pm$4 & 3.50 & ... & ... & ... & 5.0 & 7.8$\pm$0.5 & 3.95 & ... & ... & ... & ... & ... & ... & ... & ... & ... \\
   28$_{(4,24)}$-28$_{(3,25)}$ & 6.3 & 23$\pm$3 & 2.65 & 12.4 & 27$\pm$1 & 1.85 & 4.0 & 8.7$\pm$0.3 & 3.76 & ... & ... & ... & ... & ... & ... & ... & ... & ... \\
\hline
\end{tabular}
\end{center}
\onecolumn
\tablefoot{The fit errors are provided by CLASS. $v$$_{\mathrm{LSR}}$ is the LSR central velocity, $\Delta$$v$ is the line width, and $T$$^{\star}_{\mathrm{A}}$ is the antenna temperature. The units of $v$$_{\mathrm{LSR}}$, $\Delta$$v$, and $T$$^{\star}_{\mathrm{A}}$ are (km s$^{-1}$), (km s$^{-1}$), and (K), respectively, in all the cases.}\\
\label{table:so2 parameters gaussianI}
\end{table}
\end{landscape}

\pagebreak

\begin{landscape}
\begin{table}
\caption{SO$_{2}$ parameters from Gaussian fits (lines at 2-3 mm).}
\begin{center}
\begin{tabular}{lllllllllllllllllll}
\hline 
\hline
 & \multicolumn{3}{c}{Plateau} & \multicolumn{3}{c}{High-velocity} & \multicolumn{3}{c}{Hot} & \multicolumn{3}{c}{Compact} & \multicolumn{3}{c}{Extended} & \multicolumn{3}{c}{20.5 km s$^{-1}$}\\ 
Transition &  & & & \multicolumn{3}{c}{plateau} & \multicolumn{3}{c}{core} & \multicolumn{3}{c}{ridge} & \multicolumn{3}{c}{ridge} & \multicolumn{3}{c}{component}\\
& $v$$_{\mathrm{LSR}}$ & $\Delta$$v$ & $T$$^{\star}_{\mathrm{A}}$ &
$v$$_{\mathrm{LSR}}$ & $\Delta$$v$ & $T$$^{\star}_{\mathrm{A}}$ &
$v$$_{\mathrm{LSR}}$ & $\Delta$$v$ & $T$$^{\star}_{\mathrm{A}}$ & 
$v$$_{\mathrm{LSR}}$ & $\Delta$$v$ & $T$$^{\star}_{\mathrm{A}}$ & 
$v$$_{\mathrm{LSR}}$ & $\Delta$$v$ & $T$$^{\star}_{\mathrm{A}}$ & 
$v$$_{\mathrm{LSR}}$ & $\Delta$$v$ & $T$$^{\star}_{\mathrm{A}}$  \\
 & (km s$^{-1}$) & (km s$^{-1}$) & (K) & &  & &  &  & &  &  &  &  &  &  &  &  & \\
\hline
2 mm\\
\hline
   2$_{(2,0)}$-2$_{(1,1)}$ & 6.0 & 22$\pm$2 & 4.00 & 11.7 & 32$\pm$2 & 3.00 & 5.5 & 9.0$\pm$0.3 & 1.20 & ... & ... & ... & ... & ... & ... & 20.5 & 7.5$\pm$0.3 & 1.15\\
   3$_{(2,2)}$-3$_{(1,3)}$ & 6.0 & 25$\pm$4 & 4.40 & 12.5 & 32$\pm$2 & 4.10 & 5.5 & 9.1$\pm$0.3 & 5.37 & ... & ... & ... & ...  & ... &... & 20.5 & 7.5$\pm$0.2 & 1.00\\
   5$_{(1,5)}$-4$_{(0,4)}$ & 6.0 & 24$\pm$1 & 6.09 & 11.7 & 33$\pm$1 & 6.40 & 5.5 & 8.0$\pm$0.2 & 2.97 & 8.0 & 3.6$\pm$0.5 & 2.79 & ... & ... & ... & 21.0 & 7.0$\pm$0.2 & 1.50 \\
   5$_{(2,4)}$-5$_{(1,5)}$ & 6.0 & 26$\pm$1 & 5.00 & 12.0 & 32$\pm$3 & 8.00 & 5.5 & 9.2$\pm$0.9 & 5.51 & ... & ... & ... & ... & ... & ... & 20.5 & 7.5$\pm$0.3 & 1.00\\
   6$_{(2,4)}$-6$_{(1,5)}$ & 6.0 & 25$\pm$1 & 5.94 & 11.6 & 33$\pm$3 & 5.50 & 5.5 & 11.1$\pm$0.3 & 3.05 & 8.0 & 6.2$\pm$0.3 & 0.40 & ... & ... & ... & 20.5 & 7.5$\pm$0.2 & 1.80\\
   7$_{(1,7)}$-6$_{(0,6)}$ & 6.0 & 25$\pm$3 & 8.00 & 12.0 & 33$\pm$2 & 10.0 & 5.8 & 9.0$\pm$0.1 & 8.16 & ... & ... & ... & ... & ... & ... & 20.5 & 7.5$\pm$0.1 & 3.00 \\
   7$_{(2,6)}$-7$_{(1,7)}$ & 6.0 & 25$\pm$1 & 6.00 & 12.3 & 34$\pm$3 & 6.60 & 5.5 & 8.8$\pm$0.1 & 6.32 & ... & ... & ... & ... & ... & ... & 21.0 & 7.5$\pm$0.3 & 2.10 \\
   8$_{(2,6)}$-8$_{(1,7)}$ & 6.0 & 23$\pm$1 & 5.70 & 12.7 & 34$\pm$3 & 6.00 & 5.7 & 7.0$\pm$0.4 & 5.48 & ... & ... & ... & 9.5 & 5.1$\pm$0.3 & 1.52 & ... & ... & ...\\
   10$_{(0,10)}$-9$_{(1,9)}$ & 6.0 & 25$\pm$1 & 11.0 & 11.9 & 35$\pm$2 & 8.00 & 4.0 & 10.0$\pm$0.5 & 5.00 & 8.0 & 3.0$\pm$0.5 & 4.49 & ... & ... & ... & 20.5 & 7.5$\pm$0.1 & 2.50\\
   12$_{(1,11)}$-12$_{(0,12)}$ & 6.0 & 25$\pm$1 & 3.92 & 12.1 & 33$\pm$3 & 4.20 & 5.5 & 11.7$\pm$0.7 & 2.46 & ... & ... & ... & 9.0 & 6.1$\pm$0.5 & 0.84 & 20.5 & 7.5$\pm$0.3 & 1.40 \\
   14$_{(1,13)}$-14$_{(0,14)}$ & 6.0 & 25$\pm$1 & 6.00 & 11.7 & 32$\pm$5 & 6.93 & 5.0 & 10.0$\pm$0.3 & 4.50 & 7.3 & 4.5$\pm$0.5 & 1.96 & ... & ... & ... & 20.5 & 7.5$\pm$0.2 & 2.00\\
   14$_{(2,12)}$-14$_{(1,13)}$ & 6.2 & 23$\pm$2 & 6.50 & 11.9 & 34$\pm$5 & 6.40 & 5.8 & 8.2$\pm$0.2 & 4.89 & ... & ... & ... & 9.0 & 5.4$\pm$0.3 & 1.44 & 20.5 & 7.5$\pm$0.2 & 2.00\\
   16$_{(2,14)}$-16$_{(1,15)}$ & 6.0 & 25$\pm$3 & 6.00 & 11.6 & 33$\pm$3 & 5.50 & 5.5 & 10.2$\pm$0.2 & 4.05 & 7.0 & 6.0$\pm$0.2 & 0.20 & ... & ... & ... & 20.5 & 7.5$\pm$0.2 & 1.80\\
   18$_{(2,16)}$-18$_{(1,17)}$ & 6.0 & 24$\pm$3 & 6.50 & 11.6 & 32$\pm$2 & 6.00 & 5.0 & 10.3$\pm$0.2 & 3.00 & 7.5 & 6.1$\pm$0.2 & 1.72 & ... & ... & ... & 20.5 & 7.5$\pm$0.3 & 1.59\\ 
\hline
3 mm\\
\hline
   8$_{(1,7)}$-8$_{(0,8)}$ & 6.0 & 22$\pm$1 & 3.67 & 11.3 & 32$\pm$1 & 4.40 & 5.5 & 9.1$\pm$0.2 & 1.15 & 8.0 & 5.5$\pm$0.4 & 2.21 & ... & ... & ... & 20.5 & 7.5$\pm$0.4 & 0.90\\
   8$_{(3,5)}$-9$_{(2,8)}$ & 6.0 & 24$\pm$2 & 0.82 & 11.7 & 30$\pm$3 & 1.10 & 5.8 & 10.0$\pm$0.2 & 0.83 & ... & ... & ... & ... & ... & ... & ... & ... & ... \\
   10$_{(1,9)}$-10$_{(0,10)}$ & 6.0 & 24$\pm$1 & 3.98 & 11.4 & 32$\pm$3 & 6.00 & 5.5 & 10.9$\pm$0.1 & 3.50 & ... & ... & ... & 8.6 & 5.5$\pm$0.4 & 1.19 & 20.5 & 7.5$\pm$0.2 & 1.50\\

\hline
\end{tabular}
\end{center}
\tablefoot{The fit errors are provided by CLASS. $v$$_{\mathrm{LSR}}$ is the LSR central velocity, $\Delta$$v$ is the line width, and $T$$^{\star}_{\mathrm{A}}$ is the antenna temperature. The units of $v$$_{\mathrm{LSR}}$, $\Delta$$v$, and $T$$^{\star}_{\mathrm{A}}$ are (km s$^{-1}$), (km s$^{-1}$), and (K), respectively, in all the cases.}\\
\label{table:so2 parameters gaussianII}
\end{table}
\end{landscape}

\begin{table*}
\caption{Rotational temperatures, $T$$_{\mathrm{rot}}$, and column densities, $N$, obtained from rotational diagrams.}             
\begin{center}  
\begin{tabular}{c c c c c c c }     
\hline\hline       
    Component        & $T$$_{\mathrm{rot}}$(SO)   & $N$(SO) $\times$10$^{15}$  & C$_{\tau}$(SO) & $T$$_{\mathrm{rot}}$(SO$_{2}$) &  $N$(SO$_{2}$) $\times$10$^{15}$  &  C$_{\tau}$(SO$_{2}$) \\
  & (K) & (cm$^{-2}$) &  & (K) & (cm$^{-2}$) &  \\
\hline                       
Extended ridge (ER)           & 107$\pm$40  & 0.017$\pm$0.001  & 1.00-1.02   & 83$\pm$40    & 0.023$\pm$0.009 & 1.04-1.15  \\  
Compact ridge (CR)            & ...         & ...              & ...         & 80$\pm$30    & 1.2$\pm$0.2     & 1.00-1.63  \\
High-velocity plateau (HVP)    & 111$\pm$15  & 3.9$\pm$0.5      & 1.01-2.12   & 110$\pm$20   & 9.5$\pm$0.6     & 1.0-1.60   \\
Plateau (PL)                  & 130$\pm$20  & 4.6$\pm$0.2      & 1.01-1.11   & 120$\pm$20   & 13$\pm$1        & 1.01-1.69  \\
Hot core  (HC)                & 288$\pm$90  & 18$\pm$1         & 1.02-1.64   & 190$\pm$60   & 33$\pm$3        & 1.00-1.71  \\
20.5 km s$^{-1}$ comp.        &  51$\pm$10  & 0.014$\pm$0.02   & 1.06-1.36   & 90$\pm$20    & 4.6$\pm$0.1     & 1.00-1.58  \\
\hline
\label{table:rotational}                  
\end{tabular}
\end{center}
\tablefoot{C$_{\tau}$ is the range of optical depths in each component.}\\
\end{table*}

\begin{table*}
\centering
\caption{Isotopologue ratios and molecular ratios.}
\begin{tabular}{llllllll}
\hline 
\hline
      & Extended  & Compact  &  High-velocity &          &  Hot  &   20.5        & Solar \\
Ratio & ridge     & ridge    &  plateau       &  Plateau & core  & km s$^{-1}$   & isotopic \\ 
      & (ER)      & (CR)    &  (HVP)           &  (PL)    & (HC1) &  component    & abundance\\    
\hline
\hline 
Isotopologues ratios\\
\hline
\hline
  SO$_{2}$/$^{34}$SO$_{2}$                & 2$\pm$1       & 2$\pm$1    & 20$\pm$13     & 16$\pm$10      &  10$\pm$8    &  5$\pm$3   & 23 \\  
  SO$_{2}$/$^{33}$SO$_{2}$                & 7$\pm$4       & 17$\pm$10  & 137$\pm$90    & 19$\pm$13      &  29$\pm$19   &  20$\pm$14 & 127 \\ 
  SO$_{2}$/SO$^{18}$O                     & 12$\pm$8      & 40$\pm$26  & 151$\pm$100   & 164$\pm$100    &  67$\pm$48   &  ...         & 500 \\ 
  SO$_{2}$/SO$^{17}$O                     & 34$\pm$27    & 180$\pm$110 & 1300$\pm$900  & 334$\pm$200    &  111$\pm$80  &  ...         & 2625 \\ 
  $^{34}$SO$_{2}$/$^{33}$SO$_{2}$         & 3$\pm$2       & 7$\pm$4    & 7$\pm$4       & 1.2$\pm$0.8    &  3$\pm$2     &  4$\pm$3   & 5.5\\  
  $^{34}$SO$_{2}$/SO$^{18}$O              & 5$\pm$3      & 17$\pm$11   & 8$\pm$5       & 10$\pm$6       &  7$\pm$5     &  ...         & ...\\ 
  $^{33}$SO$_{2}$/SO$^{18}$O              & 2$\pm$1       & 2$\pm$1    & 1.1$\pm$0.7   & 8$\pm$6        &  2$\pm$1     &  ...         & ...\\                
\hline
   SO/$^{34}$SO        & 3$\pm$1     & \textbf{...}  & 11$\pm$5       & 1.7$\pm$0.8    & 4$\pm$2        &  1.4$\pm$0.6          & 23\\  
   SO/$^{33}$SO        & ...         & \textbf{...}  & 180$\pm$80     & 10$\pm$5       & 26$\pm$14      &  8$\pm$4              & 127\\
   SO/S$^{18}$O        & ...         & \textbf{...}  & 188$\pm$90     & 50$\pm$30      & 18$\pm$9       &  ...                  & 500 \\
   $^{34}$SO/$^{33}$SO & ...         & ...           & 16$\pm$7       & 6$\pm$3        & 6$\pm$3        &  6$\pm$3              & 5.5 \\
   $^{34}$SO/S$^{18}$O & ...         & ...           & 17$\pm$8       & 30$\pm$15      & 4$\pm$2        &  ...                  & ... \\
   $^{33}$SO/S$^{18}$O & ...         & ...           & 1.1$\pm$0.6    & 5$\pm$3        & 0.7$\pm$0.5    &  ...                  & ... \\
\hline
\hline
Molecular ratios\\
\hline
   SO/SO$_{2}$                  & 0.08$\pm$0.05     & \textbf{...}    & 0.3$\pm$0.2    & 0.5$\pm$0.3       & 0.09$\pm$0.06   & 29$\pm$16  & ...\\
   $^{34}$SO/$^{34}$SO$_{2}$    & 0.07$\pm$0.04     & \textbf{...}    & 0.6$\pm$0.3    & 5$\pm$3           & 0.2$\pm$0.1     & 97$\pm$49  & ...\\
   $^{33}$SO/$^{33}$SO$_{2}$    & ...               & ...             & 0.3$\pm$0.1    & 1.0$\pm$0.6       & 0.10$\pm$0.05   & 71$\pm$40  & ...\\
   S$^{18}$O/SO$^{18}$O         & ...               & ...             & 0.3$\pm$0.1    & 2$\pm$1           & 0.3$\pm$0.2     & ...        & ...\\    
\hline
\hline
\end{tabular}
\label{table:abundancesI}
\end{table*}

\begin{table*}
\caption{Molecular abundances, $X$.}
\begin{center} 
\centering
\begin{tabular}{lllllll}
\hline 
\hline
Region & Species & $X$ \\
 &  & ($\times$10$^{-8}$) \\
\hline 
Extended       &  SO & 0.02\\
Ridge$^{(a)}$    &  SO$_{2}$ & 0.31\\

\hline
Compact        &  SO  & ...\\
Ridge$^{(b)}$    &  SO$_{2}$ & 1.60\\

\hline
               &  SO  & 2.38\\
Plateau$^{(c)}$  &  SO$_{2}$ & 4.76\\

\hline
High           &  SO  & 72.5\\
velocity       &  SO$_{2}$ & 210\\
Plateau$^{(d)}$  \\
\hline
Hot            &  SO  & 2.14\\
core$^{(e)}$     &  SO$_{2}$ & 23.8\\
\hline
20.5 km s$^{-1}$ &  SO  & 5.00\\
component$^{(f)}$  &  SO$_{2}$ & 0.17\\
\hline 
\hline
\end{tabular}
\label{table:molecular_abundances}
\end{center} 

\tablefoot{Derived molecular abundances, $X$, assuming:\\ 
\tablefoottext{a}{$N$$_{\mathrm{H_2}}$=7.5$\times$10$^{22}$ cm$^{-2}$,}
\tablefoottext{b}{$N$$_{\mathrm{H_2}}$=7.5$\times$10$^{22}$ cm$^{-2}$,}
\tablefoottext{c}{$N$$_{\mathrm{H_2}}$=2.1$\times$10$^{23}$ cm$^{-2}$,}
\tablefoottext{d}{$N$$_{\mathrm{H_2}}$=6.2$\times$10$^{22}$ cm$^{-2}$,}
\tablefoottext{e}{$N$$_{\mathrm{H_2}}$=4.2$\times$10$^{23}$ cm$^{-2}$,}
\tablefoottext{f}{$N$$_{\mathrm{H_2}}$=1.0$\times$10$^{23}$ cm$^{-2}$.}
}
\end{table*}

\begin{table*}
\caption{Column density upper limits for undetected sulfur-bearing molecules in Orion KL.}             
\begin{center}
\begin{tabular}{c c c c}     
\hline      
Molecule & Column density            & Dipole  & References for spectroscopic \\ 
         & $\le$$N$$\times$10$^{14}$ & moment  & constants          \\ 
         & (cm$^{-2}$)               &  (D)    &          \\ 
\hline
   SO$^{+}$           & 2.5  & 2.30                                               & 1        \\  
   ($cis$)-HOSO$^{+}$ & 0.36 & $\mu$$_{a}$=1.74 $\mu$$_{b}$=0.49                  & 2        \\
   SSO                & 7.6  & $\mu$$_{a}$=0.87 $\mu$$_{b}$=1.18                  & 3, 4   \\
   OSiS               & 0.63 & 1.47                                               & 5        \\
   S$_{3}$            & 10   & 0.51                                               & 6        \\
   S$_{4}$            & 7.0  & 0.87                                               & 6        \\
   H$_{2}$SO$_{4}$    & 1.3  & $\mu$$_{c}$=2.96                                   & 7        \\
   CH$_{3}$SOCH$_{3}$ & 1.0 & $\mu$$_{b}$=3.94 $\mu$$_{c}$=0.4                    & 8, 9  \\
   H$_{2}$CSO         & 0.3 & $\mu$$_{a}$=2.95 $\mu$$_{b}$=0.50                   & 10, 11 \\  
   HNSO               & 4.1  & $\mu$$_{a}$=0.89 $\mu$$_{b}$=0.18                  & 12, 13 \\
   o-H$_{2}$S$_{2}$   & 1.6  & 0.69                                               & 14       \\ 
   SSH                & 0.71 & $\mu$$_{a}$=1.16 $\mu$$_{b}$=0.83                  & 1       \\
   CH$_{3}$SSH        & 2.6  & $\mu$$_{a}$=1.08 $\mu$$_{b}$=1.22 $\mu$$_{c}$=0.76 & 15       \\
   ($tr$)-HCSSH       & 0.36 & 1.53                                               & 16       \\
\hline 
\label{table:upper_limits}                 
\end{tabular}
\end{center}
\tablebib{ (1) CDMS catalog; (2) Lattanzi et al. (2011); (3) Thorwirth et al. (2006); (4) Meschi et al. (1959); (5) Thorwirth et al. (2011); (6) Thorwirth et al. (2005); (7) Sedo et al. (2008); (8)  Dreizler \& Dendl (1964); (9) Margules et al. (2010); (10) Joo et al. (1995); (11) Penn \& Olsen (1976); (12) Joo et al. (1996); (13) Kirchhoff (1969); (14) Behrend et al. (1990); (15) Tyblewski et al. (1997); (16) Bak et al. (1978).}\\
\end{table*}

\pagebreak

\begin{figure}
   \centering 
   \includegraphics[angle=-90,width=7.7cm]{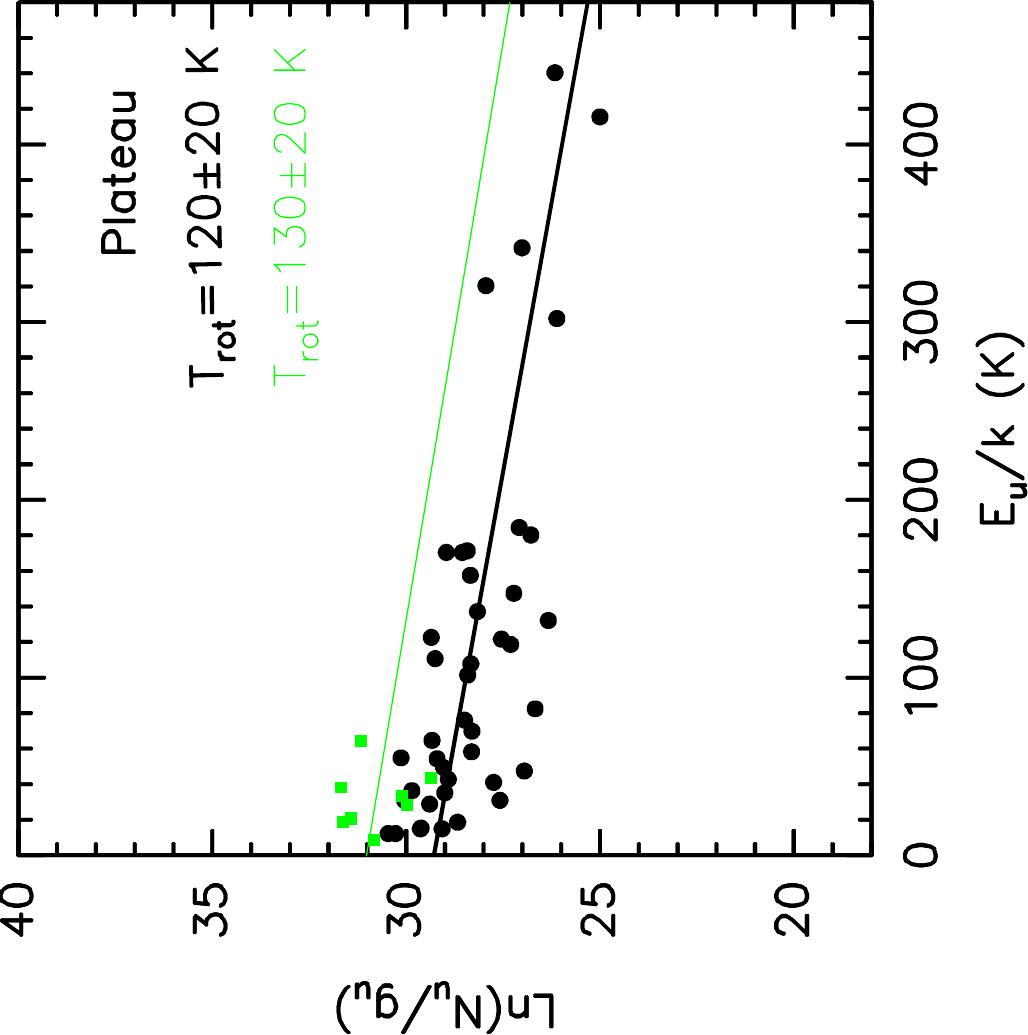}
   \includegraphics[angle=-90,width=7.7cm]{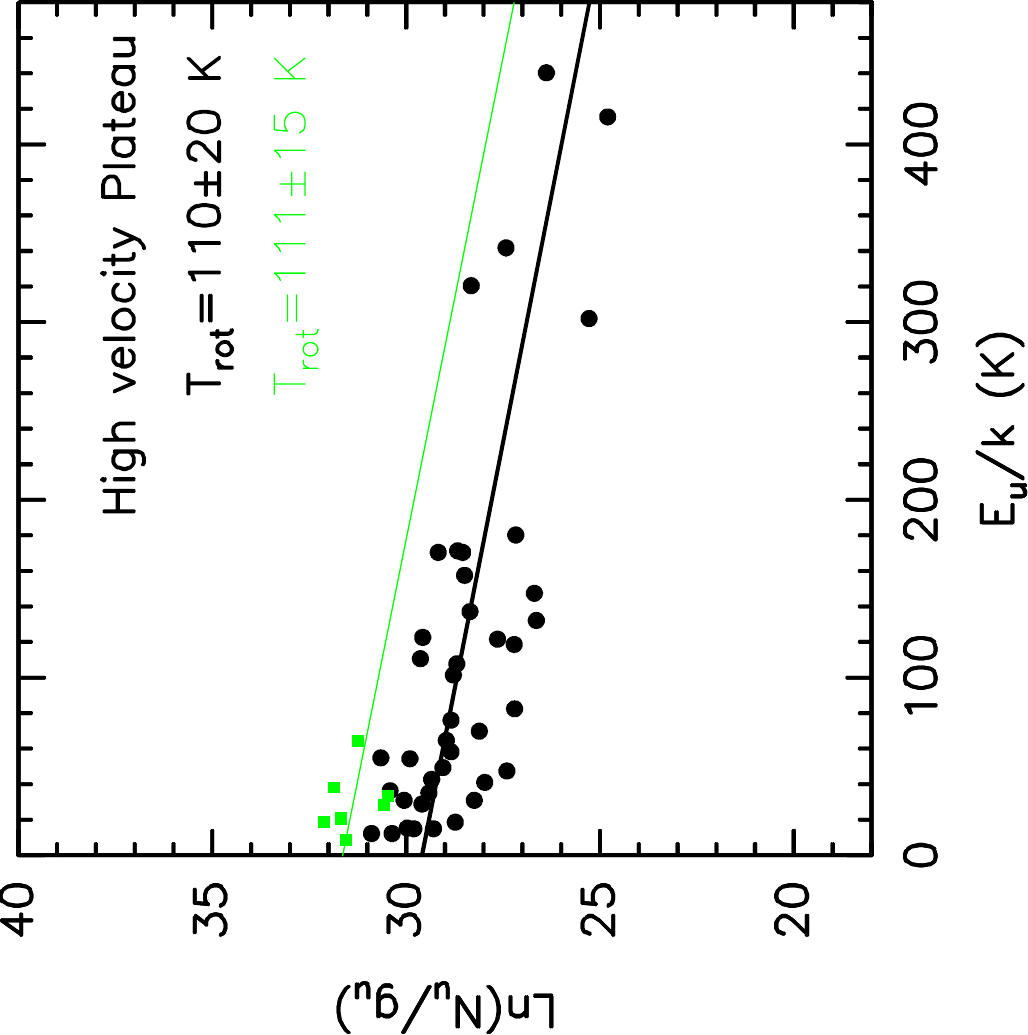}
   \includegraphics[angle=-90,width=7.7cm]{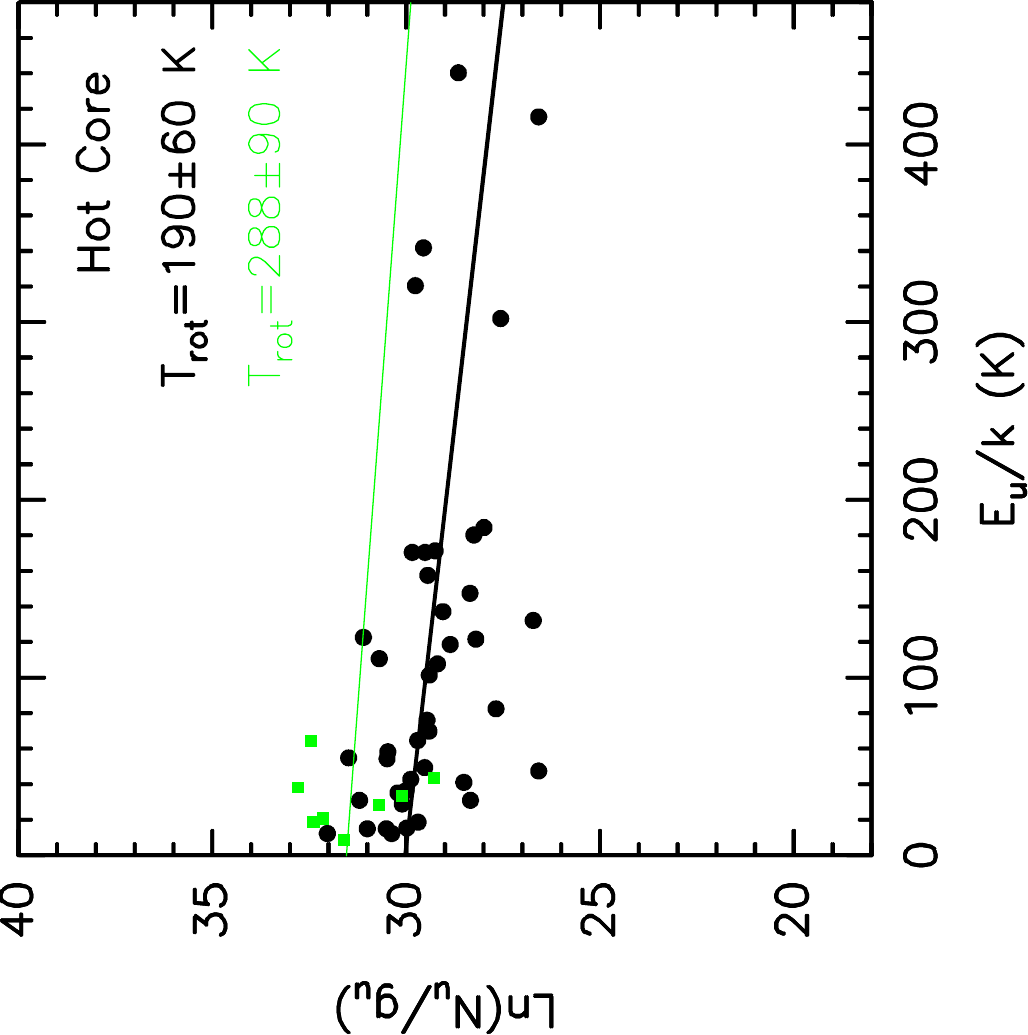}
   \caption{Rotational diagrams for the plateau, high-velocity plateau, and hot core components. Black dots for SO$_{2}$ and green dots for SO. The black and green lines are the best linear fits to the SO$_{2}$ and SO points, respectively.}
    \label{figure:rotational diagrams I}
   \end{figure}

\begin{figure}
   \centering 
    \includegraphics[angle=-90,width=7.8cm]{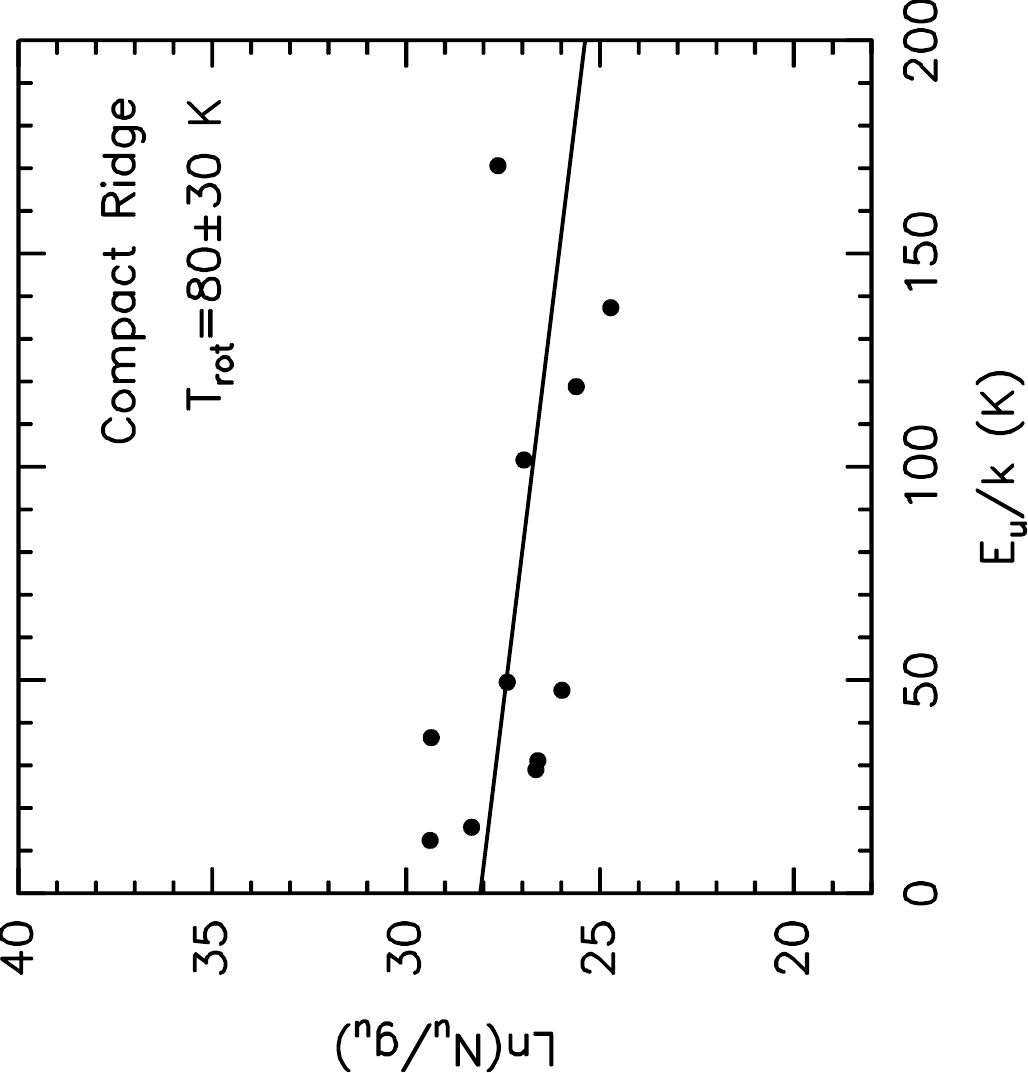}
    \includegraphics[angle=-90,width=7.8cm]{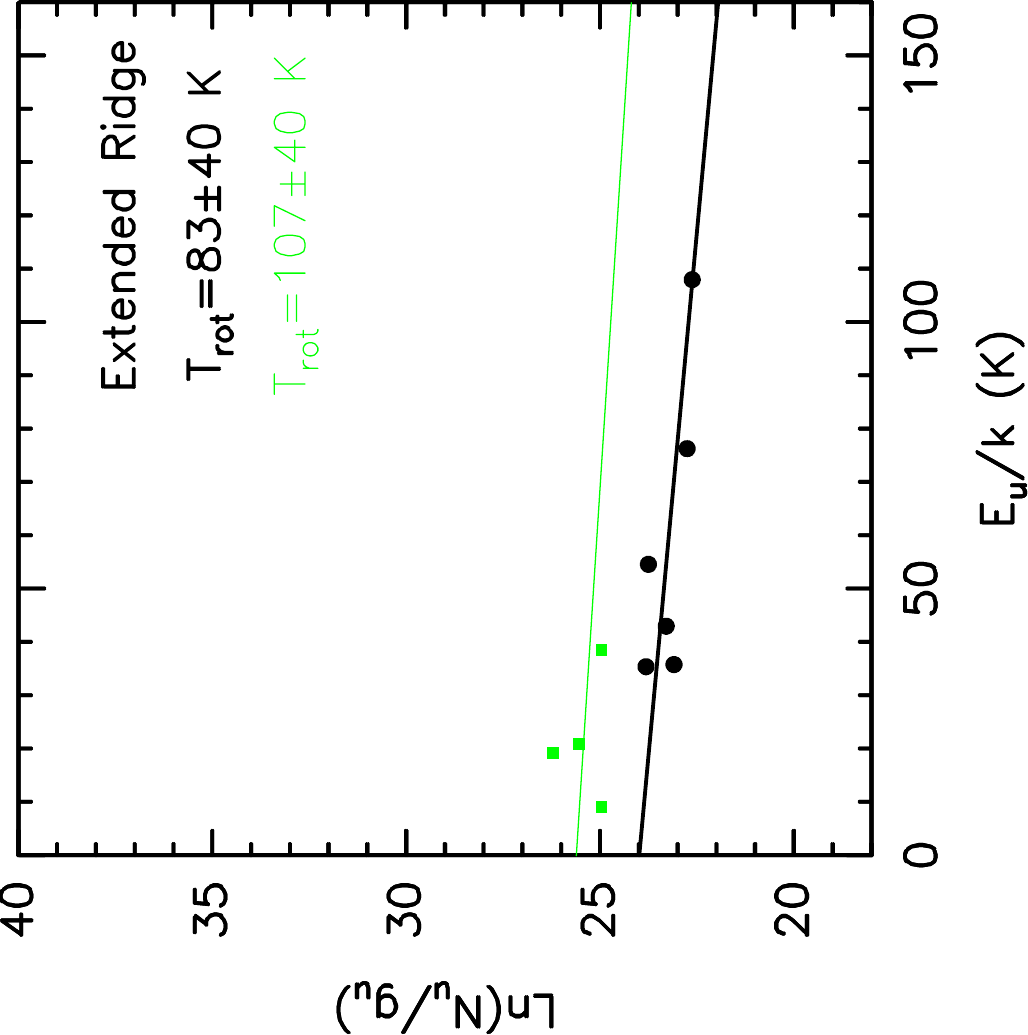}
    \includegraphics[angle=-90,width=7.8cm]{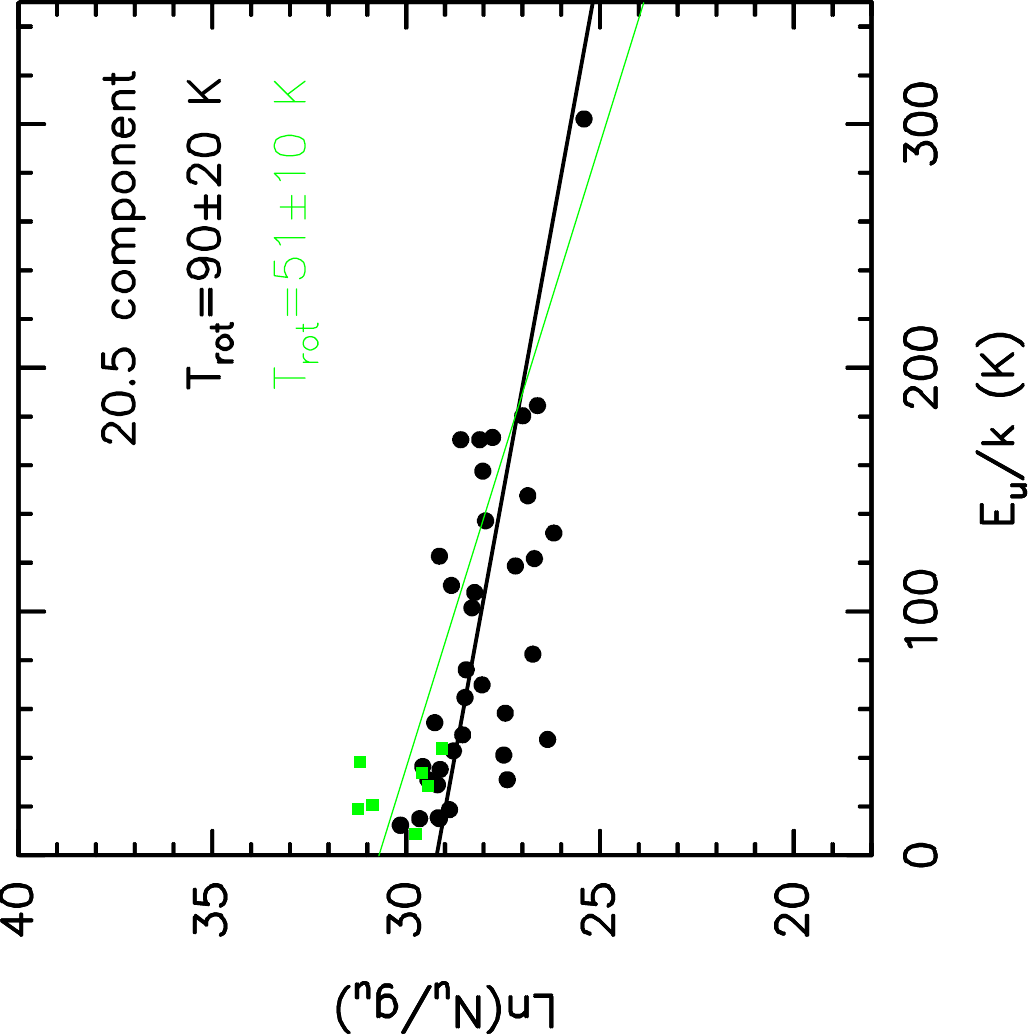}
   \caption{Rotational diagrams for the compact ridge, extended ridge, and 20.5 km s$^{-1}$ component. Black dots for SO$_{2}$ and green dots for SO. 
The black and green lines are the best linear fits to the SO$_{2}$ and SO points, respectively.}
   \label{figure:rotational diagrams II}
   \end{figure}

 \begin{figure*}
   \centering
   \includegraphics[angle=0,width=16cm]{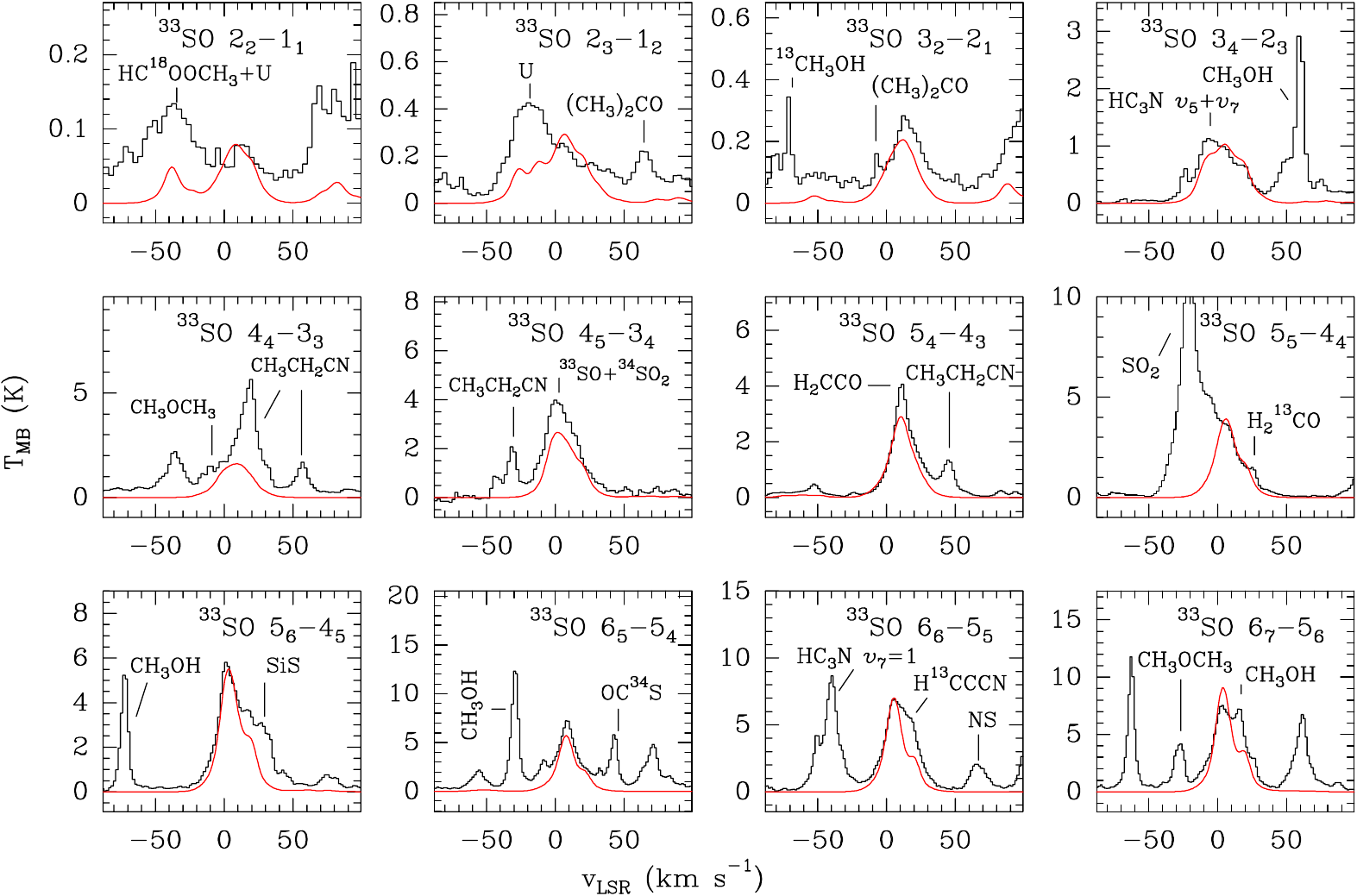}
   \caption{Observed lines of $^{33}$SO (black histogram) and best fit LVG model results (red).}                              
   \label{figure:$^{33}$SO LVG}
   \end{figure*}

 \begin{figure*}
   \centering
   \includegraphics[angle=0,width=15cm]{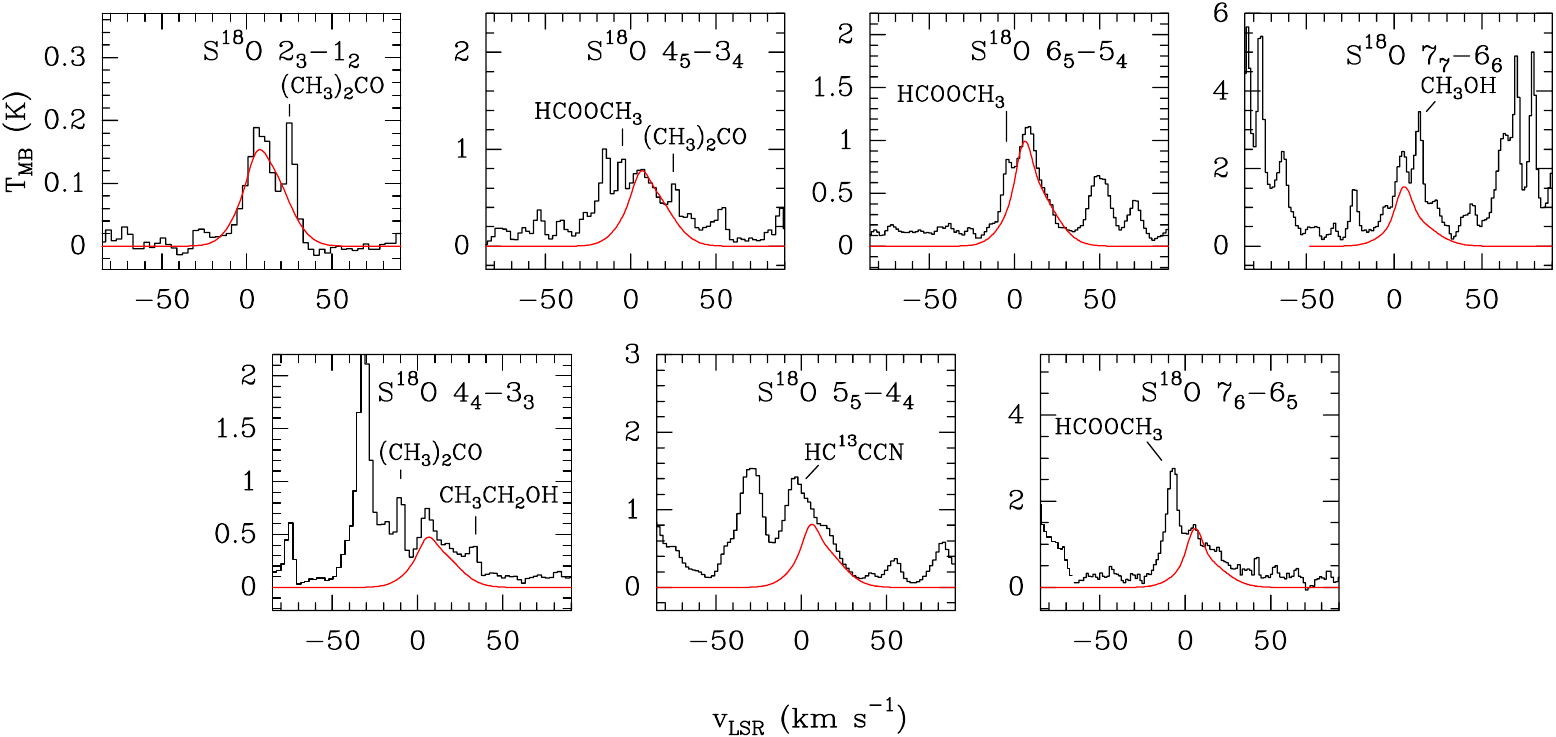}
   \caption{Observed lines of S$^{18}$O (black histogram) and best fit LVG model results (red).}
   \label{figure:S$^{18}$O LVG}
   \end{figure*}

 \begin{figure*}
   \centering
   \includegraphics[angle=0,width=16.5cm]{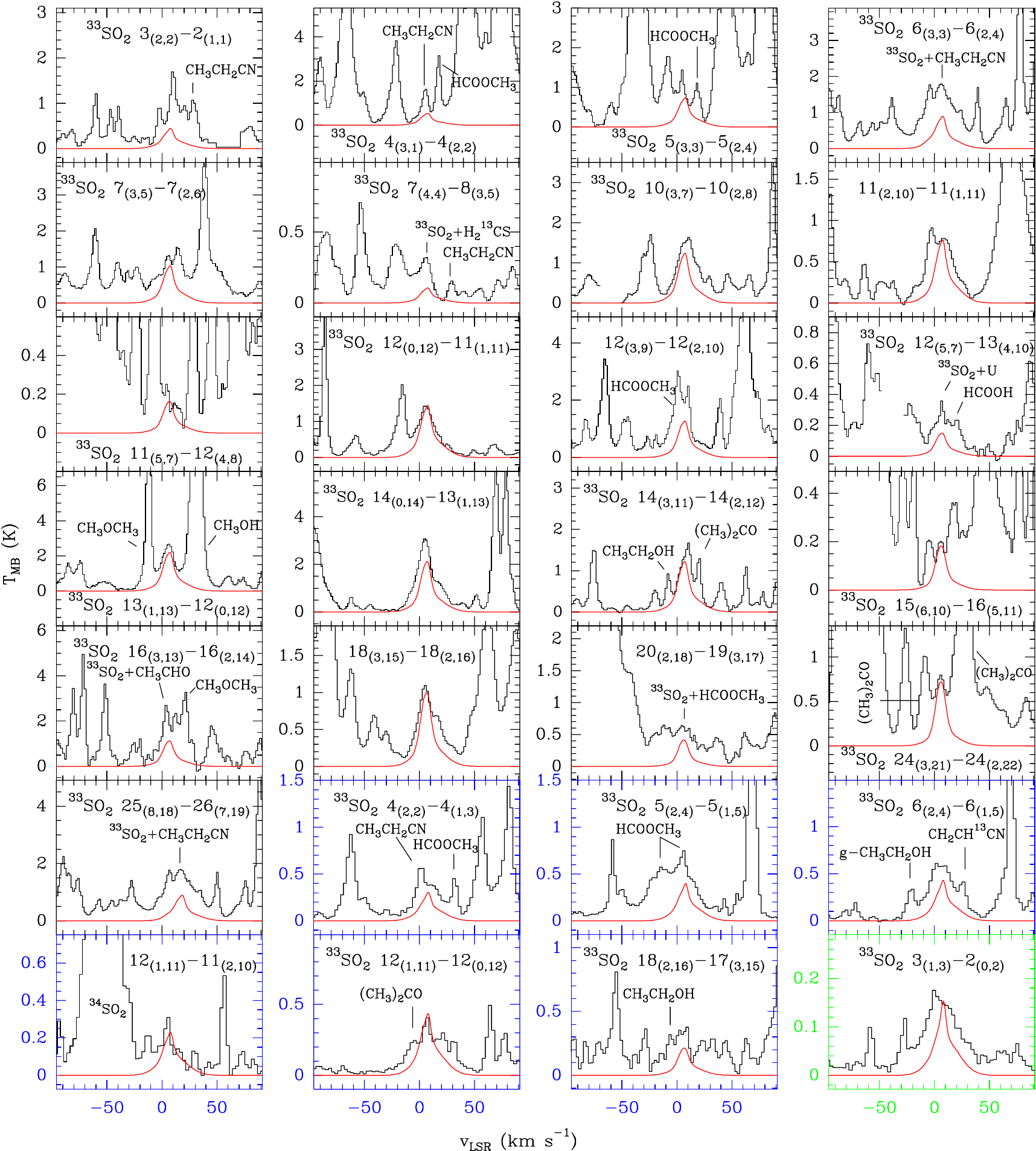}
   \caption{Observed lines of $^{33}$SO$_2$ (black histogram) and best fit LTE model (red). Boxes in black, blue, and green correspond to frequencies at 1.3, 2, and 3 mm, respectively.}
   \label{figure:$^{33}$so2 LVG}
   \end{figure*}

 \begin{figure*}
   \centering
   \includegraphics[angle=0,width=16.5cm]{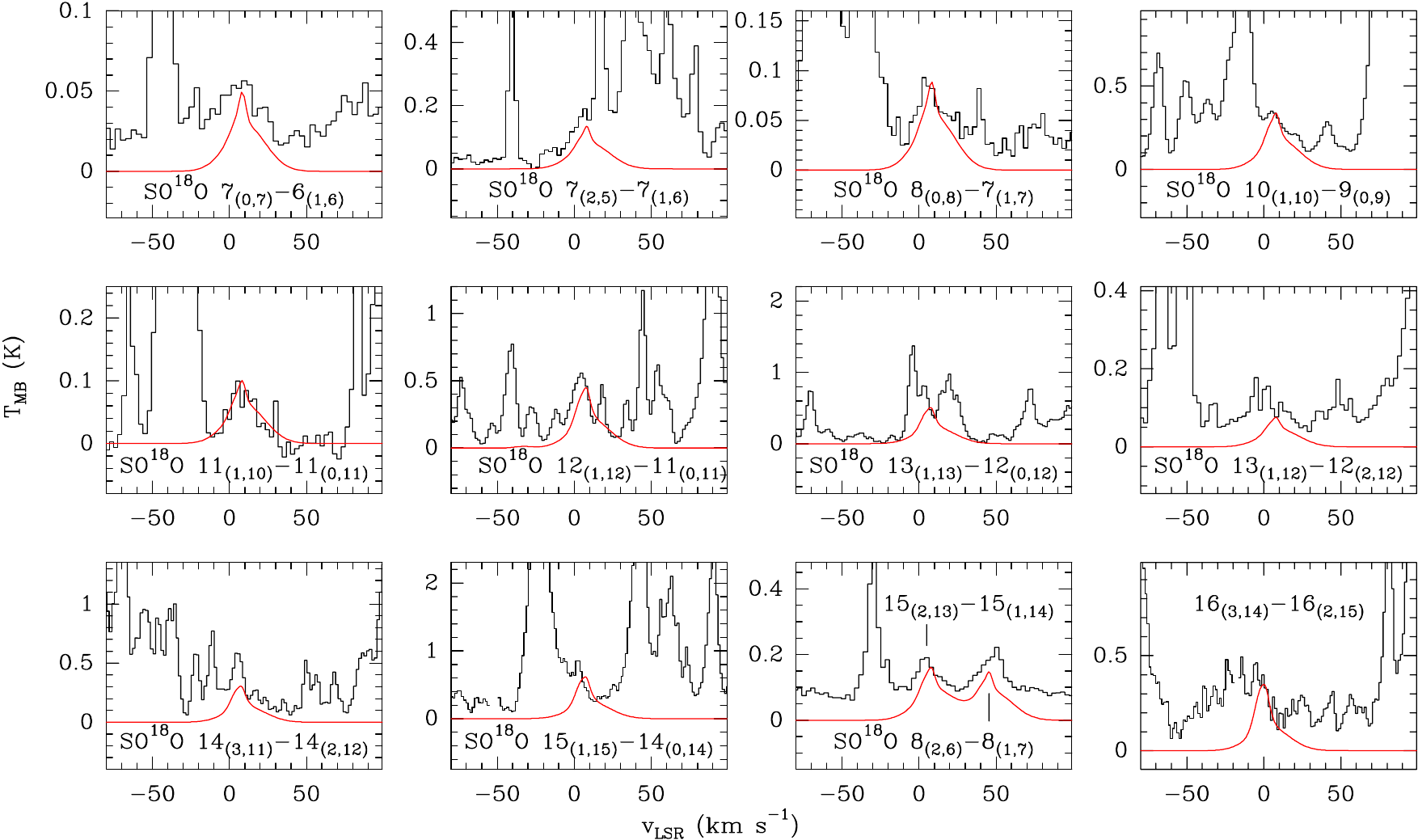}
   \caption{Observed lines of SO$^{18}$O (black histogram) and best fit LTE model (red).}
   \label{figure:SO$^{18}$O LVG}
   \end{figure*}

 \begin{figure*}
   \centering
   \includegraphics[angle=0,width=16.5cm]{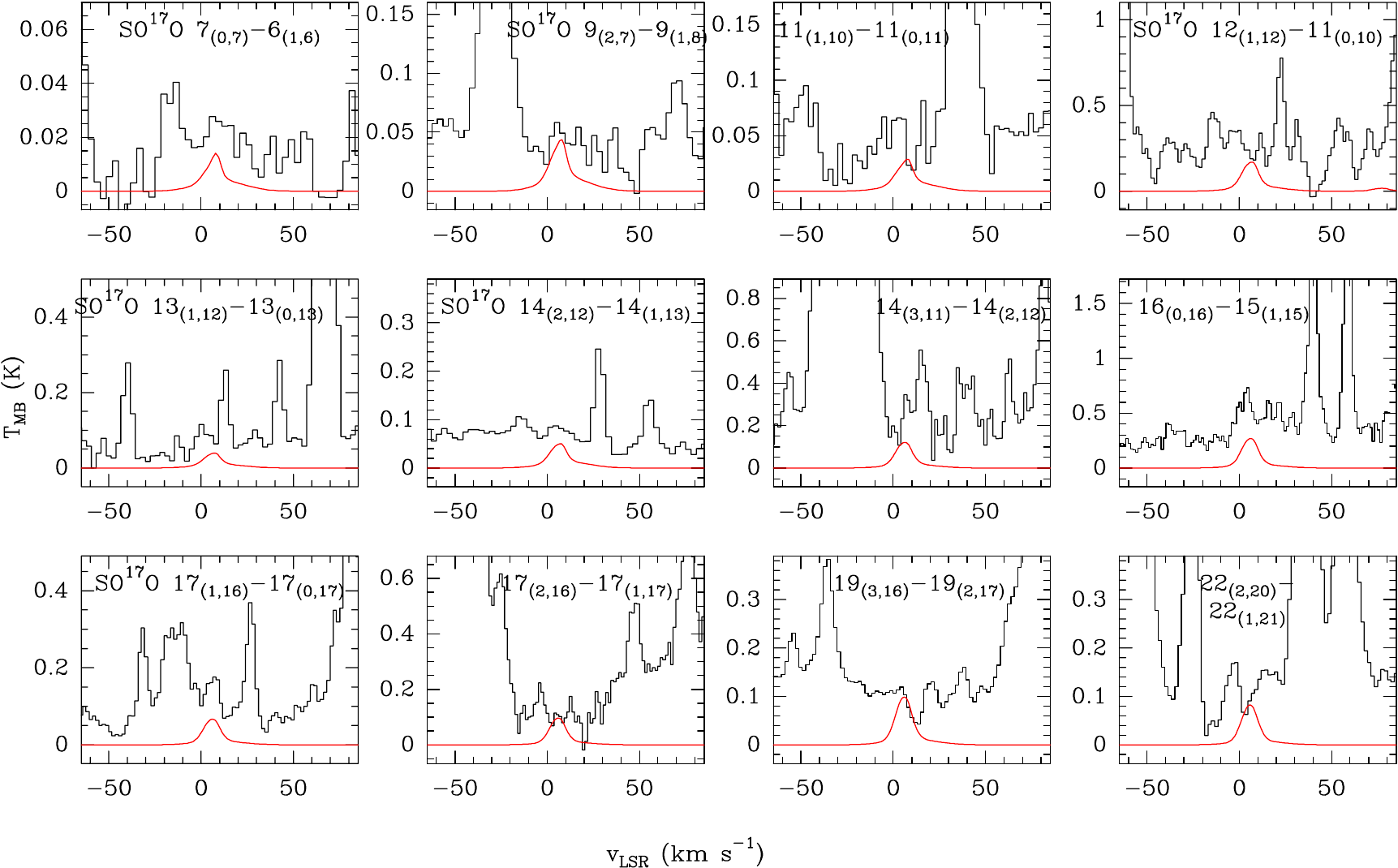}
   \caption{Observed lines of SO$^{17}$O (black histogram) and best fit LTE model (red).}
   \label{figure:SO$^{17}$O_LVG}
   \end{figure*}

 \begin{figure*}
   \centering
   \includegraphics[angle=0,width=15.0cm]{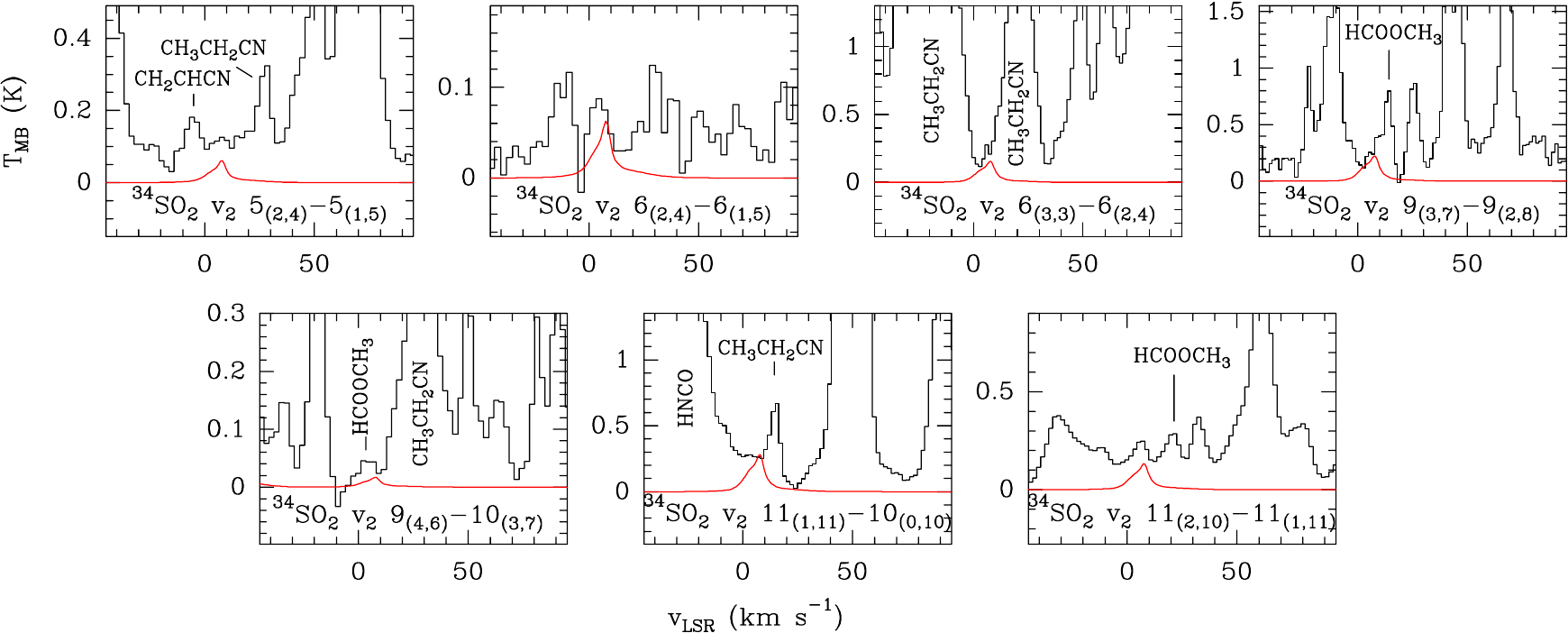}
   \caption{Observed lines of $^{34}$SO$_{2}$ $\nu$$_{2}$=1 (black histogram) and best fit LTE model (red).}
   \label{figure:34so2v2 LVG}
   \end{figure*}

\pagebreak


\longtab{8}{
\begin{longtable}{lrlrlrlrl}
\caption{Observed lines of SO and its isotopologues. \label{table:tab_so}}\\
\hline\hline                      
Species & Transition		    & Predicted   &  $S$$_{\mathrm{ij}}$ & $E$$_{\mathrm{u}}$    & Observed  & Observed       & Observed   &  Blended \\
        & $N$$_{J}$-$N$'$_{J'}$     & freq.       &                      &  (K)              & freq.     & $v$$_{\mathrm{LSR}}$    & $T^*_{A}$  &   with \\
        &                           &  (MHz)      &                      &                       & (MHz)     &  (km s$^{-1}$) & (K)        &         \\
\hline
\endfirsthead
\caption{continued.}\\
\hline\hline
Species & Transition		    & Predicted   &  $S$$_{\mathrm{ij}}$ & $E$$_{\mathrm{u}}$  & Observed    & Observed                  & Observed      &  Blended \\
        & $N$$_{J}$-$N$'$_{J'}$     & freq.       &                      &  (K)            & freq.     & $v$$_{\mathrm{LSR}}$    & $T^*_{A}$  &   with   \\
        &                           &  (MHz)      &                      &                 & (MHz)     &  (km s$^{-1}$) & (K)        &         \\

\hline
\endhead
\hline
\endfoot
SO & 2$_{2}$-1$_{1}$          &  86093.958 &  1.50   & 19.3 & 86094.4  & 7.4 & 11.73 &\\
SO & 2$_{3}$-1$_{2}$          &  99299.887 &  2.93   & 38.6 & 99300.4  & 7.4 & 22.52 &\\
SO & 5$_{4}$-4$_{4}$          & 100029.550 &  0.36   & 38.6 & 100030.5 & 6.2 & 5.90 & CH$_{3}$CH$_{2}$CN\\
SO & 3$_{2}$-2$_{1}$          & 109252.181 &  1.51   & 21.1 & 109252.5 & 8.1 & 18.38 &\\
SO & 6$_{5}$-5$_{5}$          & 136634.660 &  0.28   & 50.7 & 136636.4 & 5.2 & 7.98 & CH$_{3}$OCH$_{3}$\\
SO & 3$_{4}$-2$_{3}$          & 138178.654 &  3.94   & 15.9 & 138178.9 & 8.5 & 29.4 &\\
SO & 4$_{3}$-3$_{2}$          & 158971.811 &  2.69   & 28.7 & 158972.6 & 7.5 & 30.99 &\\
SO & 4$_{4}$-3$_{3}$          & 172181.403 &  3.75   & 33.8 & 172182.7 & 6.7 & 39.33 &\\
\hline
\end{longtable}
\tablefoot{Emission lines of SO and its isotopologues in the
frequency range of the 30-m Orion KL survey. Column 1 indicates the
species, Col.2 the quantum numbers of the line transition, Col. 3 gives the
assumed rest frequencies, Col. 4 the line strength, 
Col. 5 the energy of the upper level, Col. 6 observed frequency
assuming a $v$$_{\mathrm{LSR}}$ of
9.0 km s$^{-1}$, Col. 7 the observed radial velocities, Col. 8 the peak line antenna temperature, and Col. 9 the blended species.}
}

\longtab{9}{
\begin{longtable}{lrlrlrl}
\caption{Observed lines of SO$_2$, its isotopologues, and vibrationally excited states\label{table:tab_so2}.}\\
\hline\hline 
Species/Transition		& Predicted  & $S$$_{\mathrm{ij}}$ & $E$$_{\mathrm{u}}$ & Observed   & Observed                & Observed   \\
$J_{K_a,K_c}-J'_{K'_a,K'_c}$    & freq.      &                     & (K)                & freq.      &  $v$$_{\mathrm{LSR}}$   & $T^*_{A}$  \\
                                &  (MHz)     &                     &                    & (MHz)      &  (km s$^{-1}$)          & (K)        \\
\hline
\endfirsthead
\caption{continued.}\\
\hline\hline
Species/Transition		& Predicted  & $S$$_{\mathrm{ij}}$ & $E$$_{\mathrm{u}}$ & Observed   & Observed                & Observed \\
$J_{K_a,K_c}-J'_{K'_a,K'_c}$    & freq.      &                     & (K)   & freq.      & $v$$_{\mathrm{LSR}}$    & $T^*_{A}$ \\
                                &  (MHz)     &                     &       & (MHz)      &  (km s$^{-1}$)          & (K)       \\
\hline
\endhead
\hline
\endfoot
SO$_2$ 34$_{8,26}$-35$_{7,29}$   &  82409.542 & 5.20 &  704.4 &  82410.5                      & 5.5 &0.08\\ 
SO$_2$ 13$_{4,10}$-14$_{3,11}$   &  82951.936 & 1.93 &  123.0 &  82952.5                      & 7.0 &2.18\\ 
SO$_2$ 8$_{1,7}$-8$_{0,8}$       &  83688.092 & 6.38 &   36.7 &  83688.5                      & 7.5 &11.0\\ 
SO$_2$ 32$_{5,27}$-31$_{6,26}$   &  84320.877 & 5.05 &  549.4 &  84321.5                      & 6.8 &0.25\\ 
SO$_2$ 43$_{7,37}$-42$_{8,34}$   &  85247.002 & 6.77 &  992.8 &  85248.5                      & 3.7 &0.03\\ 
SO$_2$ 39$_{9,31}$-40$_{8,32}$   &  86153.761 & 5.97 &  916.2 &  \tablefootmark{1}            & ... &... \\ 
SO$_2$ 8$_{3,5}$-9$_{2,8}$       &  86639.090 & 1.13 &   55.2 &  86639.4                      & 7.9 &2.65\\ 
SO$_2$ 20$_{2,18}$-21$_{1,21}$   &  86828.940 & 0.27 &  207.8 &  86830.4\tablefootmark{2}     & 4.0 &0.25\\
\hline
\end{longtable}
\tablefoot{Emission lines of SO$_2$, its isotopologues, and its
vibrationally excited states present in the
frequency range of the 30-m Orion KL survey. Column 1 indicates the
species and the quantum numbers of the line transition, Col. 2 gives the
assumed rest frequencies, Col. 3 the line strength, 
Col. 4 the energy of the upper level, Col. 5 observed frequency
assuming a $v$$_{\mathrm{LSR}}$ of
9.0 km s$^{-1}$, Col. 6 the observed radial velocities, and 
Col. 7 the peak line antenna temperature.\\
\tablefoottext{1}{Blended with HCOOCH$_3$ $\nu_t$=1.}                          
\tablefoottext{2}{Blended with CH$_2$DCN.}                                     
\tablefoottext{3}{Blended with HNCO.}                                          
\tablefoottext{4}{Blended with CH$_3$CH$_2$CN.}                                
\tablefoottext{5}{Blended with HCOOCH$_3$.}                                    
\tablefoottext{6}{Blended with CH$_3$CH$_2$CN $\nu_{13}$/$\nu_{21}$.}          
\tablefoottext{7}{Blended with CH$_3$OCH$_3$.}                                 
\tablefoottext{8}{Blended with the previous transition.}                       
\tablefoottext{9}{Blended with g$^+$-CH$_3$CH$_2$OH.}                          
\tablefoottext{10}{Blended with CH$_3$C$^{15}$N.}                              
\tablefoottext{11}{Blended with SiS.}                                          
\tablefoottext{12}{Blended with NH$_2$CHO.}                                    
\tablefoottext{13}{Blended with CH$_3$CN.}                                     
\tablefoottext{14}{Blended with OCS.}                                          
\tablefoottext{15}{Blended with H$_2$CO.}                                      
\tablefoottext{16}{Blended with CH$_3$OH.}  
\tablefoottext{17}{Blended with U line.}
\tablefoottext{18}{Blended with $^{33}$SO$_2$.}
\tablefoottext{19}{Blended with CH$_3$$^{13}$CH$_2$CN.}
\tablefoottext{20}{Blended with SO$^{18}$O.}
\tablefoottext{21}{Blended with S$^{18}$O.}
\tablefoottext{22}{Blended with $^{33}$SO.}
\tablefoottext{23}{Blended with $^{34}$SO$_2$.}
\tablefoottext{24}{Blended with CH$_3$CN $\nu_8$=1.}
\tablefoottext{25}{Blended with HC$_3$N $\nu_7$=2.}
\tablefoottext{26}{Blended with CH$_3$CHO.}
\tablefoottext{27}{Blended with NO.}
\tablefoottext{28}{Blended with $^{30}$SiO.}
\tablefoottext{29}{Blended with NS.}
\tablefoottext{30}{Blended with $^{13}$CH$_3$CH$_2$CN.}
\tablefoottext{31}{Blended with $^{13}$CH$_3$OH.}  
\tablefoottext{32}{Blended with CH$_2$CHCN.}   
\tablefoottext{33}{Blended with HC$^{13}$CCN.}
\tablefoottext{34}{Blended with (CH$_3$)$_2$CO.}   
\tablefoottext{35}{Blended with H$_2$CS.}  
\tablefoottext{36}{Blended with t-CH$_3$CH$_2$OH.}  
\tablefoottext{37}{Blended with H$^{15}$NCO.}
\tablefoottext{38}{Blended with HCC$^{13}$CN.}
\tablefoottext{39}{Blended with $^{13}$CH$_3$CN.}   
\tablefoottext{40}{Blended with CH$_2$$^{13}$CHCN.}
\tablefoottext{41}{Blended with CH$_3$CH$_2$C$^{15}$N.}
\tablefoottext{42}{Blended with OC$^{33}$S.}
\tablefoottext{43}{Blended with CH$_3$CH$_2$CN $\nu_{20}$=1.}
\tablefoottext{44}{Blended with SO.}
\tablefoottext{45}{Blended with H$_2$CCO.}
\tablefoottext{46}{Blended with CCH.}
\tablefoottext{47}{Blended with CH$_2$CHCN $\nu_{11}$=1.}
\tablefoottext{48}{Blended with NH$_2$CHO $\nu_{12}$=1.}
\tablefoottext{49}{Blended with SO$_2$.}
\tablefoottext{50}{Blended with H$_2$C$^{34}$S.}
\tablefoottext{51}{Blended with $^{34}$SO.}
\tablefoottext{52}{Blended with H$^{13}$CN.}
\tablefoottext{53}{Blended with H$^{13}$COOCH$_3$.}  
\tablefoottext{54}{Blended with CH$_2$CHCN $\nu_{11}$=2.}  
\tablefoottext{55}{Blended with H$_2$$^{13}$CS.}
\tablefoottext{56}{Blended with HC$_3$N.}
\tablefoottext{57}{Blended with CH$_3$$^{13}$CN.}
\tablefoottext{58}{Blended with g$^-$-CH$_3$CH$_2$OH.}  
\tablefoottext{59}{Blended with SO$_2$ $\nu_2$=1.}
\tablefoottext{60}{Blended with CH$_2$CHCN $\nu_{15}$=1.}
\tablefoottext{61}{Blended with H$_{30}$$\alpha$.}
\tablefoottext{62}{Blended with HC$_3$N $\nu_7$=1.}
\tablefoottext{63}{Blended with HN$^{13}$CO.} 
\tablefoottext{64}{Blended with H$^{13}$CO$^+$.}    
\tablefoottext{65}{Blended with CS $v$=1.}  
\tablefoottext{66}{Blended with HC$_3$N $\nu_6$=1.}
\tablefoottext{67}{Blended with H$_{65}$$\epsilon$.}
\tablefoottext{68}{Blended with HCOO$^{13}$CH$_3$.}
\tablefoottext{69}{Blended with the next transition.}  
\tablefoottext{70}{Blended with CH$_3$CH$_2$$^{13}$CN.}
\tablefoottext{71}{Blended with HNC$^{18}$O.}
\tablefoottext{72}{Blended with DCOOCH$_3$.}
\tablefoottext{73}{Blended with CH$_3$OD.}
\tablefoottext{74}{Blended with SiO.}
\tablefoottext{75}{Blended with H$^{13}$CCCN.}
\tablefoottext{76}{Blended with SHD.}
\tablefoottext{77}{Blended with c-C$_2$H$_4$O.}
\tablefoottext{78}{Blended with CH$_3$OH $\nu_t$=1.}
\tablefoottext{79}{Blended with HCC$^{13}$CN $\nu_7$=1.}
\tablefoottext{80}{Blended with HC$_3$CN $\nu_7$+$\nu_6$.}
\tablefoottext{81}{Blended with HC$^{18}$OOCH$_3$.}
\tablefoottext{82}{Blended with HC$_3$$^{15}$N.}
\tablefoottext{83}{Blended with O$^{13}$CS.}
\tablefoottext{84}{Blended with CO.}
\tablefoottext{85}{Blended with OC$^{34}$S.}
\tablefoottext{86}{Blended with c-C$_3$H$_2$.}
\tablefoottext{87}{Blended with HCOOH.}
\tablefoottext{88}{Blended with C$_3$S.}
\tablefoottext{89}{Blended with g$^+$-g$^-$-CH$_3$CH$_2$OH.}     
\tablefoottext{90}{Blended with $^{13}$CH$_3$CCH.}
\tablefoottext{91}{Blended with HDCS.}
}
}                                                                    

\end{appendix}

\end{document}